\documentclass[longauth]{aa}
\usepackage[varg]{txfonts}
\usepackage{graphicx}
\usepackage[draft=False,colorlinks=true,linkcolor=blue,citecolor=blue,urlcolor=blue,linktoc=all]{hyperref}

\title{Fragmentation, rotation and outflows in the high-mass star-forming region IRAS 23033+5951}
\subtitle{A case study of the IRAM NOEMA large program CORE\thanks{Based on observations carried out with the IRAM NOrthern Extended Millimeter Array
(NOEMA). IRAM  is supported by INSU/ CNRS (France), MPG (Germany), and IGN (Spain).}}

\author{F.~Bosco\inst{\ref{inst:1},\ref{inst:imprs}}
	\and H.~Beuther\inst{\ref{inst:1}}
	\and A.~Ahmadi\inst{\ref{inst:1},\ref{inst:imprs}}
	\and J.~C.~Mottram\inst{\ref{inst:1}}
    	\and R.~Kuiper\inst{\ref{inst:6},\ref{inst:1}}
    	\and H.~Linz\inst{\ref{inst:1}}
    	\and L.~Maud\inst{\ref{inst:7}}
	\and J.~M.~Winters\inst{\ref{inst:3}}
	\and T.~Henning\inst{\ref{inst:1}}
	\and S.~Feng\inst{\ref{inst:4a},\ref{inst:4}}
	\and T.~Peters\inst{\ref{inst:5}}
	\and D.~Semenov\inst{\ref{inst:1},\ref{inst:18}}
	\and P.~D.~Klaassen\inst{\ref{inst:8}}
	\and P.~Schilke\inst{\ref{inst:9}}
	\and J.~S.~Urquhart\inst{\ref{inst:17}}
	\and M.~T.~Beltr{\'a}n\inst{\ref{inst:10}}
	\and S.~L.~Lumsden\inst{\ref{inst:11}}
	\and S.~Leurini\inst{\ref{inst:16}}
	\and L.~Moscadelli\inst{\ref{inst:10}}
	\and R.~Cesaroni\inst{\ref{inst:10}}
	\and {\'A}.~S{\'a}nchez-Monge\inst{\ref{inst:12}}
	\and A.~Palau\inst{\ref{inst:13}}
	\and R.~Pudritz\inst{\ref{inst:14}}
	\and F.~Wyrowski\inst{\ref{inst:9}}
	\and S.~Longmore\inst{\ref{inst:15}}
	\and the~CORE~team
	}

\institute{Max-Planck-Institut f\"ur Astronomie, K\"onigstuhl 17, 69117 Heidelberg, Germany, \email{bosco@mpia.de} \label{inst:1}
	\and {Fellow of the International Max Planck Research School on Astronomy and Cosmic Physics at the University of Heidelberg} \label{inst:imprs}
	\and Institute of Astronomy and Astrophysics, University of Tübingen, Auf der Morgenstelle 10, 72076 Tübingen, Germany \label{inst:6}
	\and Leiden Observatory, Leiden University, 2300 RA Leiden, Netherlands \label{inst:7}
  \and IRAM, 300 rue de la Piscine, Domaine Universitaire, F-38406 Saint Martin d'H{\`e}res, France, \label{inst:3}
  \and East Asian Core Observatories Association \label{inst:4a}
	\and National Astronomical Observatory of China, Datun Road 20, Chaoyang, Beijing, China \label{inst:4}
	\and Max-Planck-Institut f\"ur Astrophysik, Karl-Schwarzschild-Str. 1, 85748 Garching, Germany \label{inst:5}
	\and Department of Chemistry, Ludwig Maximilian University, Butenandtstr. 5-13, 81377 Munich, Germany \label{inst:18}
	\and UK Astronomy Technology Centre, Royal Observatory Edinburgh, Blackford Hill, Edinbugh EH9 3HJ, UK \label{inst:8}
	\and Max-Planck-Institut für Radioastronomie, Auf dem Hügel 69, 53121 Bonn, Germany \label{inst:9}
	\and Centre for Astrophysics and Planetary Science, University of Kent, Canterbury, CT2 7NH, UK \label{inst:17}
	\and INAF - Osservatorio Astrofisico di Arcetri, Largo E. Fermi 5, 50125 Firenze, Italy \label{inst:10}
	\and School of Physics and Astronomy, The University of Leeds, Woodhouse Lane, Leeds LS2 9JT, UK \label{inst:11}
	\and INAF - Osservatorio Astronomico di Cagliari, via della Scienza 5, 09047, Selargius (CA), Italy \label{inst:16}
	\and I. Physikalisches Institut, Universität zu Köln, Zülpicher Strasse 77, 50937 K\"oln, Germany \label{inst:12}
	\and Instituto de Radioastronom{\'i}a y Astrof{\'i}sica, Universidad Nacional Aut{\'o}noma de M{\'e}xico, P.O. Box 3-72, 58090, Morelia, Michoac{\'a}n, M{\'e}xico \label{inst:13}
	\and McMaster University, 1280 MAIN ST W, L8S 4M1 Hamilton, ON, Canada \label{inst:14}
	\and Astrophysics Research Institute, Liverpool John Moores University, 146 Brownlow Hill, Liverpool L3 5RF, UK \label{inst:15}
	}

\date{Received 20 Feb 2019 /  Accepted 8 Jul 2019}

\abstract{The formation process of high-mass stars (> 8\,M$_\sun$) is poorly constrained, particularly, the effects of clump fragmentation creating multiple systems and the mechanism of mass accretion onto the cores.}
{We study the fragmentation of dense gas clumps, and trace the circumstellar rotation and outflows by analyzing observations of the high-mass ($\sim 500\,\mathrm{M}_\sun$) star-forming region IRAS 23033+5951.}
{Using the Northern Extended Millimeter Array ({NOEMA}) in three configurations and the {IRAM} 30-m single-dish telescope at 220\,GHz, we probe the gas and dust emission at an angular resolution of $\sim$0.45\arcsec, corresponding to 1900\,au.}
{In the mm continuum emission, we identify a protostellar cluster with at least four mm-sources, where three of them show a significantly higher peak intensity well above a signal-to-noise ratio of 100. Hierarchical fragmentation from large to small spatial scales is discussed. Two fragments are embedded in rotating structures and drive molecular outflows, traced by $^{13}$CO (2--1) emission. The velocity profiles across two of the cores are similar to Keplerian but are missing the highest velocity components close to the center of rotation, which is a common phenomena from observations like these, and other rotation scenarios are not excluded entirely. Position-velocity diagrams suggest protostellar masses of $\sim 6$ and 19\,M$_\sun$. Rotational temperatures from fitting CH$_3$CN ($12_K - 11_K$) spectra are used for estimating the gas temperature and by that the disk stability against gravitational fragmentation, utilizing Toomre's $ Q $ parameter. Assuming that the candidate disk is in Keplerian rotation about the central stellar object and considering different disk inclination angles, we identify only one candidate disk to be unstable against gravitational instability caused by axisymmetric perturbations.}
{The dominant sources cover different evolutionary stages within the same maternal gas clump. The appearance of rotation and outflows of the cores are similar to those found in low-mass star-forming regions.}


\begin{document}

\maketitle

\section{Introduction}
High-mass stars, with masses exceeding 8~M$_\sun$, contribute a significant fraction of luminosity  of star clusters and galaxies and thus shape their visual appearance \citep[e.g.][]{Motte17}. Also, they play a key role in the internal dynamics of stellar clusters, as they affect the motion of lower-mass stars by tidal interactions, which presumably leads to the removal of low-mass stars from the cluster centers, the so called effect of mass segregation. Still, the formation process of these objects is poorly constrained by observations, which is due to the fact that high-mass stars typically form in a clustered mode and high-mass star-forming regions are on average located at large distances, on the order of a few kpc, limiting the linear spatial resolution \citep{Zinnecker07,Beuther07,Tan14}. Also, they are rare, have short formation time scales and reach the main sequence still deeply embedded in their parental molecular clump.

In the past, different models assuming high-mass star formation (HMSF) to be a consequence of a spherical collapse of a molecular gas clump \citep[e.g.,][]{Kahn74,Wolfire87} suggested a halting of the mass accretion due to high stellar radiation pressure once the mass of the high-mass protostellar object (HMPO) exceeds 40\,M$_\sun$. However, there is consensus in the ongoing discussion that the mass accretion onto the core is not halted immediately by the outward acting stellar radiation pressure. Instead, HMPOs continue the mass accretion via circumstellar disks and the radiation can escape via outflow cavities \citep[e.g.,][]{Yorke02,Arce07,Vaidya09,Krumholz09,Kuiper10,Kuiper11,Kuiper15,Kuiper16,Kuiper18,Klassen16}. \citet{Beltran16} summarize observational properties of accretion disks in high-mass star formation (HMSF) which are embedded in flattened rotating structures ($10^3 - 10^4$\,AU) and thereby circumvent the radiation pressure problem for ongoing mass accretion. This suggests that the formation scenario of high-mass objects is analogous to a {scaled-up version of the low-mass star-forming process}, i.e. with non-spherical mass accretion via circumstellar disks \citep[e.g.][]{Johnston13,Johnston15,Cesaroni14,Cesaroni17}. Such an accretion scenario is presumably traced by an ordered molecular outflow, which is launched from the disk surface \citep[e.g.,][]{Arce07}, as long as the accretion disk is stable. However, there is evidence from simulations \citep[e.g.][]{Peters10a,Peters10c,Klassen16,Rosen16, Harries17,Meyer17,Meyer18} that the rotating structures around high-mass stars are unstable against self-gravity and tend to form spiral arms, which at sufficient local density will fragment to form companion objects \citep[cf. the observations by][]{Ilee18}. Recent work additionally suggests that also episodic accretion events in non-isotropic regions may explain the growth of HMPOs to masses exceeding the assumed mass limit of 40\,M$_\sun$ \citep[e.g.,][]{CarattiOGaratti17,Hunter17,Motte17}. Such episodic accretion events in turn may on the one hand be explained by the accretion of disk fragments, as shown e.g. by \cite{Kratter06} or \cite{Meyer17}. On the other hand, they may also be introduced by infall events of larger scale gas streams \citep[e.g.][]{Gomez14,Vazquez-Semadeni19}. E.g. \cite{Peretto13} observed that HMSF regions tend to reside at the hubs of larger-scale ($\sim1-100$\,pc) filaments, along which material flows into the proto-cluster -- predominantly onto the heaviest cores, which accrete the material competitively \citep[see e.g.][for reviews]{Tan14,Andre14,Motte17}.

The Northern Extended Millimeter Array (NOEMA) large program CORE \citep{Beuther18} addresses open questions on the fragmentation and disk formation during HMSF by investigating a sample of 20 high-mass star-forming regions at high spatial resolution in the cold dust and gas emission. The survey focuses on the early protostellar phase which is accompanied by molecular outflows and accretion disks within the dense cores of the parental clump. The sample sources were selected to have luminosities exceeding $10^4$\,L$_\odot$ and are located within a maximum distance of 6\,kpc, enabling a linear spatial resolution $ \leq 3000$\,au for an angular resolution $ \lesssim 0.5 \arcsec $. As pilot studies, \citet{Beuther12,Beuther13} investigated the HMSF regions NGC~7538~IRS1 and NGC~7538S. The overview paper \citep{Beuther18} presents the source sample, the spectral setup and the goal of the survey, as well as the full continuum data from which they derive the fragmentation statistics in the 1.37\,mm dust continuum emission. Mottram et~al. (subm.) investigate the connection of cores to the extended environments in form of gas inflows by merging the interferometric with single-dish data from the IRAM 30-m telescope for W3IRS4. Furthermore, \citet{Ahmadi18} describe, in detail, the variety of molecular transitions, covered with the survey setup, using the example of the chemically rich HMSF region W3(H$_2$O), primarily focusing on the disk properties. Gieser et~al. (subm.) investigate in detail physical structure and chemical composition in the young, early hot core region AFGL2591 by combining IRAM 30-m and NOEMA data with 1D physical-chemical model.

In this paper, we report the investigation of the high-mass star-forming region \object{IRAS 23033+5951}, also listed as \object{G110.0931-00.0641} in the RMS survey catalogue \citep{Lumsden13}. This target is associated with the Cepheus Far molecular cloud complex \citep{Harju93}. \citet{Lumsden13} derived a kinematic distance of 4.3~kpc from the source velocity of $-53.1$\,km\,s$^{-1}$ and a bolometric luminosity of $L_\mathrm{bol} = 1.7\times10^4 \mathrm{L}_\sun$. We note that we discard the older distance estimate of 3.5\,kpc by \citet{Harju93} which was assumed in previous publications on this region, since this estimate is based only on the association to the Cepheus Far group located at the less precise distance estimate of 3.5\,kpc.

\cite{Maud15} have estimated the clump mass to be $\sim$\,700\,M$_\sun$ based on SCUBA 850\,$\mu$m data \citep{DiFrancesco08} and BOLOCAM data \citep{Ginsburg13}, where \cite{Beuther18} derived a clump mass of $\sim$\,500\,M$_\sun$ from the same SCUBA data. The differences are due to different assumptions on the gas temperature and opacity values. The region is associated with spatially distinct water and methanol masers \citep{Beuther02maser,Schnee09}, and with molecular gas emission \citep[e.g.][]{Wouterloot86}. In particular, \cite{Beuther02outflow} report the detection of a molecular outflow in CO emission.

\citet{Reid08} present interferometric observations obtained from the Berkeley-Illinois-Maryland Association (BIMA) array at 3\,mm wavelength and angular resolution of $\leq 5.5\arcsec$, revealing the fragmentation of the source into two mm clumps, denoted as MMS1 and MMS2. This fragmentation is also seen in SCUBA 850\,$\mu$m data \citep{DiFrancesco08}. \citet{Schnee09} reveal further fragmentation of MMS1 into two major mm sources which we will denote as MMS1a in the north, coincident with an Midcourse Space Experiment (MSX) infrared point source \citep{Reid08}, and MMS1b in the south (cf. Fig.~\ref{fig:continuum_emission}). Emission in the cm regime was detected only towards a small region in MMS1 \citep{Sridharan02,Rodriguez12}, coincident with MMS1a, where \citeauthor{Rodriguez12} report the fragmentation of the 3.6\,cm source into three regions, labeled VLA1--3. They discuss this finding as either condensations of an ionized jet or ultra-compact (UC) \textsc{H\,ii} regions from different ionization sources. Both possibilities indicate the presence of at least one evolved HMPO. However, the conversion of their flux densities yields 8\,GHz luminosities on the order of $\lesssim 1.1 \times 10^{12}$\,W\,Hz$^{-1}$ and a comparison of these values to other cm~sources from Fig.~6 of \cite{Hoare07} shows that these values are an order of magnitude too low to stem from an \textsc{UCH\,ii} region. It therefore suggests that the cm~emission traces a high-mass YSO wind or jet. \citet{Rodriguez12} analyze the positions and velocities of water maser emission towards MMS1b and infer disk rotation. Modeling the data with an inclined disk model, they obtain a radius of $0.03\arcsec$ (135\,au) and a position angle of $65\degr \pm 1\degr$. The best-fit disk is inclined by $83\degr \pm 1\degr$ towards the line of sight and orbits about a central body with a mass of 19\,M$_\sun$. \citet{Schnee09} associate the methanol maser emission at 95\,GHz with the emission peak in MMS2.

This paper is organized as follows. We report on the observations and the data reduction in Sect.~\ref{sec:observations}. The observational results from the continuum and spectral line emission are presented in Sect.~\ref{sec:results}. In Sect.~\ref{sec:PVdiagrams}, we estimate protostellar masses from the disk kinematics and perform extensive tests in Appendix~\ref{sec:discussion_pv_fit}.
We discuss our observational results in the context of clump fragmentation, outflows and disk stability in Sect.~\ref{sec:discussion} and draw our conclusions in Sect.~\ref{sec:conclusion}. We provide a summary sketch of our interpretation of the data in Fig.~\ref{fig:summary}, in the concluding section.

\section{Observations and data reduction}
\label{sec:observations}

\subsection{NOEMA}
We observed the high-mass star-forming region IRAS~23033+5951 during five epochs between June 2014 and March 2016, using NOEMA at Plateau de Bure (France) in the A, B, and D array-configurations, with a phase center at RA 23$^\mathrm{h}$\,05$^\mathrm{m}$\,25\fs00 and Dec. +60\degr\,08\arcmin\,15\farcs49 (J2000.0). These observations are part of the NOEMA large program CORE \citep{Beuther18}. The  projected  baselines  range  from  20\,m up  to  750\,m and an example uv-coverage is presented in the survey overview \citep{Beuther18}. The receivers were tuned to the 1.3\,mm (220\,GHz) band. The key molecular transition lines in this spectral window are summarized in Table~\ref{tab:obs_spec_lines}.
\begin{table*}
	\centering
	\caption{Molecular line transitions analyzed in this paper.}
	\label{tab:obs_spec_lines}
	\begin{tabular}{ccccccc}
	\hline \hline
	Molecule 		& Transition 				& Frequency\tablefootmark{a}  	& Velocity resolution 	& $E_\mathrm{u}/k_\mathrm{B}$ 	& Critical density\tablefootmark{b} 	& rms noise\tablefootmark{c} \\
				& 						& (MHz) 					& (km s$^{-1}$) 		&  (K) & $n_\mathrm{crit}$ (cm$^{-3}$) & (mJy beam$^{-1}$) \\ \hline
	$^{13}$CO\tablefootmark{d}		& 2 -- 1 					& 220398.68 				& 2.7				& 15.9 				& 9.87 $\times 10^{3}$					& 1.99\tablefootmark{d} \\
	H$_2$CO		& 3$_{0,3}$ -- 2$_{0,2}$ 		& 218222.19 				& 0.5 			& 21.0 				& 3.36 $\times 10^{6}$					& 5.29 \\
				& 3$_{2,2}$ -- 2$_{2,1}$ 		& 218475.63 				& 0.5 			& 68.1 				& 2.96 $\times 10^{6}$					& 4.58 \\
	CH$_3$OH	& 4$_{+2,2,0}$ -- 3$_{+1,2,0}$	& 218440.05 				& 0.5				& 45.5 				& 7.81 $\times 10^{7}$					& 4.95 \\
	CH$_3$CN 	& 12$_0$ -- 11$_0$ 			& 220747.26 				& 0.5				& 68.9 				& 4.46 $\times 10^{6}$					& 4.06 \\
				& 12$_1$ -- 11$_1$ 			& 220743.01 				& 0.5 			& 76.0 				& 4.13 $\times 10^{6}$					& 4.06 \\
				& 12$_2$ -- 11$_2$ 			& 220730.26 				& 0.5 			& 97.4 				& 4.19 $\times 10^{6}$					& 4.06 \\
				& 12$_3$ -- 11$_3$ 			& 220709.02 				& 0.5 			& 133.1 				& 4.27 $\times 10^{6}$					& 4.06 \\
	 			& 12$_4$ -- 11$_4$ 			& 220679.29 				& 0.5 			& 183.1 				& 3.96 $\times 10^{6}$					& 4.06 \\
				& 12$_5$ -- 11$_5$ 			& 220641.08 				& 0.5 			& 247.3 				& 3.62 $\times 10^{6}$					& 4.06 \\
				& 12$_6$ -- 11$_6$ 			& 220594.42 				& 0.5 			& 325.8 				& 3.68 $\times 10^{6}$					& 4.06 \\
	DCN			& 3$_{0,0}$ -- 2$_{0,0}$ 		& 217238.54 				& 2.7				& 20.9				& 1.82 $\times 10^{7}$								& 1.76 \\
	SO			& 5$_6$ -- 4$_5$			& 219949.44 				& 2.7				& 35.0				& 2.31 $\times 10^{6}$					& 2.58 \\
	\hline
	\end{tabular}
	\tablefoot{
		\tablefoottext{a}{The rest frequencies were extracted from the Cologne Database for Molecular Spectroscopy (CDMS). }
		\tablefoottext{b}{The critical density was estimated from the approximation $n_\mathrm{crit} \approx A/ \Gamma$ \citep{Shirley15}, for collision rates at $T=100$\,K. Both, the Einstein $ A $ and the  $ \Gamma $ coefficients were taken from the Leiden Atomic and Molecular Database (LAMDA). DCN is approximated by the corresponding values for HCN in the database.}
		\tablefoottext{c}{The rms noise of the images was estimated from line-free channels. The beam size is $0.43\arcsec \times 0.35\arcsec$.}
		\tablefoottext{d}{Only for $^{13}$CO we also imaged the interferometric data merged with the single-dish data. The resulting cube has an rms noise of 2.15\,mJy\,beam$^{-1}$, with a synthesized beam of $0.8\arcsec \times 0.67\arcsec$.}
		}
\end{table*}
The spectral resolution of the broadband correlation units is 1.95\,MHz or 2.7\,km\,s$^{-1}$ at 1.37\,mm and we have eight narrow band units achieving 0.312\,MHz or 0.42\,km\,s$^{-1}$ for a subset of the transitions. The complete set of spectral lines covered with this setup is described in detail in \citet{Ahmadi18} and Mottram et~al. (subm.).
The NOEMA observations were carried out in track-sharing mode with IRAS~23151+5912 and the calibration sources listed in Table~\ref{tab:calibration_sources}.

\subsection{IRAM 30-m telescope}
Furthermore, the region was observed with the IRAM 30-m telescope at Pico Veleta (Spain) in March 2016 in the on-the-fly mode. Merging these single-dish data with the interferometric visibilities yields coverage of the $uv$-plane in the inner 15\,m. This is necessary to recover the extended emission which is filtered out by the interferometric observations. The single dish data by themselves have an angular and spectral resolution of $\sim11\arcsec$ and 0.195\,MHz or 0.27\,km\,s$^{-1}$ at 1.37\,mm. In this work, only the $^{13}$CO~(2 -- 1) data were used for the combination. The process of merging the interferometric with the single-dish data is described in detail in Mottram et~al. (subm.).

\subsection{Data reduction}
The data were calibrated using the \textsc{gildas/clic}\footnote{Grenoble Image and Line Data Analysis Software, \url{www.iram.fr/IRAMFR/GILDAS/}.} software. The respective sources for the calibration process are listed in Table~\ref{tab:calibration_sources}.
\begin{table*}
	\centering
    \caption{Calibration sources for the interferometer data.}
    \label{tab:calibration_sources}
    \begin{tabular}{llllll}
    	\hline \hline
        Observation date	& Array	& $N_\mathrm{ant}$	& \multicolumn{3}{c}{calibration sources}						\\\cline{4-6}
        	&			&		& amplitude \& phase				& bandpass			& flux				\\
        \hline
    	4 Jun 2014	& D			& 5		& 0059+581, J2201+508 				& 0059+581			& MWC349			\\
        12 Mar 2015	& A			& 6		& 0059+581, J2201+508, J0011+707 	& 0059+581, 3C84	& MWC349, LKHA101	\\
        26 Mar 2015	& B			& 7		& J2201+508, J0011+707 				& J0011+707			& MWC349			\\
        25 Jan 2016	& A			& 6		& J2201+508, J0011+707, J2223+628 	& 1749+096			& MWC349			\\
        21 Mar 2016	& B			& 7		& J2201+508, J0011+707 				& 1928+738			& MWC349			\\
        \hline
    \end{tabular}
\end{table*}
The imaging and deconvolution processes were conducted using \textsc{gildas/mapping}. The continuum data were imaged with uniform weighting to obtain a high spatial resolution, where the low number of channels containing spectral line emission were excluded by hand. We applied self-calibration on the continuum data using \textsc{casa} \citep[version 4.7.2,][]{McMullin07}, to reduce the rms noise from 0.46 to 0.28\,mJy\,beam$^{-1}$ \citep[cf.][]{Beuther18}. For the spectral line cubes, we subtracted the continuum emission from the high-resolution narrow band data in the $uv$-domain and resampled to a spectral resolution of 0.5\,km\,s$^{-1}$. The resulting tables were imaged applying uniform weighting (robust weighting parameter of 0.1), and \textsc{clean}ed with the \textsc{hogbom} algorithm. With this procedure, we achieved an angular resolution of $0.45\arcsec \times 0.37\arcsec$ (position angle, PA~47\degr) for the continuum image and $0.43\arcsec \times 0.35\arcsec$ (PA~61\degr) for the spectral line cubes. The continuum image has an rms noise of 0.28\,mJy\,beam$^{-1}$ and the corresponding values for the spectral line cubes are estimated from line-free channels and listed in Table~\ref{tab:obs_spec_lines}. The merged data cube for $^{13}$CO~(2 -- 1)  has an angular resolution of $0.8\arcsec \times 0.67\arcsec$ (PA~52\degr) and a rms noise of 2.15\,mJy\,beam$^{-1}$.

\section{Observational results}
\label{sec:results}
\subsection{Continuum emission at 1.37\,mm}
\label{sec:continuum}

\begin{figure*}
	\centering
	\resizebox{0.9\hsize}{!}{\includegraphics{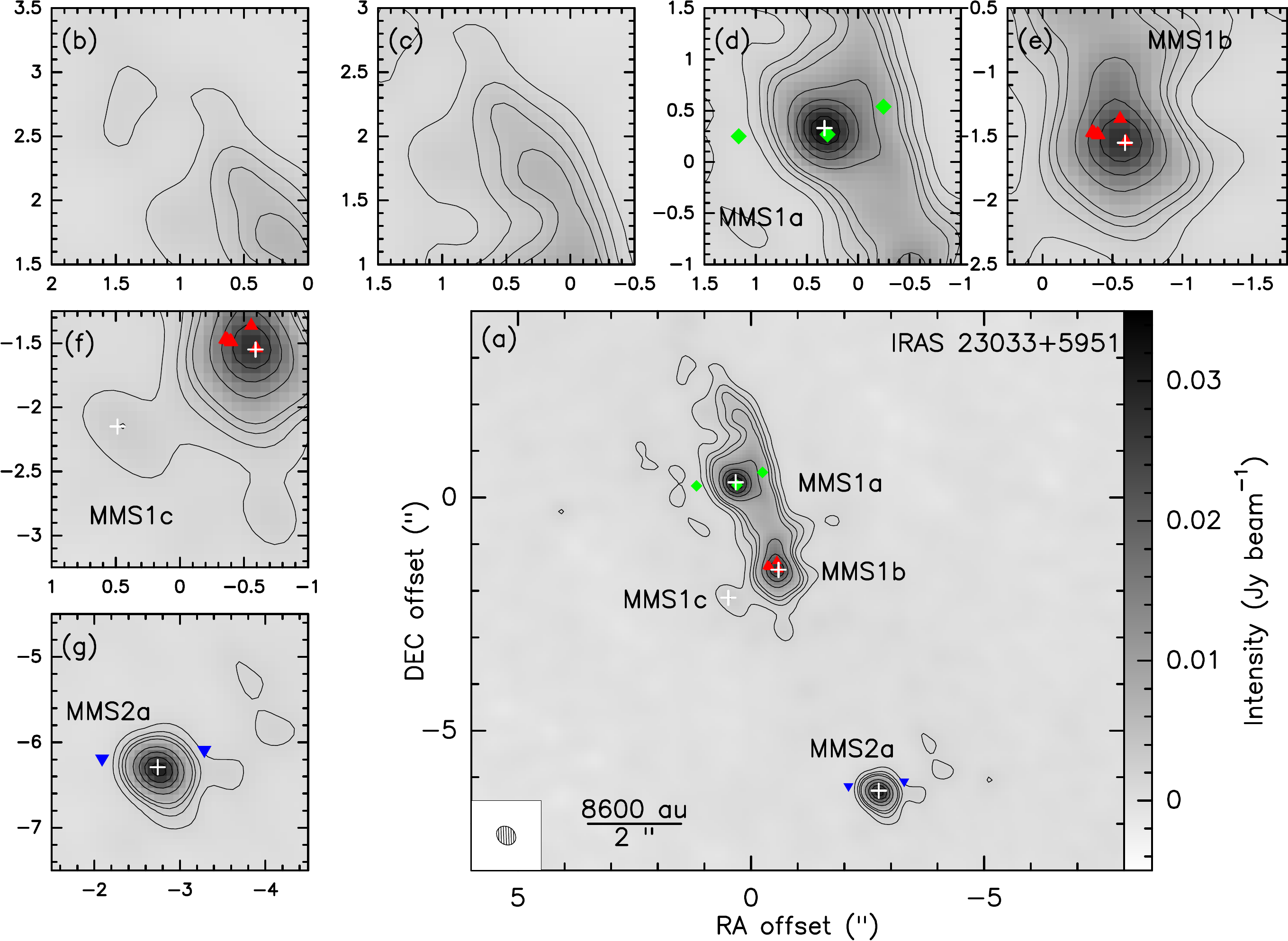}}
	\caption{Continuum emission towards IRAS~23033+5951 at 1.37\,mm, with a phase center at RA 23$^\mathrm{h}$\,05$^\mathrm{m}$\,25\fs00 and Dec. +60\degr\,08\arcmin\,15\farcs49 (J2000.0). The black contours indicate the 5, 10, 15, 20\,$\sigma$ levels and increase further in steps of 20\,$\sigma$, where $\sigma$ = 0.28\,mJy\,beam$^{-1}$. The corresponding negative values are absent in the presented region. The white + and labels mark the mm sources identified by \textsc{clumpfind}. The green, red and blue symbols are cm point sources, H$_2$O masers \citep[both][]{Rodriguez12} and methanol masers \citep{RodriguezGarza17}, respectively. The shaded ellipse in the bottom left shows the synthesized beam of $0.45\arcsec \times 0.37\arcsec$ (PA~47\degr). The small panels zoom in on (\emph{b,c}) the elongation of the core MMS1a, (\emph{d}) MMS1a, (\emph{e}) MMS1b, (\emph{f}) MMS1c and (\emph{g}) MMS2a.}
	\label{fig:continuum_emission}
\end{figure*}

The continuum emission at 1.37\,mm towards IRAS~23033+5951, presented in Fig.~\ref{fig:continuum_emission}, shows three strong mm cores surrounded by a group of smaller structures. Following the approach conducted for the whole CORE sample in \cite{Beuther18}, we analyze this using the \textsc{clumpfind} algorithm \citep{Williams94}, with a detection threshold of 10\,$\sigma = 2.8$\,mJy\,beam$^{-1}$, as applied in \citet{Beuther18}. This algorithm is capable of disentangling substructures within the given threshold contour. The positions of all sources, as identified by \textsc{clumpfind}, are summarized in Table~\ref{tab:core_positions} along with their inferred masses and column densities (see Sect.~\ref{sec:mass_estimates}).

This analysis reveals the fragmentation of the mm~sources from \cite{Reid08}: MMS1 hosts two of the major cores, denoted in the following as MMS1a in the north and MMS1b in the south, and one smaller condensation above the 10\,$\sigma$ detection threshold, denoted as MMS1c. The two major cores are coincident with the northern two sources detected at 3\,mm by \citet{Schnee09}. The southern clump MMS2 shows one major core, MMS2a, being coincident with the third source at 3\,mm by \citet{Schnee09}.
There are groups of 5\,$\sigma$ detections towards the east and towards the north and east from MMS1a (panels a~\&~b) and to the west of MMS2a (panel g). Furthermore the core MMS1a shows some elongation towards the north (panel c).
We discuss evidence of further fragmentation along with the insets of Fig.~\ref{fig:continuum_emission} in Sect.~\ref{sec:discussion_fragmentation}.

\begin{table*}
	\centering
	\caption{Position, mass and column density estimates for the mm sources.}
	\label{tab:core_positions}
	\label{tab:core_masses}
	\begin{tabular}{lccrrrrrrr}
		\hline \hline
		Core	& \multicolumn{2}{c}{Position}	& $I_\mathrm{peak}$ & $S/N$\tablefootmark{a}	& $S_\nu$\tablefootmark{b}	& $T_\mathrm{core}$	&  $M_\mathrm{core}$\tablefootmark{c}	& $N_\mathrm{H_2}$\tablefootmark{e} \\
				& RA (J2000)    & DEC (J2000) 	& (mJy beam$^{-1}$) & 				& (mJy)  	 				& (K)					& (M$_\sun$) 							&  ($10^{24}$ cm$^{-2}$) \\ \hline
		MMS1a 	&
		23$^\mathrm{h}$\,05$^\mathrm{m}$\,25\fs044 & 60\degr\,08\arcmin\,15\farcs82
		& 33.4	& 119.3						& 151.9	& 55	& 30.5	& 3.68	\\
		& & & & & & 70\tablefootmark{d}	& 23.5	& 2.81	\\
       		MMS1b	&
       		23$^\mathrm{h}$\,05$^\mathrm{m}$\,24\fs921 & 60\degr\,08\arcmin\,13\farcs94
        	& 28.2	& 100.7						& 95.0	& 55	& 19.1	& 3.10	\\
		& & & &	& & 100\tablefootmark{d}	& 10.0	& 1.62	\\
		MMS1c	&
		23$^\mathrm{h}$\,05$^\mathrm{m}$\,25\fs065 & 60\degr\,08\arcmin\,13\farcs34
		& 2.9	& 10.4						& 5.9	& 55	& 1.2	& 0.32	\\
		MMS2a  	&
		23$^\mathrm{h}$\,05$^\mathrm{m}$\,24\fs633 & 60\degr\,08\arcmin\,09\farcs20
		& 38.9	& 138.9						& 56.8	& 55	& 11.4	& 4.28	\\		\hline
	\end{tabular}
	\tablefoot{
	\tablefoottext{a}{We derived the signal-to-noise ratio $S/N$ from the peak intensity $I_\mathrm{peak}$ in units of $\sigma$ = 0.28\,mJy\,beam$^{-1}$.}
	\tablefoottext{b}{The flux densities were integrated inside the 5\,$\sigma$ contour levels of the core.}
	\tablefoottext{c}{The core mass was estimated from the flux density $S_\nu$ using Eq.~\eqref{eq:core_mass} for the clump-averaged temperature of $T_\mathrm{gas} = 55$\,K, see text.}
	\tablefoottext{d}{The second estimates for MMS1a \& b were calculated from the core-averaged rotational temperature estimate from CH$_3$CN (see Fig.~\ref{fig:temperature_maps}).}
	\tablefoottext{e}{The column density was estimated from the peak intensity $I_\mathrm{peak}$ using Eq.~\eqref{eq:column_density}.}
	Uncertainties of the mass and column density estimates are discussed in Sect.~\ref{sec:mass_estimates} and \ref{sec:XCLASS_fits}. We stress that both quantities are only lower limits due to the flux filtering effect of $\sim65$\% \citep{Beuther18}.
		}
\end{table*}

\subsection{Core mass and column density estimates}
\label{sec:mass_estimates}
The dust continuum flux density $ S_\nu $ of a core is proportional to the core mass $ M $ \citep{Hildebrand83} via the following expression:

\begin{equation}
	M_\mathrm{core} = \frac{S_\nu \cdot d^2 \cdot R}{B_\nu(T_\mathrm{dust}) \cdot \kappa_\nu} \quad,
	\label{eq:core_mass}
\end{equation}
with the distance $d$  towards the source, the gas-to-dust mass ratio $R$, the Planck function $B_\nu$ as a function of the dust temperature $T_\mathrm{dust}$, and the dust opacity $\kappa_\nu$. We integrate the flux density within the $5\sigma$ contours and compute core masses using the distance estimate of 4.3\,kpc from the RMS survey \citep{Lumsden13}. We note that we \citep[consistent with][]{Beuther18} increase the canonical value of 100 for the gas-to-dust mass ratio to 150 to account for the contribution of elements heavier than H \citep[Table 23.1]{Draine11}, which is furthermore reasonable since IRAS~23033+5951 resides at a galactocentric radius of $\gtrsim 10$\,kpc where higher gas-to-dust mass ratios are expected \citep{Giannetti17}. We assume the dust opacity to be $\kappa_{1.37\,\mathrm{mm}} = 0.9\ \mathrm{cm}^2\ \mathrm{g}^{-1}$, which is in agreement with the estimate of \citet{Ossenkopf94} for dust grains with thin ice mantles for gas densities of $10^6$\,cm$^{-3}$ of $\kappa_{1.3\,\mathrm{mm}} = 0.899\ \mathrm{cm}^2\ \mathrm{g}^{-1}$. \cite{Beuther18} have estimated the clump-averaged gas temperature from H$_2$CO spectral line fits and found $T_\mathrm{gas} = 55$\,K. Assuming that the dust is thermally coupled to the gas, i.e. $T_\mathrm{dust} \approx T_\mathrm{gas}$, we use this temperature to estimate core masses and derive additional estimates for the two cores MMS1a~\&~b from the respective core-averaged CH$_3$CN rotational temperatures of $T_\mathrm{rot} \approx 70$\,K and 100\,K, see Sect.~\ref{sec:XCLASS_fits}.
The core mass estimates are presented in Table~\ref{tab:core_masses}, along with the beam averaged H$_2$ column densities, which we calculated from the peak intensity $ I_\mathrm{peak} $, using the following equation \citep[][]{Schuller09}:

\begin{equation}
	N_{\mathrm{H}_2} =\frac{I_\mathrm{peak} \cdot R}{B_\nu(T_\mathrm{dust}) \cdot \kappa_\nu \cdot \mu m_\mathrm{H} \cdot \Omega_\mathrm{beam}} \quad ,
	\label{eq:column_density}
\end{equation}
where $\mu m_\mathrm{H}$ is the product of the mean molecular weight, assumed to be 2.8 \citep[see Appendix of][]{Kauffmann08}, and the mass of atomic hydrogen $m_\mathrm{H}$, $\Omega_\mathrm{beam}$ is the beam solid angle. We furthermore compute H$_2$ column density maps, presented in Fig.~\ref{fig:column_density_maps}, using the map of mm~continuum intensity $I_\nu$ (Fig.~\ref{fig:continuum_emission}) and the maps of rotational temperature $T_\mathrm{rot}$ (Fig.~\ref{fig:temperature_maps}, Sect.~\ref{sec:XCLASS_fits}). All over the regions, where CH$_3$CN emission is available for tracing gas (and dust) temperatures, the H$_2$ column density values are above $10^{23}$\,cm$^{-2}$.

We note that the inferred mass and column density estimates are only lower limits due to the flux filtering effect of interferometric observations. \cite{Beuther18} estimate the percentage of missing flux to be 65\% for this source and find the absolute flux scale over the entire survey to be correct within 20\%, cf. their Table~3 and Sect.~4. Furthermore, the estimates are based on the assumption of optically thin continuum emission at 1.37\,mm, which was confirmed by comparing brightness temperatures ($\sim 2$\,K) to rotational temperatures from fitting CH$_3$CN spectra in Sect.~\ref{sec:XCLASS_fits} (see Fig.~\ref{fig:temperature_maps}) and the gas temperature of $T_\mathrm{gas}=22$\,K, obtained by \cite{Maud15} from the analysis of C$^{18}$O emission.
In addition to this, we assume that there is no contribution from free-free emission, which may not necessarily be the case towards MMS1a with cm emission (see Sect.~\ref{sec:cm_emission}).

\subsection{Spectral line emission: spectra and moment maps}
\label{sec:moment_maps}

\begin{figure}
	\centering
	\resizebox{.96\hsize}{!}{\includegraphics{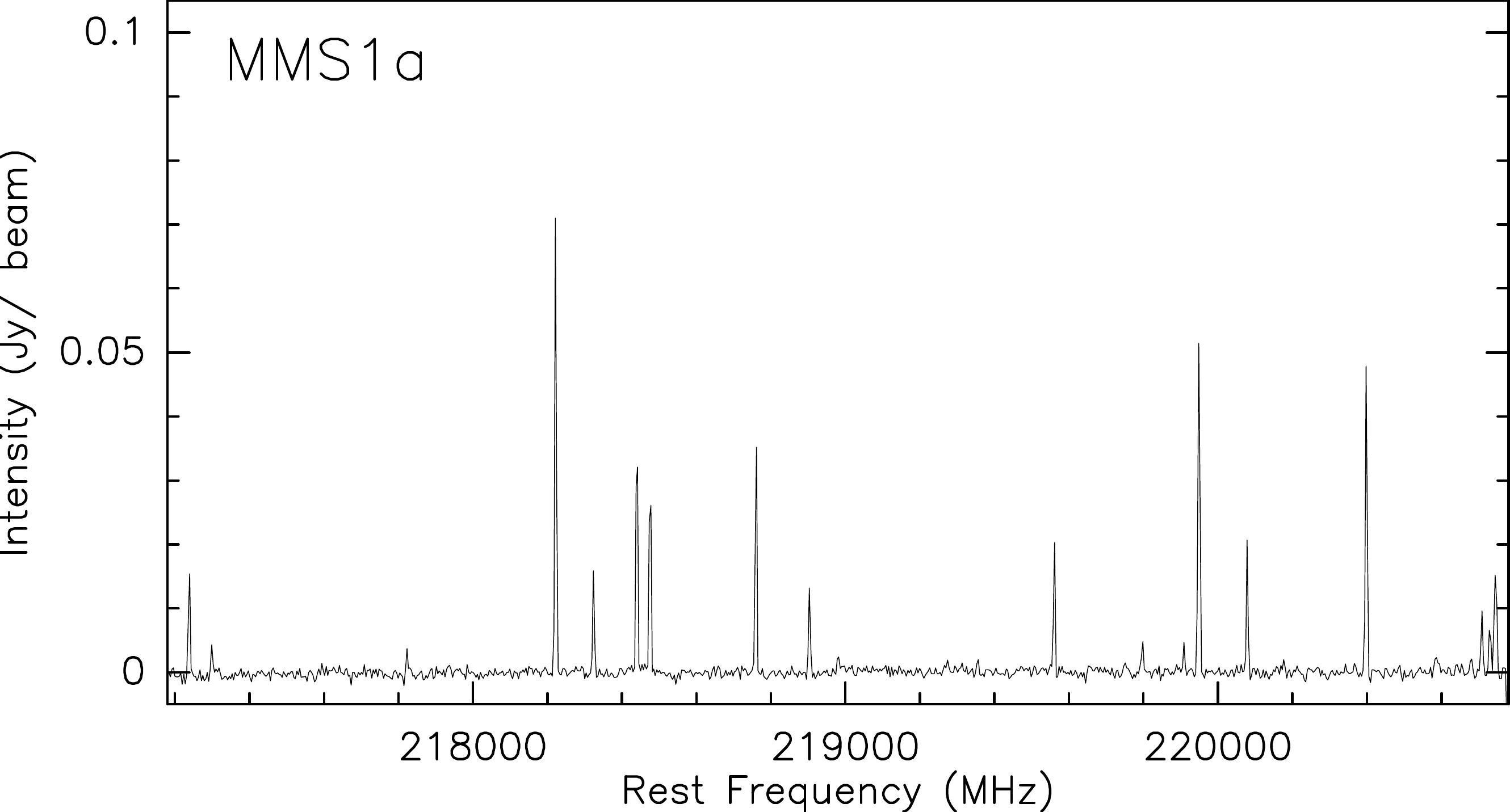}}
	\resizebox{.96\hsize}{!}{\includegraphics{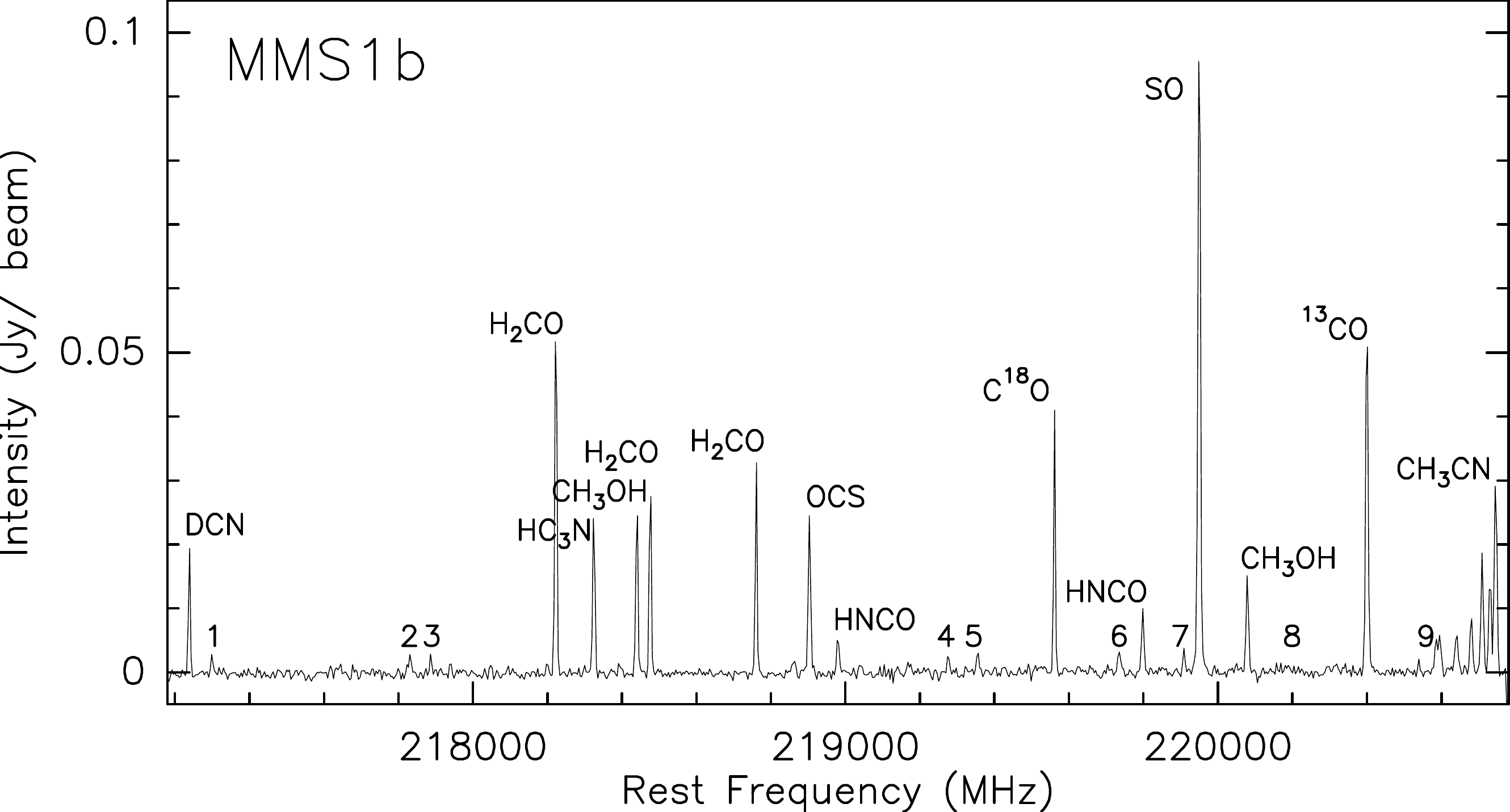}}
	\resizebox{.96\hsize}{!}{\includegraphics{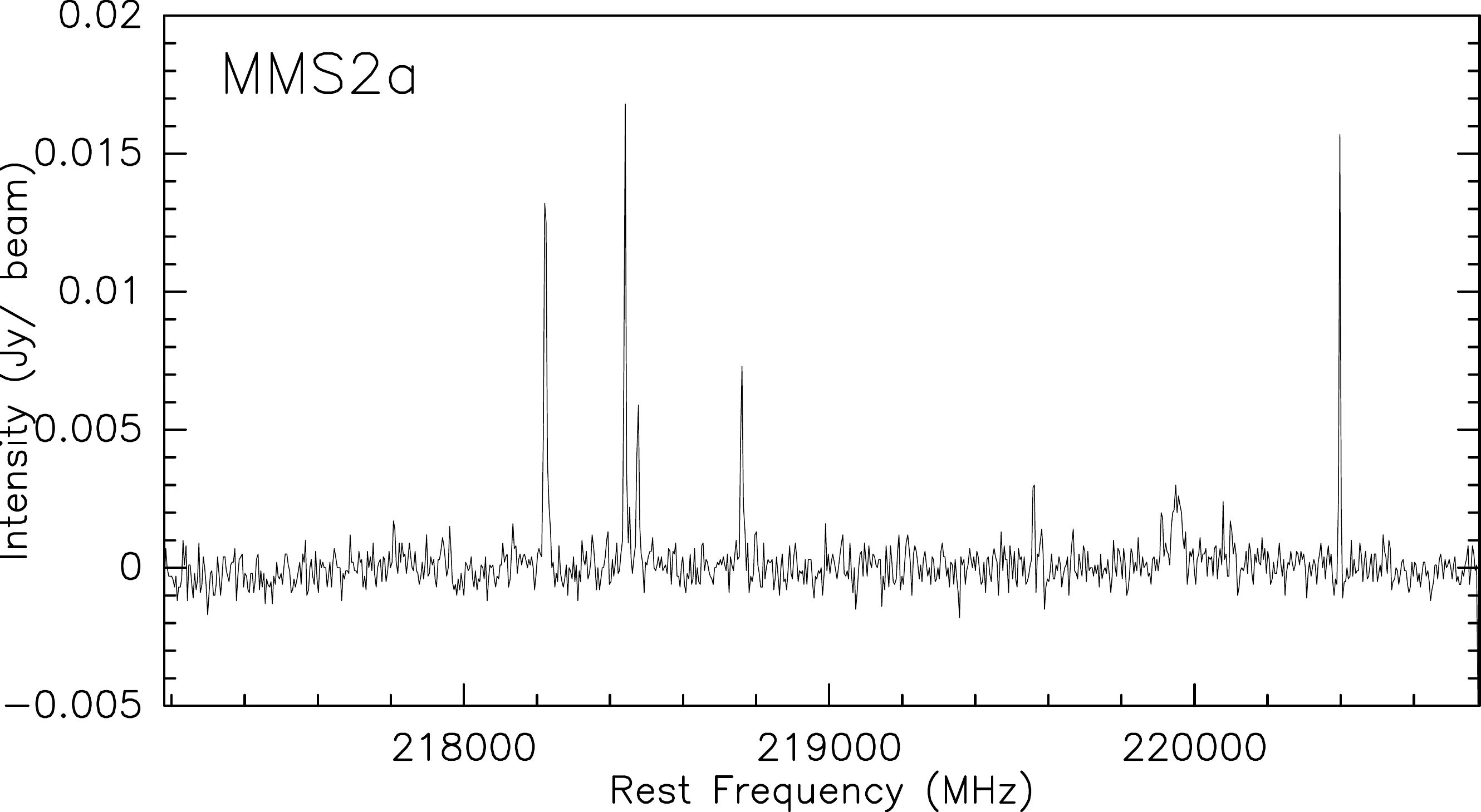}}
	\caption{Spectra towards the three major cores from the broad band correlator unit data, obtained from averaging the central 0.5\arcsec around the continuum emission peak. The spectral resolution in these bands is $\sim$\,2.7\,km\,s$^{-1}$  or $\sim$\,1.95\,MHz.
	The molecule labels are given for the chemically rich core MMS1b. Note that the flux scale is smaller for MMS2a. Lines labelled with numbers are: (1)~and (3)~unidentified, (2)~$^{33}$SO, (4)~SO$_2$, (5)~$^{34}$SO$_2$, (6)~and (9)~HNCO, (7)~H$_2^{13}$CO and (8)~CH$_2$CO.}
	\label{fig:Spectra}
\end{figure}

The spectra of the three dominant cores MMS1a,b and 2a in Fig.~\ref{fig:Spectra} were averaged over the central 0.5\arcsec of each core to estimate the respective chemical composition.
The parameters in Table~\ref{tab:obs_spec_lines} are given only for the transitions, which were analyzed in the context of kinematics, and were extracted from the CDMS\footnote{Cologne Database for Molecular Spectroscopy, \url{www.cdms.de}} \citep{Muller01, Muller05,Endres16} and LAMDA\footnote{Leiden Atomic and Molecular Database, \url{www.strw.leidenuniv.nl/~moldata}} \citep{Schoier05} databases.
The spectrum of MMS2a only shows spectral lines from comparably simple molecules like $^{13}$CO, H$_2$CO and CH$_3$OH, all having upper energy levels $\lesssim70$\,K. Compared to the emission from the other two cores, MMS2a emits only weakly < 0.03\,Jy\,beam$^{-1}$. In contrast to this, the northern cores show a variety of species with stronger emission ($\lesssim 0.1$\,Jy\,beam$^{-1}$). Besides the C-bearing species from above, we also detect N-bearing species such as DCN, HC$_3$N, HNCO and CH$_3$CN, and S-bearing species as SO and OCS towards these cores.

\begin{figure*}
	\centering
	\resizebox{0.9\hsize}{!}{\includegraphics{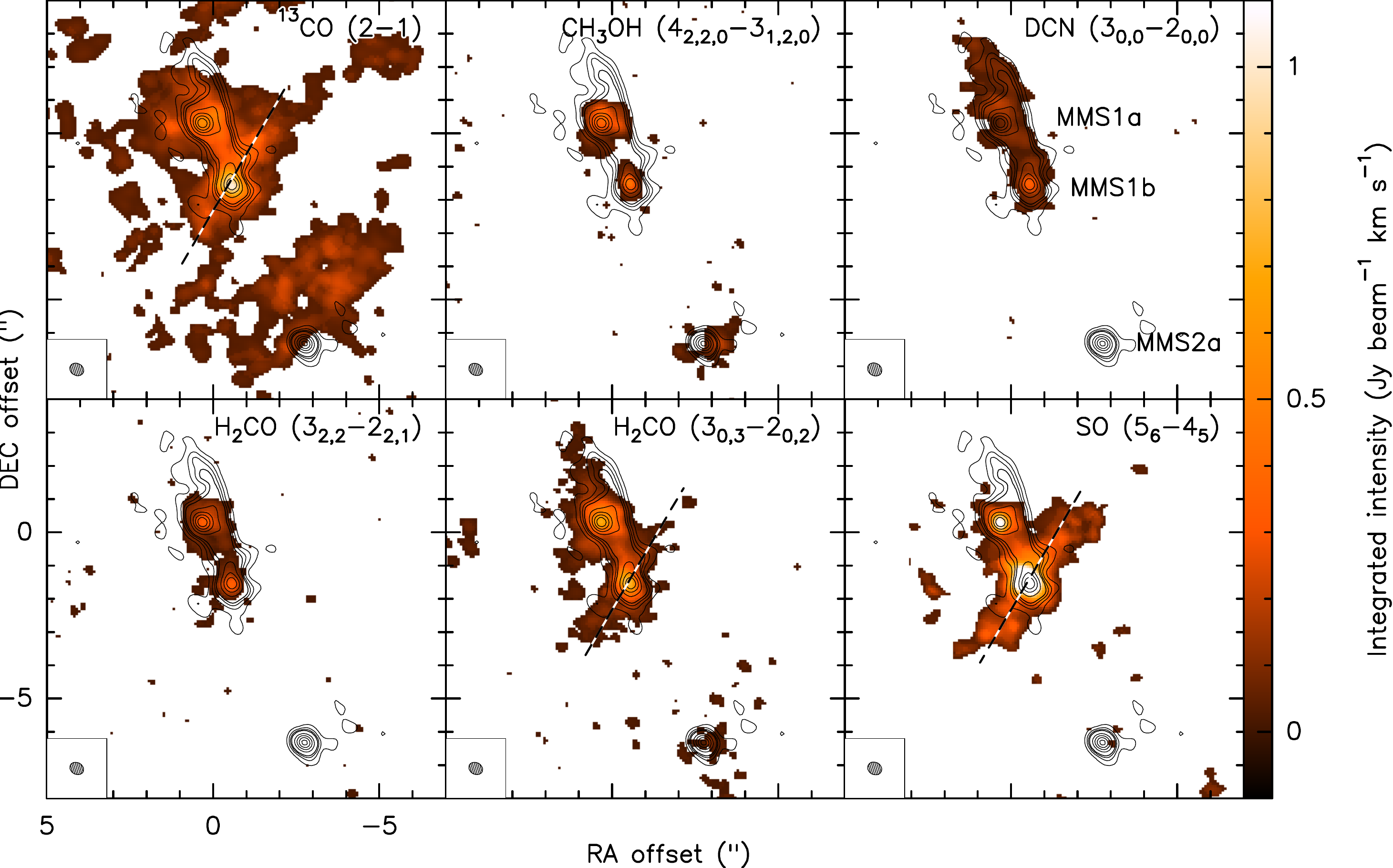}}
	\caption{Spectral line emission from selected molecular species towards IRAS~23033+5951. For the integrated intensity (zeroth order moment) maps, the intensity was integrated over the $\pm$\,10\,km\,s$^{-1}$ around the $v_\mathrm{LSR}=-53.1$\,km\,s$^{-1}$ and above the 5\,$\sigma$ threshold of each respective molecule. The black contours indicate the 5, 10, 15, 20\,$\sigma$ continuum emission levels and increase further in steps of 20\,$\sigma$, where $\sigma$ = 0.28\,mJy\,beam$^{-1}$. The dashed ellipses in the lower left corner of each panel show the respective synthesized beam $\approx 0.43\arcsec \times 0.35\arcsec$ (PA~61\degr). The dashed lines indicate an elongated structure through MMS1b.}
	\label{fig:line_emission}
\end{figure*}

\begin{figure*}
	\centering
	\resizebox{0.49\hsize}{!}{\includegraphics{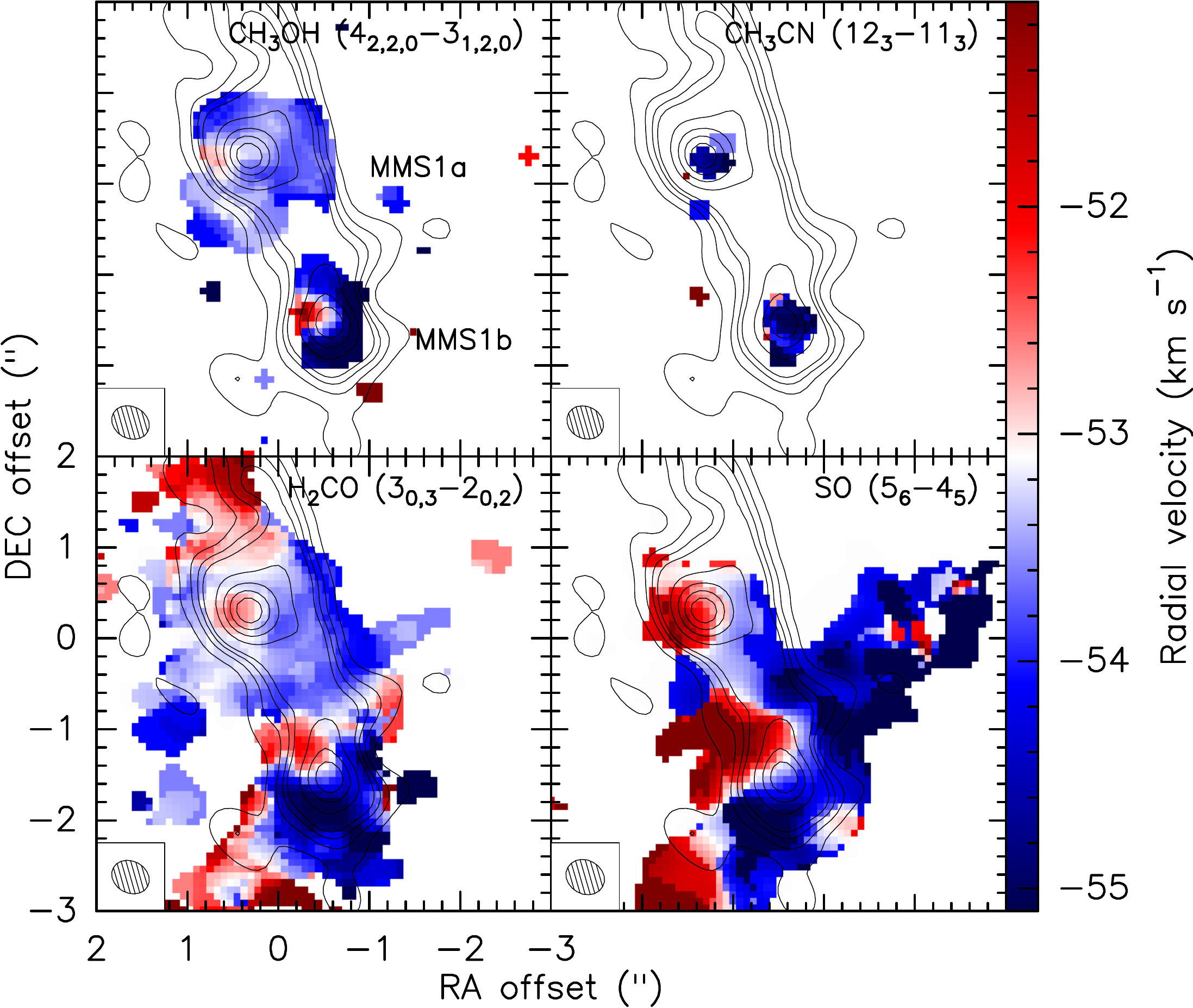}}
	\resizebox{0.49\hsize}{!}{\includegraphics{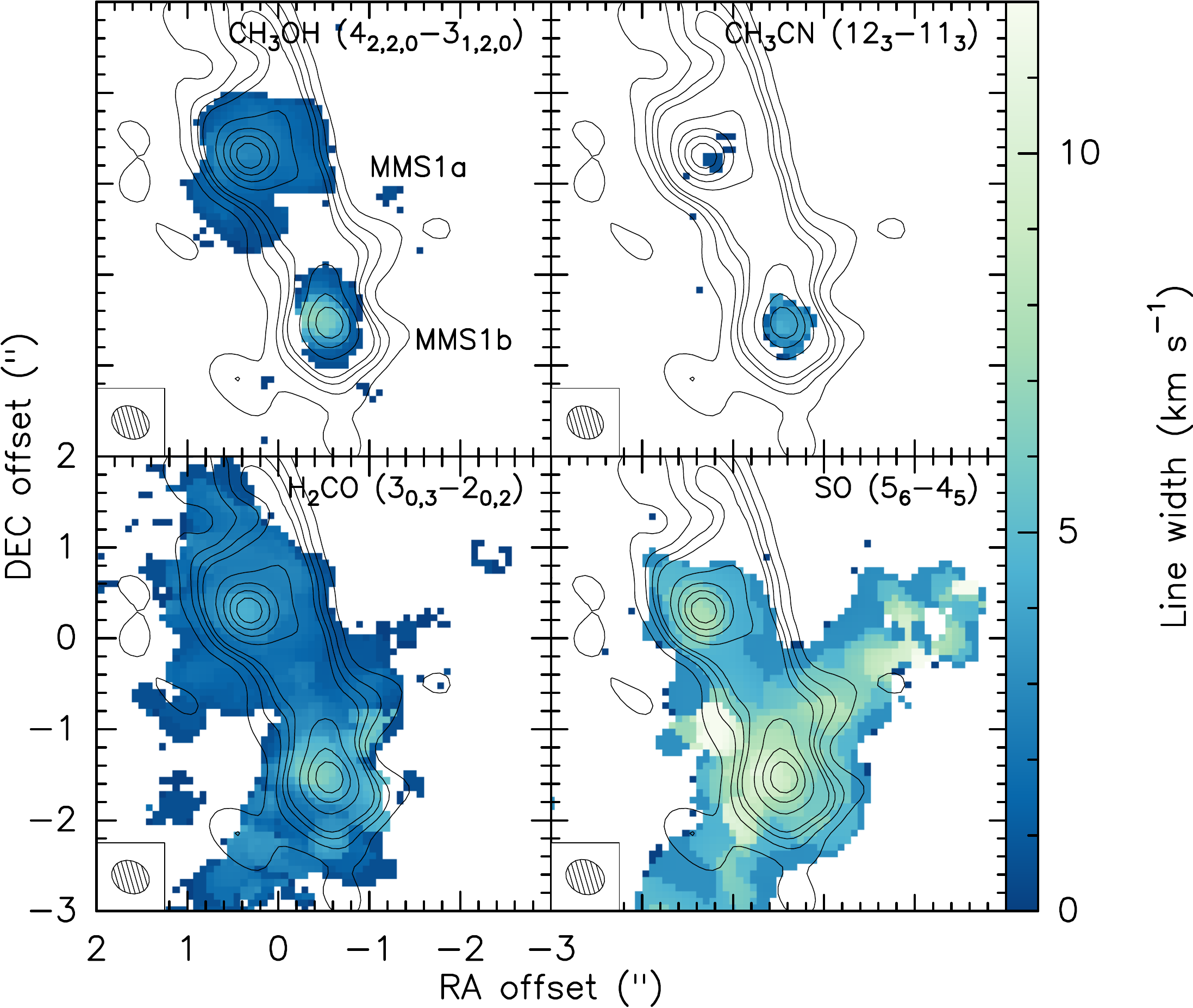}}
	\caption{Spectral line emission towards IRAS~23033+5951. For the first (intensity-weighted peak velocity, \textit{left panels}) and second order moment maps (line width, \textit{right panels}), the flux was integrated over the velocity range $v_\mathrm{LSR}\pm$ 10\,km\,s$^{-1}$, with $v_\mathrm{LSR}=-53.1$\,km\,s$^{-1}$, considering emission above the 5\,$\sigma$ threshold of each respective molecule.
	The black contours indicate the 5, 10, 15, 20\,$\sigma$ continuum emission levels and increase further in steps of 20\,$\sigma$, where $\sigma$ = 0.28\,mJy\,beam$^{-1}$. The dashed ellipses in the lower left corner of each panel show the respective synthesized beam $\approx 0.43\arcsec \times 0.35\arcsec$ (PA~61\degr).}
	\label{fig:line_emission_zooms}
\end{figure*}

For the analysis of the spatial distribution, we present integrated line intensity maps (zeroth order moment) only for the transitions $^{13}$CO (2 -- 1), CH$_3$OH (4$_{+2,2,0}$ -- 3$_{+1,2,0}$), DCN (3$_{0,0}$ -- 2$_{0,0}$), H$_2$CO (3$_{2,2}$ -- 2$_{2,1}$) and (3$_{0,3}$ -- 2$_{0,2}$), and SO (5$_6$ -- 4$_5$) in Fig.~\ref{fig:line_emission}, which all have a high signal-to-noise ratio and trace different density regimes, cf. Table~\ref{tab:obs_spec_lines}. Other transitions are not shown, as they do not provide further spatial information. The distribution of CH$_3$CN is implicitly shown in Fig.~\ref{fig:line_emission_zooms}, where the emission from all transitions is stacked and blanked below the 6\,$\sigma$ level.
As expected, we detect those transitions with a lower critical density in the regions far away from the main dense cores, as well as towards the dense cores where we also detect those transitions with higher critical densities. DCN follows the elongation of MMS1a to the north (panel c in Fig.~\ref{fig:continuum_emission}). In the transitions $^{13}$CO (2--1), SO (5$_6$ -- 4$_5$) and H$_2$CO (3$_{0,3}$ -- 2$_{0,2}$) we identify an elongated structure through MMS1b at a position angle $\phi \approx 315\degr$ from the north-south axis, marked by the dashed line in Fig.~\ref{fig:line_emission}. We will discuss this structure in the context of molecular outflows in Sect.~\ref{sec:outflows}.

We probe the kinematics of the clump MMS1 utilizing maps of the intensity weighted peak velocities (first order moment), with closer zooms in Fig.~\ref{fig:line_emission_zooms}. Towards MMS1b, we identify a velocity gradient in SO (5$_6$ -- 4$_5$), H$_2$CO (3$_{0,3}$ -- 2$_{0,2}$), and CH$_3$OH (4$_{+2,2,0}$ -- 3$_{+1,2,0}$) from north-east to south-west ($\phi_\mathrm{grad} \approx 130\degr$), from red to blue-shifted emission of $\sim$\,5\,km\,s$^{-1}$ over an angular distance of $\sim2\arcsec$, corresponding to 8600\,au. While CH$_3$CN ($12_3 - 11_3$) remains inconlusive, a similar gradient is seen in the combined map of the CH$_3$CN ($12_K - 11_K$) transitions (cf. Fig.~\ref{fig:kinematics}). The gradient is oriented roughly perpendicular to the elongated structure, seen in the $^{13}$CO, H$_2$CO and SO transitions mentioned above. Towards the northern core MMS1a, we identify a velocity gradient in east-west orientation for SO and CH$_3$CN with $\phi_\mathrm{grad} \approx 270\degr$. In contrast to this, the remaining maps for the CH$_3$OH, H$_2$CO and DCN transitions do not show specific gradients.

In the right panel of Fig.~\ref{fig:line_emission_zooms}, we present the spectral line width (second order moment). For CH$_3$OH (4$_{+2,2,0}$ -- 3$_{+1,2,0}$) and H$_2$CO (3$_{0,3}$ -- 2$_{0,2}$), these maps show larger line widths of 5--7\,km\,s$^{-1}$ towards the center of the cores, which indicates that the cores are fed by streams of gas and dust material or that YSOs in the cores launch outflows. In the SO ($5_6 - 4_5$) map, we additionally detect larger line widths $\gtrsim 10$\,km\,s$^{-1}$ along the elongated structure, where several velocity components from Fig.~\ref{fig:channel_maps} are overlapping.

Towards the third core MMS2a, we only report the detection of $^{13}$CO, H$_2$CO and CH$_3$OH transitions, as in the core-averaged spectra, at a similar systemic velocity. Due to the overall low signal-to-noise ratio, we will not work on the kinematics of this core.

\subsection{Molecular outflows}
\label{sec:outflows}
\citet{Beuther02outflow} report a large-scale outflow structure. They observed the emission of CO (2 -- 1) towards IRAS~23033+5951 with the IRAM 30~m telescope, with an angular resolution of $11\arcsec$, corresponding to $ \sim 47\,300 $\,au. Their data show red-shifted emission towards the south-east and blue-shifted emission centered around the northern part of the clump with some tailing emission towards the north-west. We use $^{13}$CO (2 -- 1) and SO ($6_5 - 5_4$) emission to identify and distinguish molecular outflows based on their collimation degree and to relate the outflows to the mm~sources.

\begin{figure*}
	\centering
	\resizebox{.9\hsize}{!}{\includegraphics{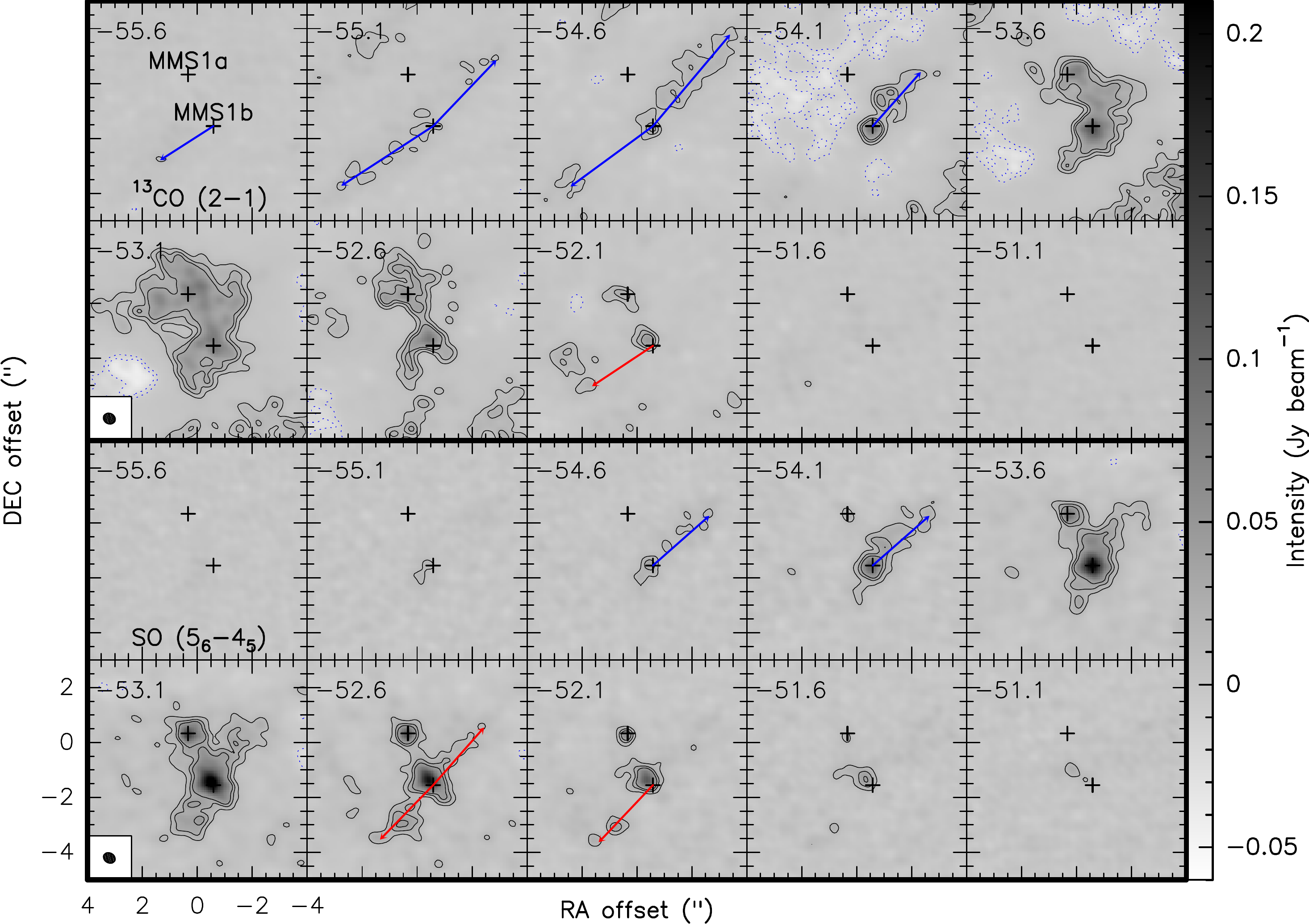}}
	\caption{Channel maps of $^{13}$CO (2 -- 1, \textit{top panels}) and SO ($5_6 - 4_5$, \textit{bottom panels}) emission in the interferometric data. The black contours are the 5, 10, and 15\,$\sigma$ levels for the respective molecule, and the blue dotted contours are the corresponding negative values. The black + signs mark the position of the mm~emission peaks of MMS1a~\&~b. The blue and red arrows shall guide the eye to the direction of the respective Doppler-shifted emission. The shaded ellipse in the lower left corners indicate the synthesized beam of $0.44\arcsec \times 0.36\arcsec$ (PA~61\degr) for both data cubes.}
	\label{fig:channel_maps}
\end{figure*}

\begin{figure*}
	\centering
	\resizebox{.9\hsize}{!}{\includegraphics{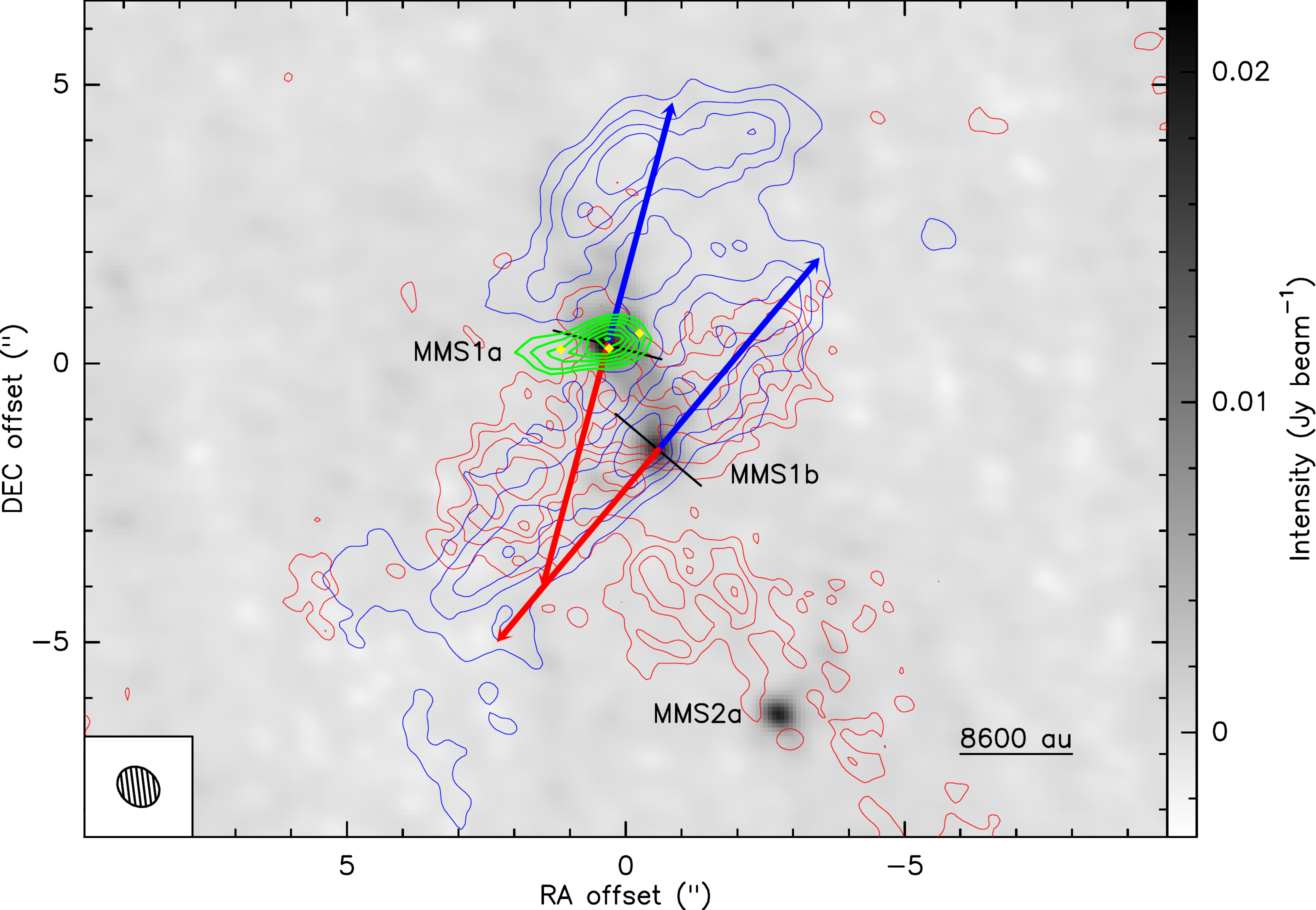}}
	\caption{Multiple outflow structures towards IRAS 23033+5951 as traced by $^{13}$CO (2 -- 1) emission in the merged data cube. The gray scale in the background shows the continuum intensity from Fig.~\ref{fig:continuum_emission}. The blue and red contours present the integrated line wing emission from $^{13}$CO for the intervals of $-70$ to $-60$\,km\,s$^{-1}$ and $-46$ to $-36$\,km\,s$^{-1}$ ($v_\mathrm{LSR} = -53.1$\,km\,s$^{-1}$), respectively, starting at 55\% and increasing in steps of 10\% of the peak intensities $I_\mathrm{peak, blue}=0.78$\,Jy\,beam$^{-1}$\,km\,s$^{-1}$ and $I_\mathrm{peak, red}=0.25$\,Jy\,beam$^{-1}$\,km\,s$^{-1}$. The green contours show the continuum emission at 3.6\,cm, reported by \citet{Beuther02maser}, starting at $4\,\sigma$ and increasing in steps of $\sigma = 41.5\,\mu$Jy\,beam$^{-1}$ with an angular resolution of $1.04\arcsec \times 0.62\arcsec$, and the yellow diamonds towards MMS1a mark the positions of the 3.6\,cm emission peaks \citep{Rodriguez12} with an angular resolution of $0.31\arcsec \times 0.25\arcsec$. The blue and red arrows are drawn by eye to guide the reader to the outflow lobes from MMS1a~\&~b. The solid lines across the cores are drawn perpendicular and indicate the corresponding inferred disk major axes. The shaded ellipse in the lower left corner indicates the synthesized beam of $ 0.8\arcsec \times 0.67\arcsec$ (PA~52\degr) of the merged $^{13}$CO (2 -- 1) data. }
	\label{fig:outflows}
\end{figure*}

In Fig.~\ref{fig:channel_maps}, we present channel maps of the two transitions $^{13}$CO (2 -- 1) and SO ($5_6 - 4_5$), tracing low-density gas.
Furthermore, we make use of our merged data cube from NOEMA interferometric and the 30-m single-dish telescope data. We integrate the blue and red shifted line wings from $-70$ to $-60$\,km\,s$^{-1}$ and from $-46$ to $-36$\,km\,s$^{-1}$, respectively, i.e. over a range of $\Delta v =10$\,km\,s$^{-1}$ each starting 7\,km\,s$^{-1}$ from the $v_\mathrm{LSR}$, and plot them over the continuum intensity, yielding Fig.~\ref{fig:outflows}. These velocity intervals contain the respective intervals for the blue and red outflow lobes from \cite{Beuther02outflow}.

In the channel maps in Fig.~\ref{fig:channel_maps} we identify in several channels an elongated structure which is connected to MMS1b in both species. This structure is highlighted by blue and red arrows for blue and red shifted emission, respectively, and is elongated in the north-west to south-east direction ($\phi_\mathrm{out} \approx 315\degr$, starting at the blue shifted side). The detection of blue and red shifted velocity components on both sides of the core suggests that we see an outflow almost perpendicular to the line of sight \citep{Cabrit86,Cabrit88}. In contrast to this, the northern core MMS1a is not connected to components with velocities higher than $\pm1$\,km\,s$^{-1}$ from the $v_\mathrm{LSR}$. However, the analysis of the merged data set, presented in Fig.~\ref{fig:outflows}, suggests the possibility that the blue shifted emission to the north belongs to a blue outflow lobe launched from MMS1a, where the respective red lobe matches to the leftover red emission to the south. This candidate outflow appears to be considerably less collimated than the one launched from MMS1b and is seen under a projected position angle of $\phi_\mathrm{out} \approx 350\degr$. This disentanglement will be further discussed with the overall gas kinematics, in Sect.~\ref{sec:kinematics}.

The overall structure is in good agreement with the extended H$_2$ emission found by \citet{Navarete15}, Fig.~A130, observing the source with an H$_2$ narrow band filter ($\lambda=2.122\mu$m, $\Delta\lambda=0.032\mu$m). Besides the north-west to south-east elongation seen in $^{13}$CO emission, they detect additional emission towards the east of MMS1, but there is no information on whether this represents another outflow launched from this clump.

\subsection{Spectral line emission: derivation of gas properties and kinematics}
\label{sec:XCLASS_fits}
The methyl cyanide CH$_3$CN ($12_K - 11_K$) transition set is a well known tracer for the rotational temperature \citep[][]{Green86,Zhang98,Araya05}. This quantity is a good approximation of the temperature of the surrounding gas and dust if the medium is in local thermodynamic equilibrium (LTE) and if CH$_3$CN is coupled to the ambient medium. This is presumably the case for densities above the critical density $n_\mathrm{crit}\sim 4 \times 10^6 $\,cm$^{-3}$ \citep{Loren84}. We use the equivalent core radii from \citet{Beuther18} to transfer the column densities from Table~\ref{tab:core_masses} into volume densities and obtain values $\sim 5.0 \times 10^{7} \mathrm{cm}^{-3}$, well above the critical value, which confirms that the medium is in LTE.

We use \textsc{xclass} \citep[eXtended \textsc{Casa} Line Analysis Software Suite,][]{Moller17} for fitting the CH$_3$CN spectra. This package solves the radiative transfer equations for a medium in LTE, utilizing the VAMDC\footnote{VAMDC consortium, \url{http://www.vamdc.org}} and CDMS\footnote{Cologne Database for Molecular Spectroscopy, \url{www.cdms.de}} databases \citep{Muller01,Muller05,Endres16}. From the \textsc{xclass} package, we utilize the \textsc{xclassmapfit} function, which fits the spectra from a data cube pixel-by-pixel, yielding best-fit parameter maps for rotational temperature, column density, peak velocity and velocity dispersion.

Since significant emission, $>6\sigma$, is detected only towards small regions around the continuum emission peaks of the two major cores MMS1a~\&~b, we consider only a $ 2.2\arcsec \times 2.2\arcsec $ area around each core and extract the corresponding regions from the CH$_3$CN ($12_K - 11_K$, $K=0-6$) data cube. Tests revealed that good fitting is achieved by a sequence of 300 iterations of the \textit{genetic} algorithm and another 50 iterations of the \textit{Levenberg-Marquardt} algorithm \citep[see][for descriptions]{Moller17}. We note that the map fits include also the isotopologues of methyl cyanide and that we furthermore assume that the beam filling factor is 1 or, equivalently, that the source size is much larger than the beam.

\begin{figure}
	\centering
	\resizebox{.96\hsize}{!}{\includegraphics{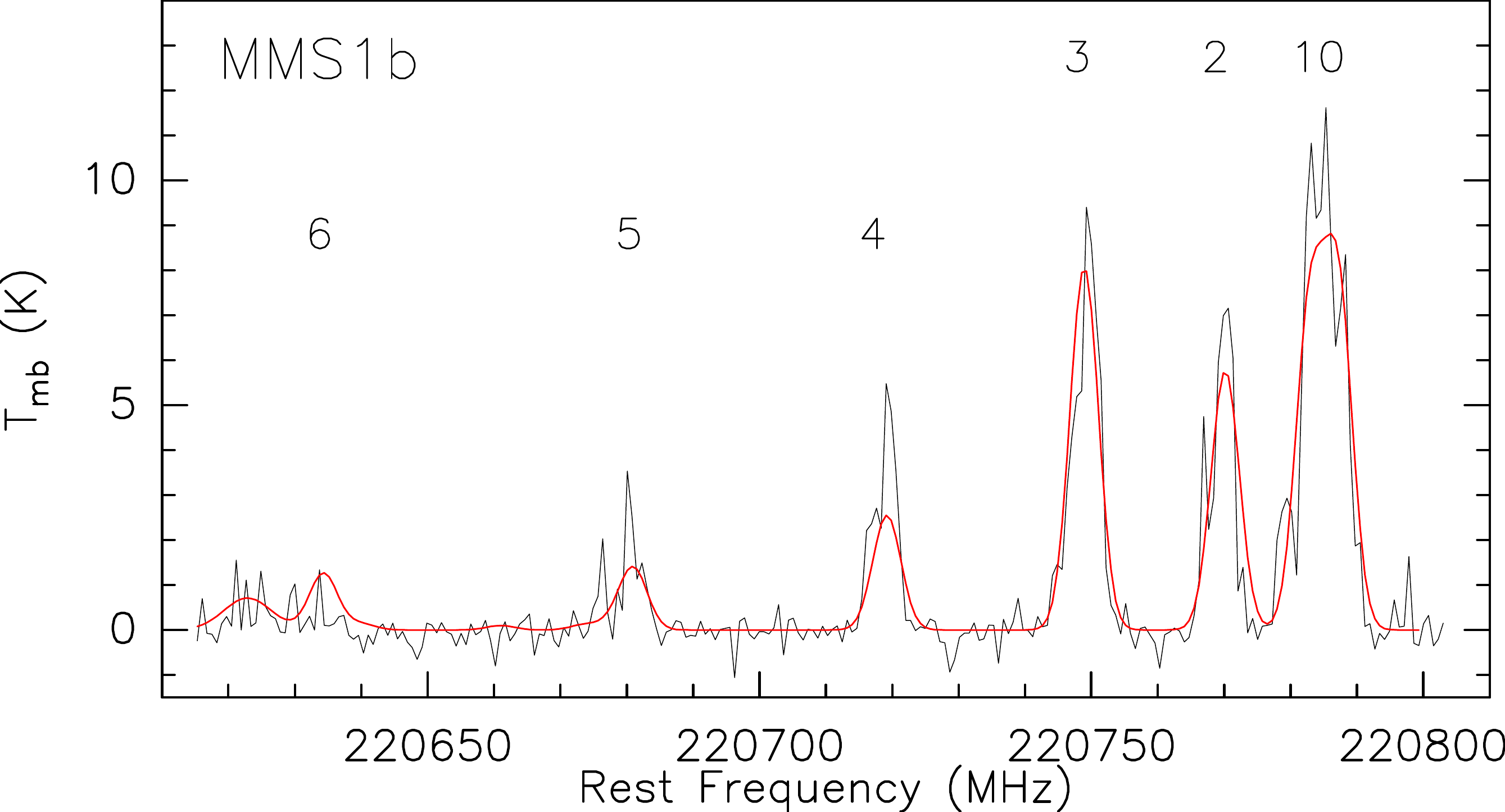}}
	\caption{Spectrum of the CH$_3$CN ($12_K - 11_K$, $K=0-6$) transitions towards the continuum emission peak in MMS1b. The red curve indicates the best fit as obtained from the \textsc{xclass} procedures, with $T_\mathrm{rot} = 85$\,K and $N_{\mathrm{H}_2} = 5.4 \times 10^{14}$\,cm$^{-2}$.}
    \label{fig:xclass_spectrum}
\end{figure}

An example spectrum from the MMS1b continuum emission peak is presented in Fig.~\ref{fig:xclass_spectrum} along with the corresponding best fit. This fit shows slight systematic deviations from the data as it underestimates the peaks of the emission lines. A possible explanation of this are the 'shoulder' features towards lower frequencies, indicating a second velocity component of the gas. However, not all spectra can be properly fitted with two distinct velocity components and we thus decided to adopt a single-component fit to obtain self-consistent parameter maps. An improvement was achieved only in a region of $5 \times 5$ pixels around the continuum emission peak, corresponding to about one synthesized beam. In this region the $\chi^2$ was reduced by $\leq30\%$.

\begin{figure*}
	\centering
	\resizebox{.46\hsize}{!}{\includegraphics{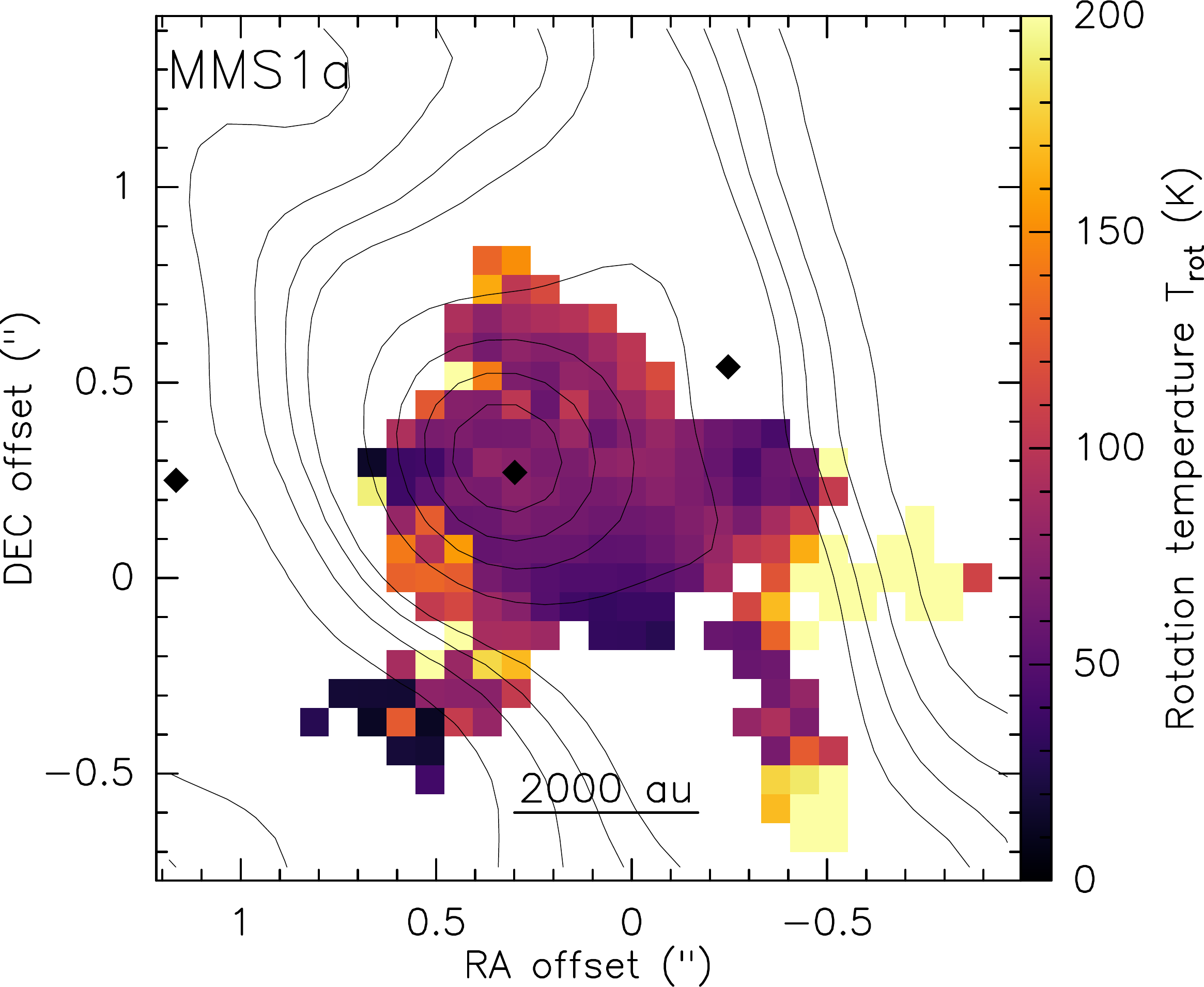}}
	\resizebox{.46\hsize}{!}{\includegraphics{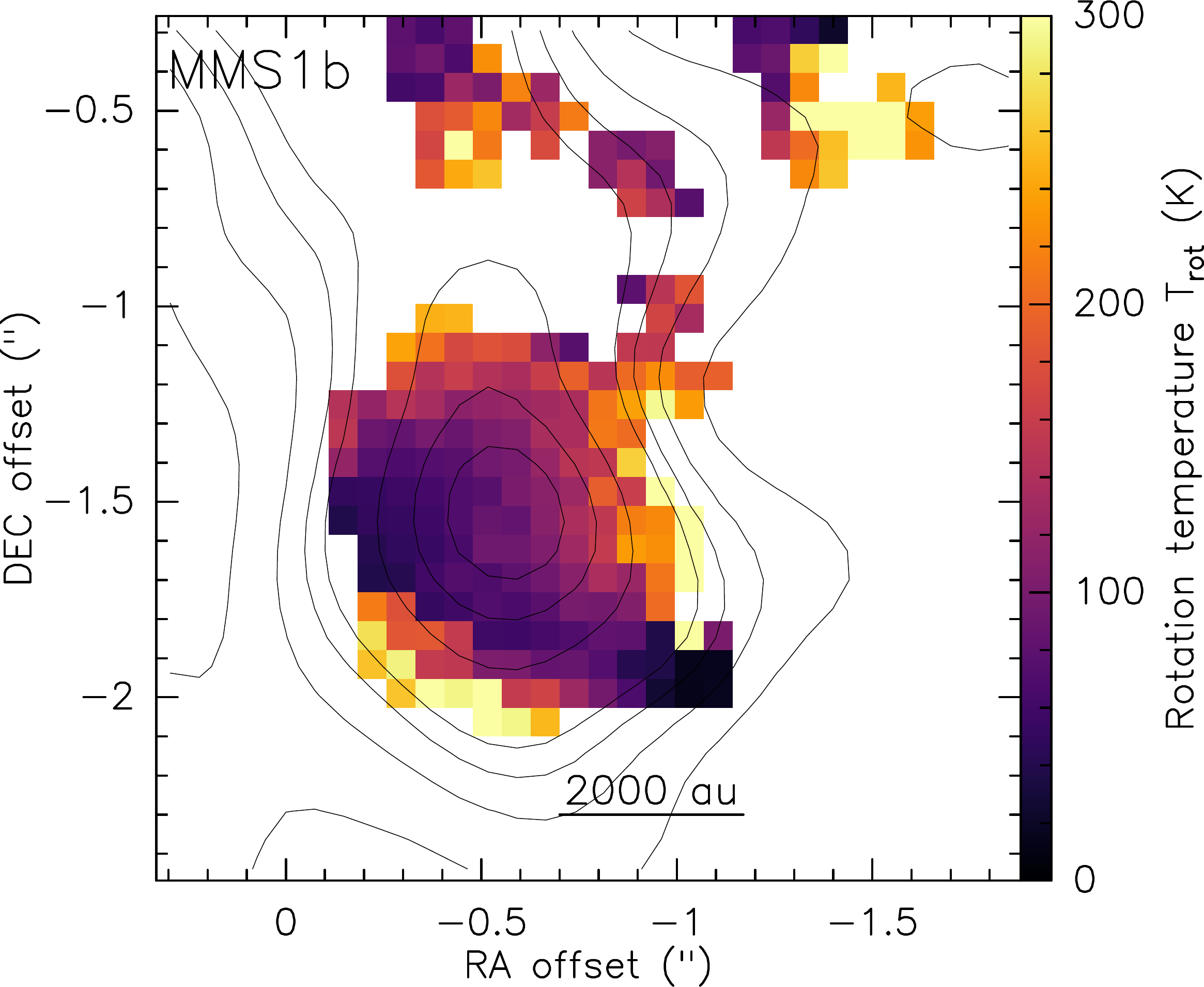}}
	\resizebox{.46\hsize}{!}{\includegraphics{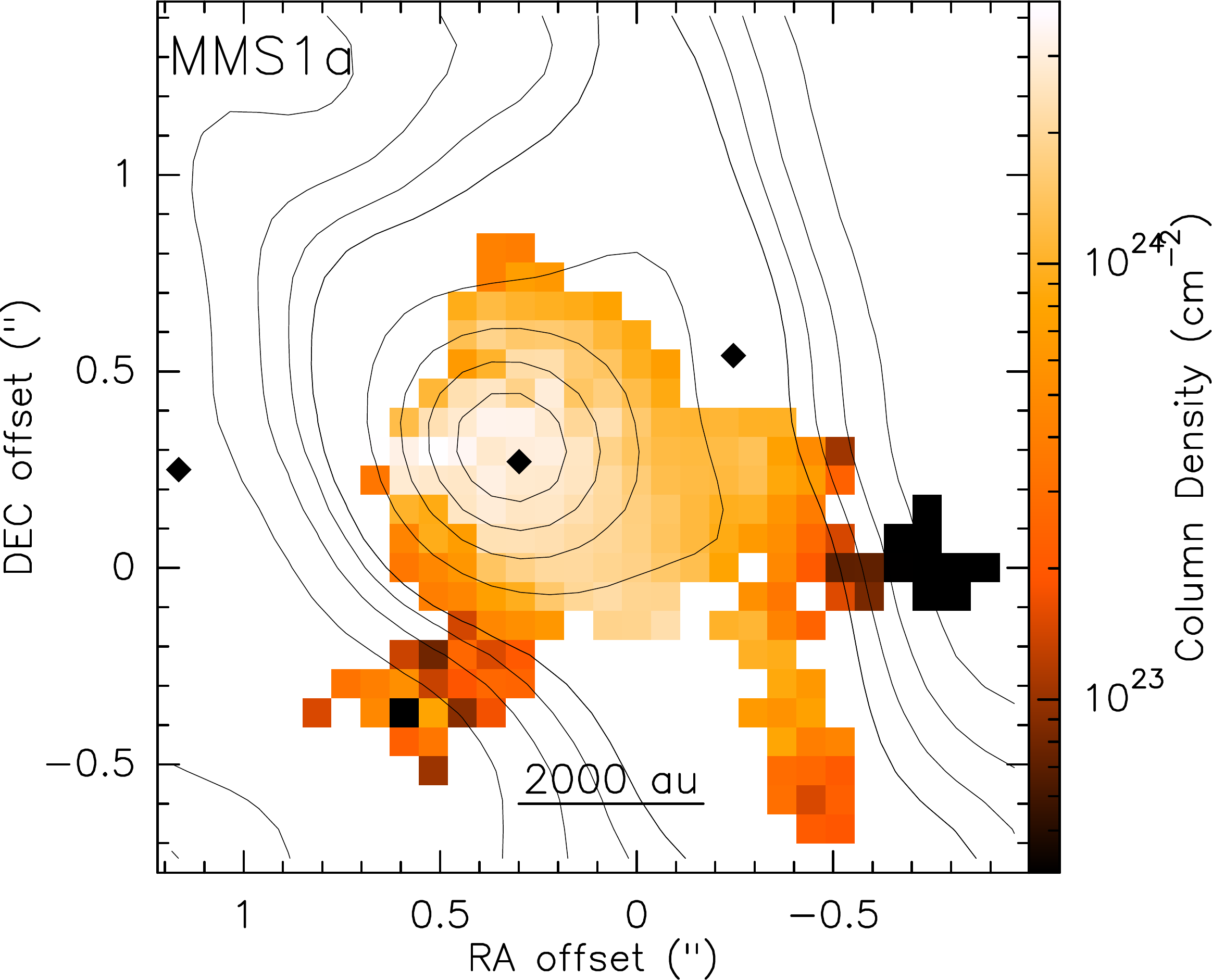}}
	\resizebox{.46\hsize}{!}{\includegraphics{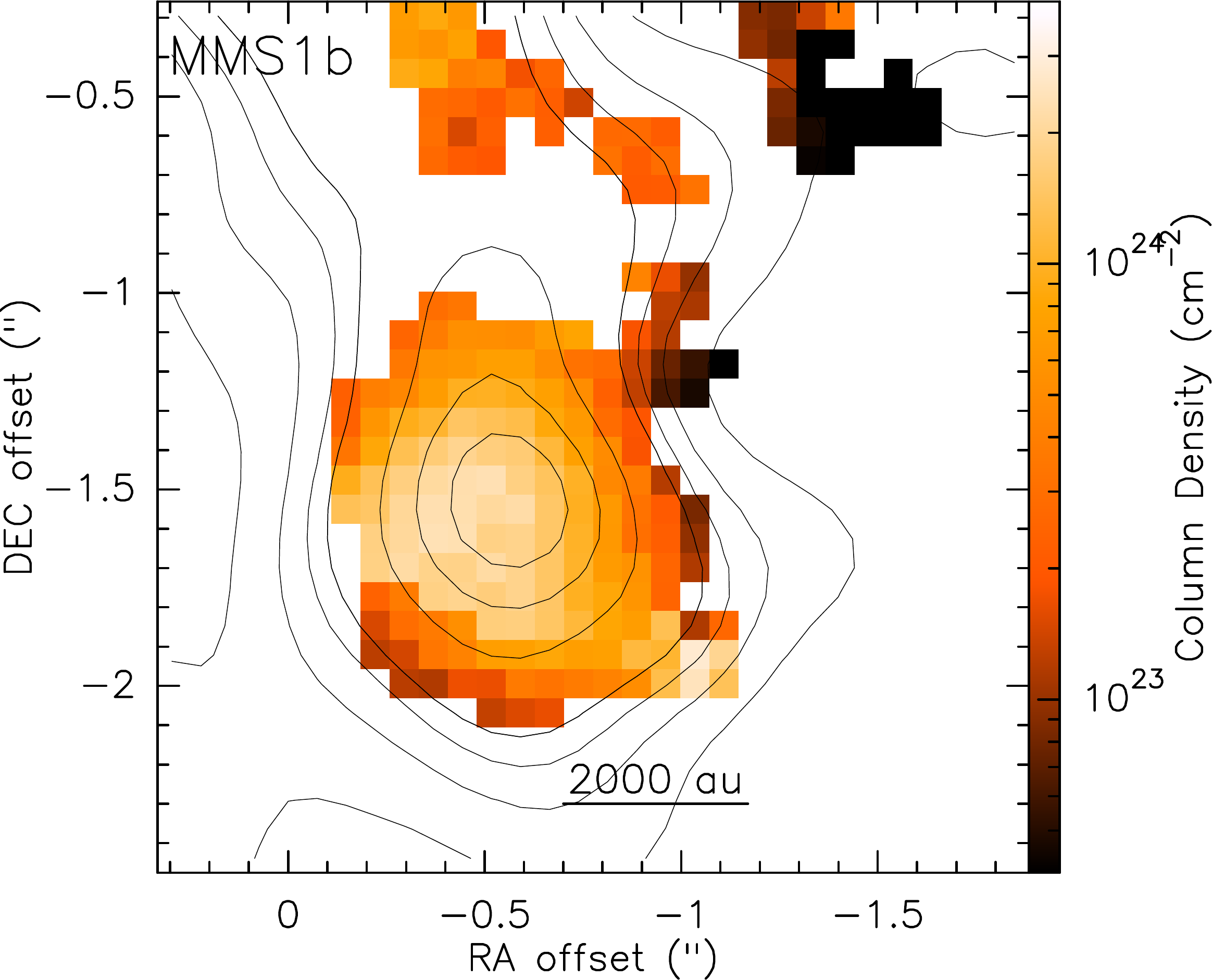}}
	\resizebox{.46\hsize}{!}{\includegraphics{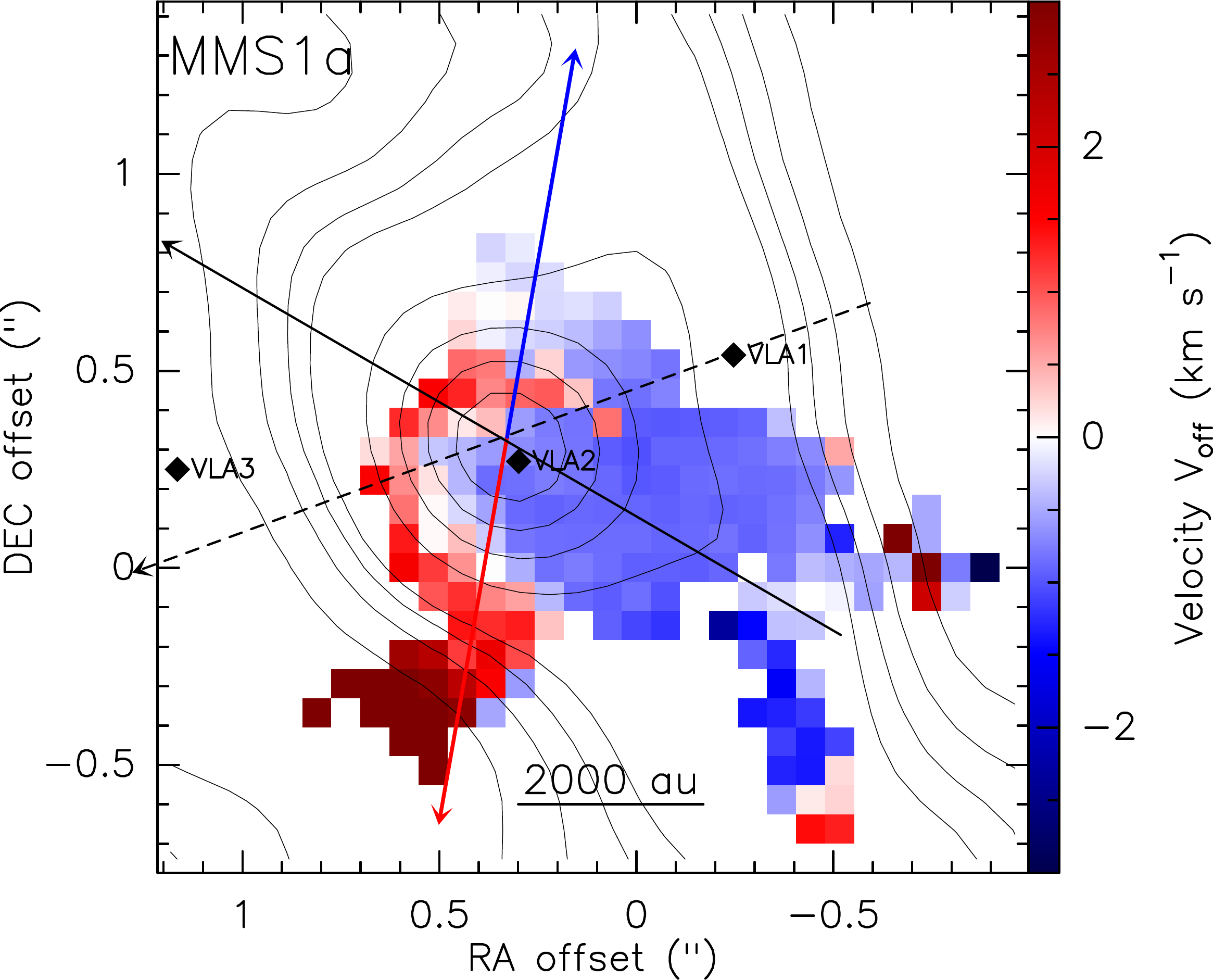}}
	\resizebox{.46\hsize}{!}{\includegraphics{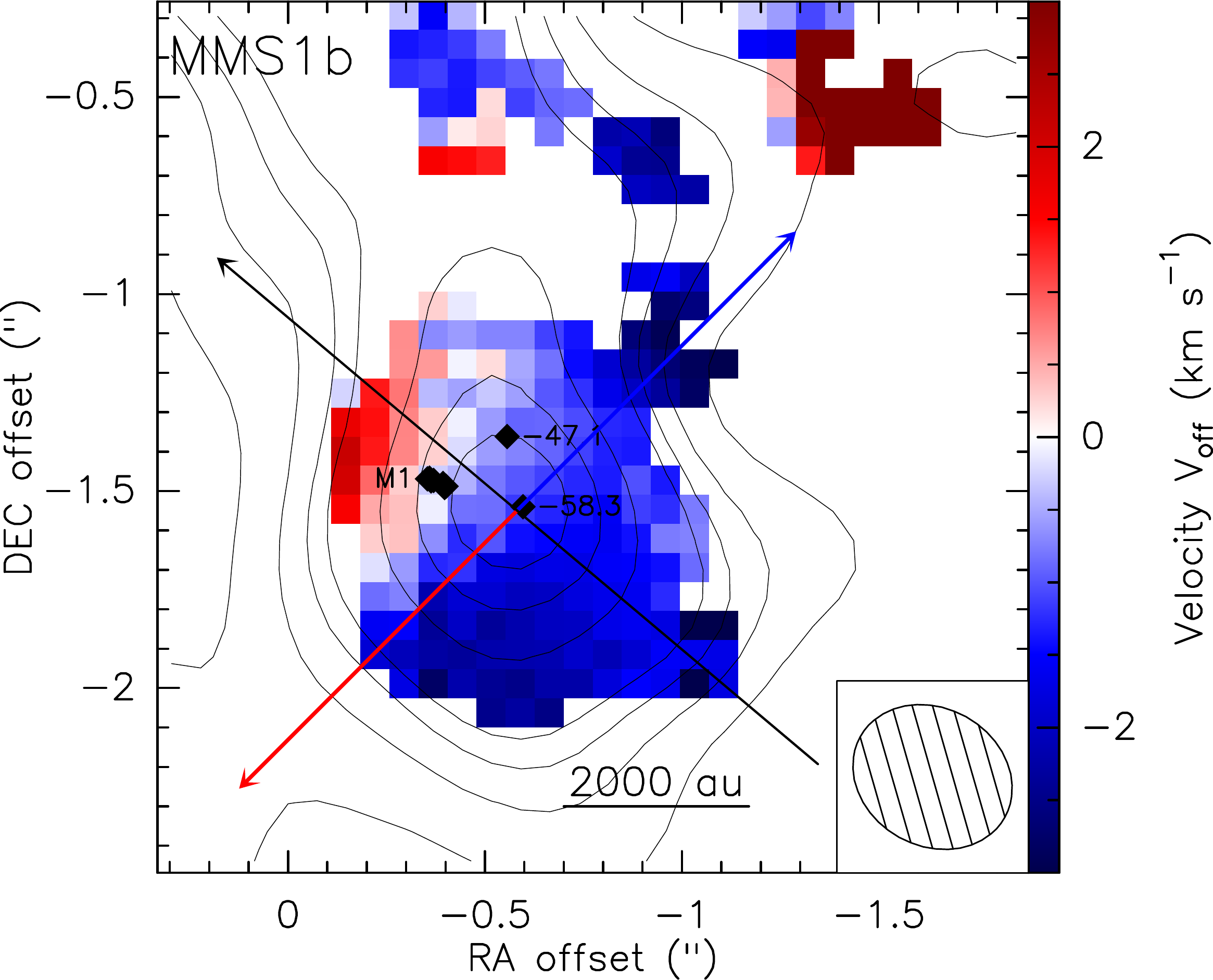}}
	\caption{
	Results of the radiative transfer modeling with \textsc{xclass} of the CH$_3$CN ($12_K - 11_K$, $K=0-6$) data. The maps cover $2.2\arcsec \times 2.2\arcsec$ extracted around the two major cores MMS1a~\&~b, where pixels with flux below $6\sigma_\mathrm{CH_3CN}$ were blanked out. The black contours indicate the 5, 10, 15, 20\,$\sigma$ continuum emission levels and increase further in steps of 20\,$\sigma$, where $\sigma$ = 0.28\,mJy\,beam$^{-1}$. The black diamonds in the MMS1a panel mark the positions of the emission peaks at 3.6\,cm \citep{Rodriguez12}.  The shaded ellipse in the lower right corner indicates the synthesized beam of the CH$_3$CN data of $0.42\arcsec \times 0.35\arcsec$ (PA~61\degr). Details to the spectral line map fit are summarized in Sect.~\ref{sec:XCLASS_fits}.
	(\emph{Top}) Rotational temperature maps from \textsc{xclass}.
	(\emph{Middle}) H$_2$ column density maps, estimated from the continuum intensity $I_\nu$ (Fig.~\ref{fig:continuum_emission}) and rotational temperature $T_\mathrm{rot}$ (\emph{top panels}) using Eq.~\eqref{eq:column_density}.
	(\emph{Bottom}) Velocity offsets from \textsc{xclass}. The colors indicate the relative velocity with respect to the $v_\mathrm{LSR}=-53.1$\,km\,s$^{-1}$. In the MMS1b panel, the additional black diamonds mark the positions of the H$_2$O masers with the corresponding velocity in km\,s$^{-1}$ from \citet{Rodriguez12}. The M1 maser group contains maser emission with velocities ranging from $-37.2$ to $-66.8$\,km\,s$^{-1}$. The blue and red arrows are the outflow axes as in Fig.~\ref{fig:outflows}. The solid and dashed black arrows are the cut directions for the PV diagrams presented in Fig.~\ref{fig:pv_diagrams}, along the presumable disk semi-major axes.
	}

	\label{fig:temperature_maps}
	\label{fig:column_density_maps}
	\label{fig:kinematics}
\end{figure*}

We present the resulting maps of the rotational temperature $T_\mathrm{rot}$ in Fig.~\ref{fig:temperature_maps}. The respective results for the relative velocity and line width are shown in the CH$_3$CN panels in both Fig.~\ref{fig:line_emission_zooms} and \ref{fig:kinematics}. The general temperature structure towards both cores, shows values as low as $\sim70$ and $\sim100$\,K in the center, for MMS1a \&~b respectively, and higher temperatures towards the edges up to $\sim300$\,K.
Such strong deviations towards larger radii from the emission peak, however, are probably due to the lower signal-to-noise ratio where the emission is just above the 6\,$\sigma$ detection threshold. This makes it more difficult for the fitting procedure to solve the parameter ambiguity, for instance between gas column density and temperature, and the procedure may tend towards the limits of the parameter range, 600\,K in this case.
As deviations from the general trend, we identify no temperature increase towards the south and west of MMS1a and only a slight increase towards the east, and towards MMS1b the temperature estimates decrease from the central values down to $\sim 40$\,K, towards the east, and to $\sim70$\,K, to the south-west.
The analysis of the central 7$\times$7 pixels of the cores yields temperatures of 70$\pm$20 and 100$\pm$40\,K for MMS1a~\&~b, respectively.
We note that, although CH$_3$CN may not be optically thin and hence trace slightly different layers of the region than the dust continuum emission, at the given high densities ($\sim 5.0 \times 10^{7} \mathrm{cm}^{-3}$) gas and dust should be coupled well. Therefore, in the following we assume coupling of dust and gas and hence the same temperature for both.

\subsection{Position-velocity diagrams: protostellar mass estimates}
\label{sec:PVdiagrams}
\begin{figure*}
	\centering
	\resizebox{.45\hsize}{!}{\includegraphics{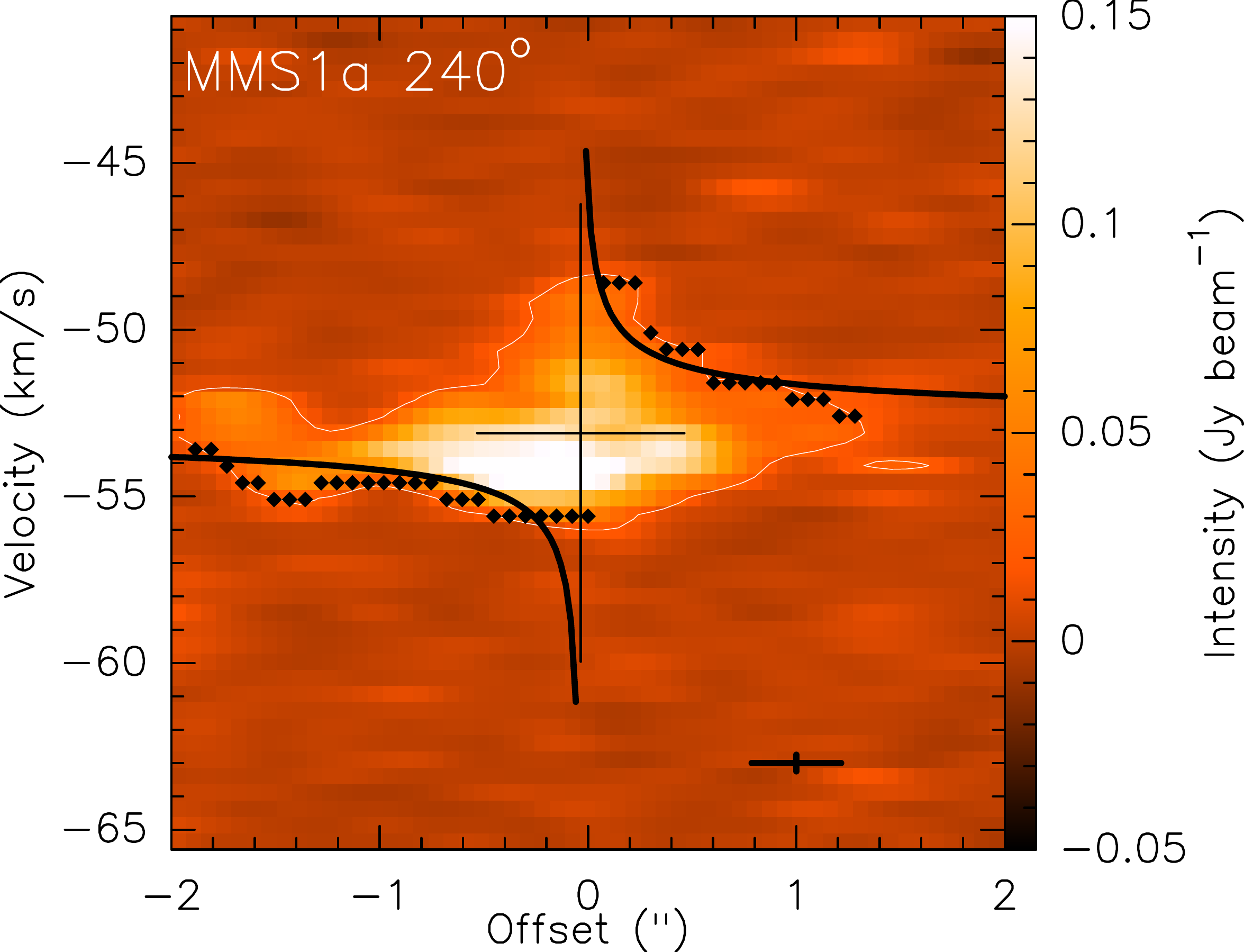}}
	\resizebox{.45\hsize}{!}{\includegraphics{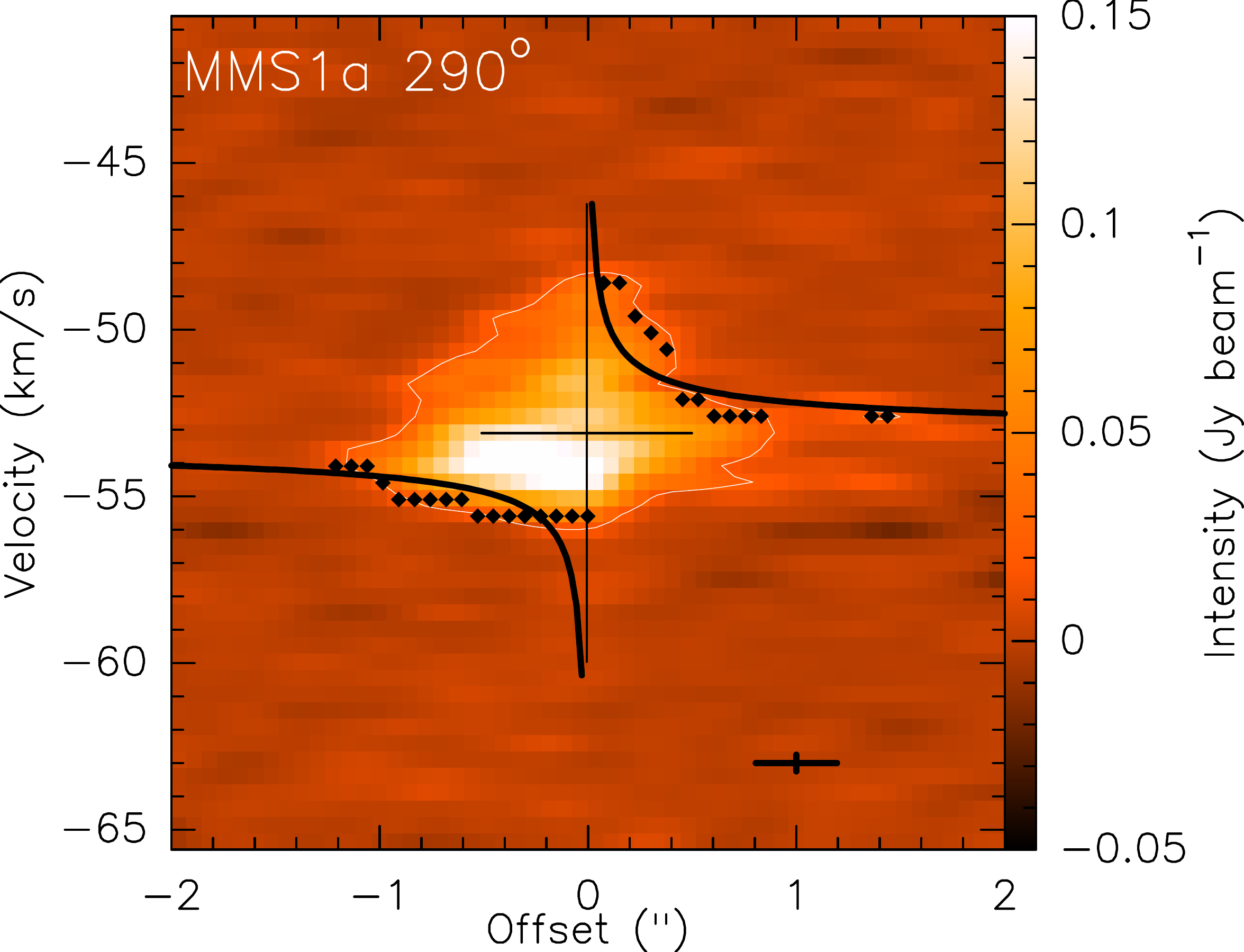}}
	\resizebox{.45\hsize}{!}{\includegraphics{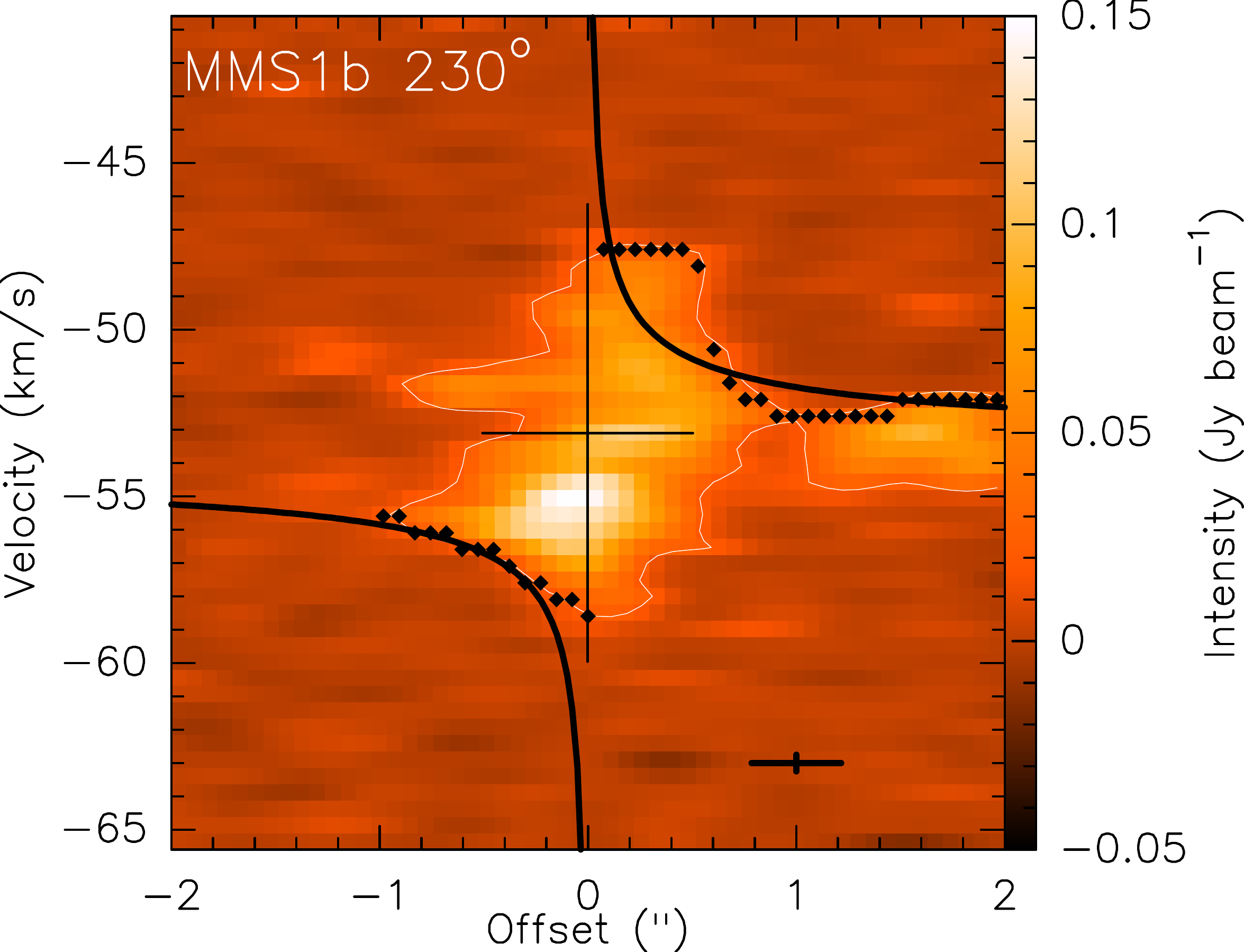}}
	\resizebox{.45\hsize}{!}{\includegraphics{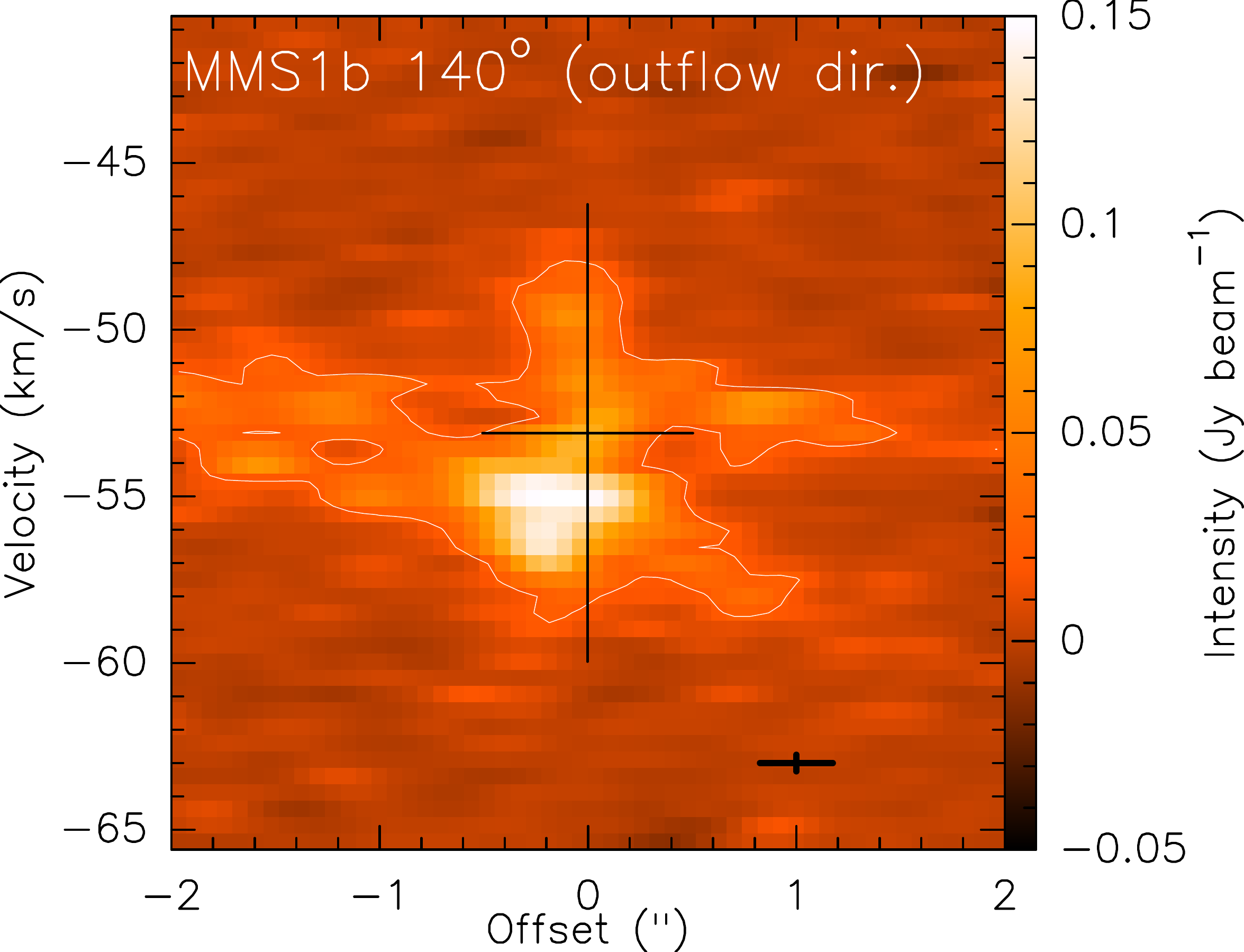}}
	\caption{Position-velocity diagrams for the H$_2$CO (3$_{0,3}$ -- 2$_{0,2}$) transition towards the two major cores MMS1a~\&~b from the NOEMA data, with a synthesized beam of $\approx 0.43\arcsec \times 0.35\arcsec$ (PA~61\degr). The cut directions are shown in Fig.~\ref{fig:kinematics}. The bottom right PV diagram is from a cut through MMS1b, perpendicular to the left cut and along the inferred outflow axis (see Sect.~\ref{sec:outflows}). The white contour shows the 4\,$\sigma$ detection threshold. The black vertical and horizontal bars indicate the position of the continuum emission peak and the systemic velocity, respectively. Their lengths correspond to $1\arcsec \approx 4300$\,au and 14\,km\,s$^{-1}$. The black dots indicate the detection with the most extreme velocity at each position. The black solid curves indicate the Keplerian velocities, corresponding to the mass estimates in Table~\ref{tab:best_fit}. The black crosses in the lower right corners indicate the resolution elements along both axes.}
	\label{fig:pv_diagrams}
\end{figure*}

We analyze the velocity structure by means of position-velocity (PV) diagrams (see Fig.~\ref{fig:pv_diagrams}) along the gradients in Fig.~\ref{fig:line_emission_zooms}, where the PV cuts follow the arrows in the kinematics overview, in Fig.~\ref{fig:kinematics}. Before deriving protostellar mass estimates, we need to characterize the velocity profiles in the corresponding PV diagrams, since the kinematic mass estimates presented below are based on the assumption that the material is in disk-like Keplerian rotation around the central young stellar object (YSO).

\subsubsection{Characterization of velocity profiles}
In Fig.~\ref{fig:pv_diagrams}, we plot PV diagrams of H$_2$CO (3$_{0,3}$ -- 2$_{0,2}$), which is a good tracer of the outer (rotating) structure. We note that we postpone the analysis of more typical high-density gas tracers such as CH$_3$CN to the Appendix (Sect.~\ref{sec:test_chemical_species}) due to their comparably weak emission towards IRAS~23033+5951. Furthermore, the strong low-$K$ emission lines of this molecule are blended, leaving only the weaker high-$K$ lines for the analysis. Due to the lack of an unambiguous velocity gradient towards MMS1a, we present in Fig.~\ref{fig:pv_diagrams} two PV diagrams for this core, with $\phi = 240\degr$ (counting from north to east, solid black line in Fig.~\ref{fig:kinematics}) and $\phi = 290\degr$ (dashed black line), where the second is motivated by the $^{13}$CO and CH$_3$CN velocity gradient and the first is roughly perpendicular to the outflow axis and to the elongation of the cm emission.

We compare the PV diagrams in Fig.~\ref{fig:pv_diagrams} to the results from \citet{Ohashi97}. In their Fig.~11, the authors show PV diagrams for rigid body rotation with and without infalling material and also the signature of Keplerian rotation without infall. In the infall-only scenario, the diagram is expected to be symmetric about both axes, which is only roughly the case in the 240\degr  plot of MMS1a. However, the emission in the lower left and upper right quadrants dominate their counterparts, which is an indicator of rotation about the central YSO and also apparent in the other plots. At this point, we note that we cannot distinguish clearly between rigid body rotation with infall and Keplerian rotation, where the both scenarios are expected to show emission almost only in two of the four quadrants. This is because of the spread of intensity due to the finite beam size -- we find the highest velocities distributed over $\sim 0.5\arcsec$ along the positional axis, which is roughly the beam size -- what indicates that emission is spread into the other quadrants and that the underlying profile thereby is obscured.

The black curves in the plots indicate Keplerian rotation for a given protostellar mass (see Sect.~\ref{sec:kinematic_mass_estimates} for details). The respective intensity distribution in the three plots is in qualitative agreement with a Keplerian rotation profile, but we do not detect emission at the highest expected velocities in the very center at the given sensitivities. This lack of the most extreme velocities may occur due to the limited sensitivity and spatial (and spectral) resolution -- as \citet{Krumholz07} showed with synthetic observations -- or due to optical depth effects, either in the continuum or the lines, close to the central position such that emission from the central region is hidden. The effect of the limited spatial resolution was mentioned above and suggests that we may detect the highest velocities from the center but spread over the size of the synthesized beam.
For both sources, we find the structure to follow the Keplerian velocity profile up to a radius of $\sim 1\arcsec$ (4300\,au). A comparison to other rotating disks and toroids around high-mass stars \citep[Table~2]{Beltran16} shows that this is a typical scale for objects of similar mass.

\subsubsection{Kinematic mass estimates}
\label{sec:kinematic_mass_estimates}
Now assuming the scenario of Keplerian disk-like rotation around a central YSO, the mass of the latter is traced by the velocity distribution of the material, since the highest velocity at a given radial distance $r$ to the center of mass is limited by the Kepler orbital velocity $v_\mathrm{Kepler}(r)$:
\begin{equation}
	v_\mathrm{Kepler}(r) = \sqrt{G\frac{M_\star+M_\mathrm{disk}(r)}{r}} \approx \sqrt{\frac{GM_\star}{r}}
	\label{eq:kepler_velocity}
\end{equation}
In this equation, $G$ is the gravitational constant, $M_\star$ is the mass of the central object(s), and $M_\mathrm{disk}(r)$ is the mass of the disk, enclosed inside the radius $r$. At this point, we assume that the gravitational potential is dominated by the central YSO and hence we neglect the disk-mass term in Eq.~\ref{eq:kepler_velocity}, where we conduct an a-posteriori check on this assumption in the last paragraph of this subsection.

\citet{Seifried16} presented a method for estimating the highest velocity at a given radial distance from the central object, by evaluating position-velocity (PV) diagrams. In Sect.~4.2 of that paper, the authors propose to start for each position at the most extreme velocities and iterate over the velocity channels until the first position-velocity pixel above a certain threshold is found. They find this method to be the most robust among those they tested on synthetic PV data where the true YSO mass was known.

We apply this method to the PV diagrams in Fig.~\ref{fig:pv_diagrams}, to estimate the highest velocity as a function of the radial distance to the central YSO (black diamonds) and to finally fit the result with a function of Keplerian tangential velocity, according to Eq.~\eqref{eq:kepler_velocity}.
For the fitting procedure, we use the \textsc{python} package \textsc{astropy.modeling}\footnote{\textsc{python} package \textsc{astropy.modeling}, \url{http://docs.astropy.org/en/stable/modeling/}} \citep{Astropy13}. We add the parameters $v_0 \approx v_\mathrm{LSR}$ and $r_0$ to the model function to account for deviations in the local standard of rest (LSR) and for the resolution-limited position of the central object. A detailed description of the fit procedure is presented in Appendix~\ref{sec:discussion_pv_fit} along with a variety of tests of this method, to explore the effects of various input parameters such as the weighting parameter, applied during the imaging process. The code is made available online.\footnote{\textsc{KeplerFit}, \url{https://github.com/felixbosco/KeplerFit}}

The curves in the PV diagrams of the H$_2$CO (3$_{0,3}$ -- 2$_{0,2}$) transition were obtained from this method choosing a detection threshold of 4\,$\sigma$. The best-fit parameters of the modeling procedure are listed in Table~\ref{tab:best_fit}. The velocity and position offsets do not differ significantly from the literature value of $v_\mathrm{LSR} = -53.1$\,km\,s$^{-1}$ and the mm~emission peak, varying only by fractions of a pixel size.

\begin{table}
	\centering
	\caption{Best fit parameters from the H$_2$CO (3$_{0,3}$ -- 2$_{0,2}$) PV diagram evaluation.}
	\label{tab:best_fit}
	\begin{tabular}{lrr}
		\hline \hline
		Core				 	& $M_\star \cdot \sin^2 i$ 	& $v_\mathrm{LSR}$  \\
							&  (M$_\sun$)			& (km\,s$^{-1}$) 	\\ \hline
        	MMS1a ($\phi = 240\degr$) & $8.6 \pm 0.6$			& $-53.4 \pm 0.1$	\\
        	MMS1a ($\phi = 290\degr$) & $6.3 \pm 0.8$			& $-53.8 \pm 0.1$	\\
		MMS1b 				& $21.8 \pm 2.0$		& $-54.3 \pm 0.1$	\\
		\hline
        	\multicolumn{3}{c}{\emph{error weighted average over all molecules:}} \\
       		\hline
        	MMS1a				& $5.8 \pm 0.3$		& \\
        	MMS1b				& $18.8 \pm 1.6$	& \\
        	\hline
	\end{tabular}
\end{table}

Mass estimates obtained from this method are uncertain (see Appendix~\ref{sec:discussion_pv_fit}). An obvious source of uncertainty is the yet unconstrained disk inclination $i$, as we obtain only a fraction of the protostellar mass $M_\mathrm{fit} = M_\star \cdot \sin^2 i$ \citep[cf. e.g. Fig.~7 in][for PV diagrams for three different inclinations]{Jankovic19}. In fact, a comparison between the $240\degr$ panel of MMS1a with the panels for inclinations of $i=30\degr$ in Fig.~6 and~7 of \cite{Jankovic19}, suggests that we detect rotation under significant inclination towards this core, since we do not detect an emission gap around the $v_\mathrm{LSR}$ between blue and red shifted emission, which is expected for edge-on disk inclinations (cf. their Fig.~6 and~7). Deducing the presence of spiral arms from the substructure of emission in the diagram, however, seems to be unreasonable due to the limited spatial resolution. This makes our mass estimate a lower limit with respect to inclination. A less accessible source of uncertainty, however, is the effect of the size of the synthesized beam or of line broadening, as mentioned above. These spread the emission in the spatial and velocity direction, respectively, thereby populating more extreme pixels in the PV diagrams, where the larger spread clearly provides larger mass estimates (cf. Fig.~\ref{fig:test_weighting_schemes}). However, we refrain from characterizing this effect as we would need a deconvolution of the image prior to selecting the PV pixels from the data cube, and postpone a more extensive study of the uncertainties to a following paper by Ahmadi et~al. (in prep.).

For making our estimate more robust, we also include the transitions CH$_3$CN ($12_3 - 11_3$) and CH$_3$OH (4$_{+2,2,0}$ -- 3$_{+1,2,0}$) and compute the average mass from all estimates for a given core, see bottom rows in Table~\ref{tab:best_fit}, where we weight with the respective model error. These provide significantly lower mass estimates than the H$_2$CO-only estimates because most of the molecules do not show such extended emission as H$_2$CO and the overall signal-to-noise is lower, especially towards the more extreme velocities. For the core MMS1a, we have included both PV cut angles in the averaging process and find that the $\phi = 290\degr$ estimate from H$_2$CO of $6.3 \pm 0.8$\,M$_\sun$ is in better agreement with the average estimate of $5.8 \pm 0.3$\,M$_\sun$, in contrast to the $\phi = 240\degr$ estimate of $8.6 \pm 0.6$\,M$_\sun$. Both estimates for the core MMS1b are in agreement with the 19\,M$_\sun$ YSO mass estimate, which \cite{Rodriguez12} obtained from modeling H$_2$O maser emission.

We conduct an a-posteriori check on the assumption from above that the contribution of the disk mass to the gravitational potential is negligible compared to the YSO mass $M_\star$. Therefore, we assume a typical gas density structure and $M_\mathrm{disk}(r) = M_\mathrm{core} \cdot \left(r / R_\mathrm{core}\right)^{3/2}$, neglecting the envelope contribution to the core mass. We assess a lower limit for the radius $r_\mathrm{dom}$ out to which the rotating structure should be dominated by the central YSO, i.e. $M_\star \leq M_\mathrm{disk}(r_\mathrm{dom})$. We use the core masses from Table~\ref{tab:core_masses} and the core radii of 5081 and 3915\,au for MMS1a \&~b, respectively \citep[Table~5]{Beuther18}. If we now compare this to our YSO mass estimates from Table~\ref{tab:best_fit}, we get an a posteriori confirmation and obtain that in MMS1b, the gravitational potential in the core should be completely dominated by the YSO, whereas in the case of MMS1a this is only true out to half the core radius, $\sim 2000$\,au. However, this is only a lower limit as the MMS1a kinematic mass estimate should only be treated as a lower limit and thereby we argue that the disk mass should indeed be negligible in the analysis of these two cores. Still, we have to note that also the core masses are only lower limits due to the flux filtering effect.

\subsubsection{Accelerating or decelerating material}
We investigate the H$_2$CO (3$_{0,3}$ -- 2$_{0,2}$) emission for signatures of in- or outflowing material. Therefore we create a PV diagram (see {bottom right panel} in Fig.~\ref{fig:pv_diagrams}) for this transition along the presumable MMS1b outflow axis ($\phi=220\degr$), obtained from the analysis in Sect.~\ref{sec:outflows}. We identify a signature in the lower-right quadrant of the corresponding PV diagram with more blue-shifted velocities towards larger radii, to the north-west. This indicates that gas is either accelerating outwards or decelerating inwards, relative to the central velocity. Thus, this PV diagram suggests that the elongated structure represents either a feeding flow onto the core or a molecular outflow (see Sect~\ref{sec:kinematics} for further discussion).

\section{Analysis and discussion}
\label{sec:discussion}

\subsection{Hierarchical fragmentation}
\label{sec:discussion_fragmentation}
From the continuum emission at 1.37\,mm in Fig.~\ref{fig:continuum_emission}, we infer the fragmentation of IRAS~23033+5951, where the angular resolution of $0.45\arcsec \times 0.37\arcsec$ corresponds to a spatial resolution element of $\sim 1900$\,au, at a distance of 4.3\,kpc. In Sect.~\ref{sec:continuum}, we report that the mm~source MMS1 \citep[detected at an angular resolution of $\sim 5\arcsec$,][]{Reid08}, fragments into at least two cores. The three major cores MMS1a, MMS1b, and MMS2a are coincident with the 3\,mm detections by \citet{Schnee09}, but the lower intensity structures, such as MMS1c, appear only in the high-angular-resolution, sensitive interferometric data presented in this work. We obtain a hierarchy of three levels of fragmentation in the HMSFR IRAS~23033+5951:
\begin{enumerate}
	\item Large scale clumps with connected 1.37\,mm-continuum emission above the 5\,$\sigma$ detection limit: Within the map of continuum emission, we have the two mm~emission clumps MMS1 and MMS2, being separated from each other by a projected distance of more than 20\,000\,au\,$\approx 0.1$\,pc.
	\item Separated cores within the major clumps: As identified by the \textsc{clumpfind} algorithm, we find the MMS1 clump to be composed of at least two cores. The projected separations of the emission peaks are on the order $5000-9000$\,au.
	\item Indications towards further fragmentation: As mentioned above, we find indications of further clump fragments, i.e. 5\,$\sigma$ detections being separated from the major cores by $\sim 0.5 - 1.0''$, corresponding to $2000-4000$\,au, see panels (b) and (g) in Fig.~\ref{fig:continuum_emission}. However, with the given data there is no clear evidence for protostellar cores within these groups of weak sources.
\end{enumerate}

\subsubsection{Comparison to Jeans fragmentation}
We derive the Jeans fragmentation length $\lambda_\mathrm{J}$ and mass $M_\mathrm{J}$, using the mean density estimate for IRAS~23033+5951 from \citet{Beuther02dust} of $3.6 \times 10^5\,\mathrm{cm}^{-3}$. We compute $\lambda_\mathrm{J}$ and $M_\mathrm{J}$ for three dust temperature regimes, i.e. for cold dust with $T_\mathrm{dust}=22$\,K \citep{Maud15}, for the 55\,K estimate from \cite{Beuther18}, and for the hot core dust temperatures $T_\mathrm{dust}\approx100$\,K, from Sect.~\ref{sec:XCLASS_fits}. We note that the above dust temperatures are inferred from gas temperatures under the assumption that that gas and dust are well coupled (i.e. $T_\mathrm{dust} \approx T_\mathrm{gas}$), which is reasonable at the given high densities. We obtain 6000\,au and 0.32\,M$_\sun$, 9500\,au and 1.27\,M$_\sun$, and 12\,500\,au and 3.1\,M$_\sun$ for the three temperature regimes, respectively.
The mass estimates of the three major cores (see Table~\ref{tab:core_masses}) are significantly larger than the corresponding Jeans masses of the hot core temperature regime while the faint core MMS1c roughly has the Jeans mass for the intermediate temperature regime. This core is relatively close to MMS1b, at a projected distance of $\sim5200$\,au, and the projected distance between MMS1a~\&~b of 9000\,au is also smaller than expected from the Jeans analysis. However, we have to note that the mass estimates from above underlie large uncertainties due to the various underlying assumptions.
A more extensive study of the fragmentation of a larger number of clumps is presented in \citet{Beuther18}.

\subsubsection{On the preferred axis of structure and fragmentation}
\begin{figure}
	\centering
	\resizebox{.96\hsize}{!}{\includegraphics{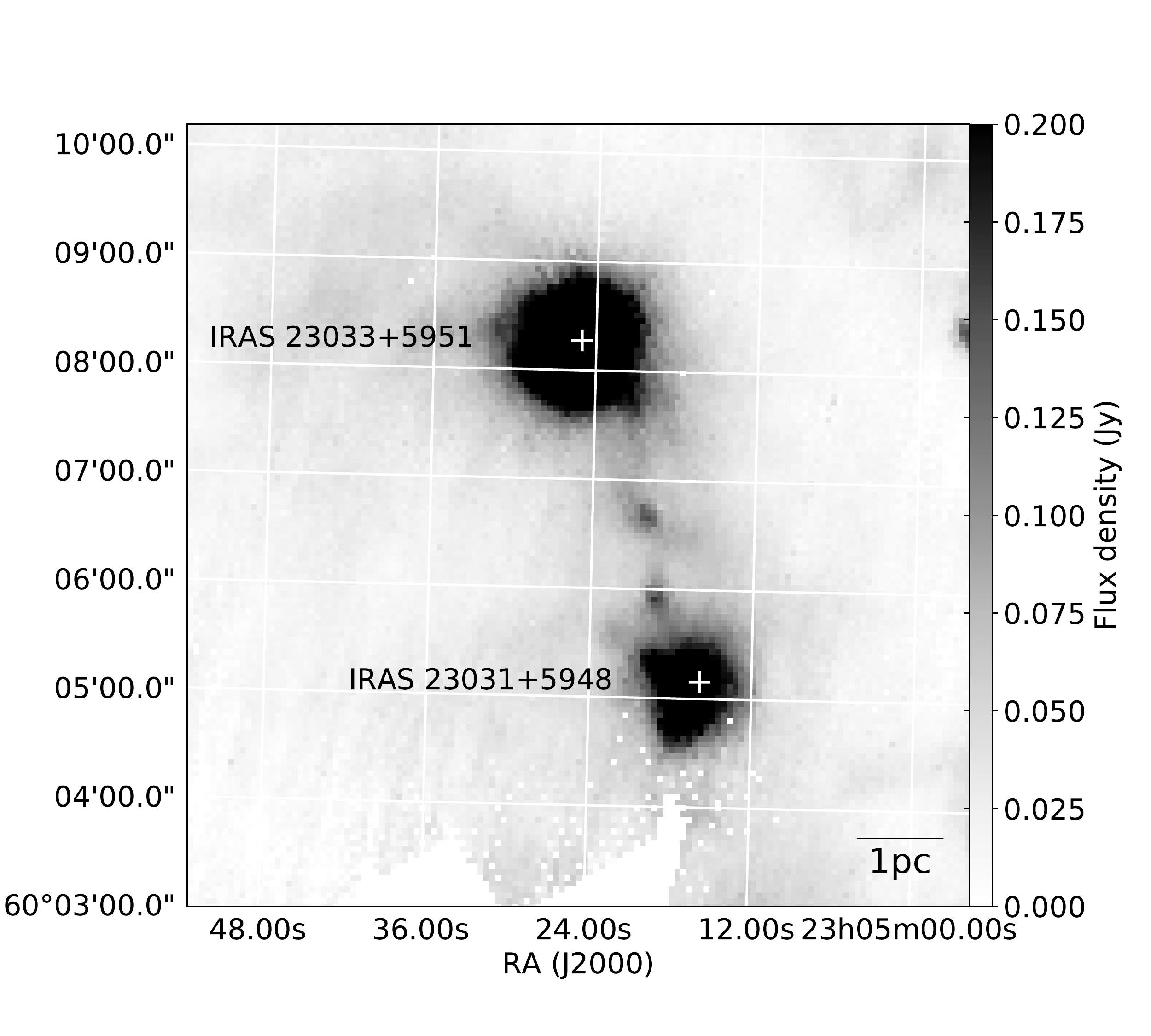}}
	\caption{Herschel/PACS observations at $160\mu$m towards IRAS~23033+5951 and IRAS~23031+5948, from the Hi-GAL survey  \citep{Molinari10}.}
	\label{fig:farinfrared}
\end{figure}

Interestingly, the structure of IRAS~23033+5951 appears to be elongated along an axis from the north-north-east to the south-south-west. All three major structures lie on this axis and the minor cores are located not further away than 2\arcsec. We seek for larger scale patterns and compare this to FIR data from the Hi-GAL survey \citep{Molinari10}, see Fig.~\ref{fig:farinfrared} for the 160\,$\mu$m emission. In these data, we find another source, IRAS~23031+5948, to be located along this axis, at about $3\arcmin$ (corresponding to $\sim 4$\,pc) to the south-southwest, with some intermediate emission forming a 'bridge' between the two IRAS sources. In the IRAM 30-m single dish observations, covering a field of view of $1.5\arcmin \times 1.5\arcmin$, we find that the $^{13}$CO and C$^{18}$O emission extend from the central clump towards the south-southwest, confirming a molecular connection to the other source.

\cite{Reid08} analyze PV data along this axis, across IRAS~23033+5951, and report on a velocity gradient of $\sim4$\,km\,s$^{-1}$ over $\sim40\arcsec$, seen in H$^{13}$CO$^+$ interferometric data. They interpret this as a large scale rotation of a flattened structure with a major axis of about 0.5\,pc. In the context of the larger-scale FIR emission, this may now also be interpreted as some kind of molecular accretion flow along a filamentary structure onto the major cores. A large scale structure with a comparable velocity gradient towards G35.20-0.74N has been discussed to resemble either a flattened rotating object or a filamentary structure, where \cite{Sanchez14} find the latter hypothesis to be the more plausible explanation for the regular fragmentation pattern. Another similar example of such a molecular flow onto high-mass protostellar cores on slightly smaller scales is presented by Mottram et~al. (subm.), reporting the flow of material along a molecular stream across several cores onto the most luminous core in W3\,IRS4. Other examples of similar accretion flows are reported by, e.g., \citet{FernandezLopez14}, \citet{Peretto14} and \citet{Tackenberg14}, or more recently by \citet{Lu18filament}, \citet{Veena18} and \citet{Yuan18}.

However, it needs further observational data covering the large-scale environment of IRAS~23033+5951 at a decent velocity resolution $\lesssim 1$\,km\,s$^{-1}$ to analyze the gas kinematics and to finally address the question, whether or not this indeed indicates a larger scale filamentary flow or a fragmentation scheme which is inherited from the larger scales.

\subsection{Kinematics of the molecular gas}
\label{sec:kinematics}
The mm~source MMS1 shows a complex kinematic structure which is likely dominated by the two cores MMS1a~\&~b, at the scale of the angular resolution of the presented data ($\sim0.45\arcsec$, corresponding to 1900\,au).

\subsubsection{Disentanglement of molecular outflows and characterization of velocity gradients}
\label{sec:outflows_discussion}
To disentangle the complex kinematic structure of this region, we consider the velocity gradients from Sect.~\ref{sec:moment_maps}, the PV diagrams from Fig.~\ref{fig:pv_diagrams} and the molecular outflow structure from Sect.~\ref{sec:outflows} along with the continuum emission at 3.6\,cm \citep{Beuther02maser,Rodriguez12} and the maser analysis by \citet{Rodriguez12} for the following cores:

\paragraph{The core MMS2a:}
Molecular outflows are often found as symmetric structures centered around their launching core. The emission of the outflow tracing species $^{13}$CO and SO, however, does not reveal any outflow structure towards the position of the core MMS2. Hence, we have no evidence that MMS2a has already launched a molecular outflow. Interesting is, however, the detection of Class I methanol masers at 44\,GHz \citep[$7_{0} - 6_{1}$,][]{RodriguezGarza17} and 95\,GHz \citep[$8_{0} - 7_{1}$,][]{Schnee09}, exclusively towards the west and north-east of MMS2a but not towards MMS1 (cf. Fig.~\ref{fig:continuum_emission}). These Class I masers indicate the presence of shocked gas \citep[e.g.][]{Leurini16}, an implication that is difficult to address for MMS2a with our data.
Furthermore, the emission of the dense-gas tracers is too weak towards this core to infer any velocity gradient in the dense gas. In the following, we therefore focus on the northern two cores MMS1a~\&~b, both hosting outflow-driving protostellar candidates.

\paragraph{The core MMS1b:}
The characterization of the core MMS1b is the least ambiguous one, since we have identified a well defined outflow axis at a projected position angle $\phi_\mathrm{out} \approx 315\degr$. The interpretation as a molecular outflow is consistent with $^{13}$CO and SO channel maps and the outwards-accelerating structure, which we found in the corresponding panel of the H$_2$CO PV diagrams along the outflow axis. Furthermore, we see a strong velocity gradient of $\sim$\,5\,km\,s$^{-1}$ over $\sim2\arcsec$ across this core (black arrow in Fig.~\ref{fig:kinematics}), which is oriented perpendicular to the presumed outflow axis at $\phi_\mathrm{grad} \approx 230\degr$. \citet{Rodriguez12} analyzed the H$_2$O maser group M1 in terms of disk rotation. The best-fit disk model has a radius of $0.03\arcsec$, corresponding to 135\,au, and a position angle of $65\degr \pm 1\degr$ which is in agreement with the inferred velocity gradient in the outer rotating structure. We note, that the position of this maser group \citep{Rodriguez12} deviates slightly from the MMS1b mm~emission peak. However, this deviation of only half the beam width is below the resolution of our data.

The collected information suggests an outflow axis roughly perpendicular to the rotating structure in the core with $\Delta \phi = 70\degr - 85\degr$. Since outflows are launched from circumstellar disks \citep[e.g.][]{Kolligan18}, these findings consistently suggest the presence of a disk-outflow system in MMS1b.

However, it is not clear yet to what extent the velocity gradient stems from a circumstellar disk in Keplerian rotation and/ or from the rotation of the protostellar envelope. To adress this question, we can use the emission of dense gas tracing molecules like CH$_3$OH and CH$_3$CN. In the zoomed-in velocity maps in Fig.~\ref{fig:line_emission_zooms}, we see the gradient for CH$_3$CN, with $\phi_\mathrm{grad} \approx 225\degr$ (cf. Fig.~\ref{fig:kinematics}), whereas the lower-density gas tracing molecules differ in the position angle. This indicates that the signal from these latter molecules has some contribution from envelope material either flowing inward to be accreted onto the core or flowing outward to be removed from the core in form of a molecular outflow.

\paragraph{The core MMS1a:}
In contrast to MMS1b, we do not clearly identify an elongated but narrow molecular emission structure towards the core MMS1a in the integrated intensity maps in Fig.~\ref{fig:line_emission} or in the channel maps in Fig.~\ref{fig:channel_maps}. If we assume that the blue and red shifted emission, which does not appear to belong to the collimated outflow from MMS1b, stems from an outflow from MMS1a, then we can reconstruct the outflow axis in Fig.~\ref{fig:outflows}. As these two lobes cover large areas, we state the position angle as $\phi_\mathrm{out} \approx 350\degr$ with large uncertainties of $\sim \pm 30\degr$. A comparison of these two lobes in Fig.~\ref{fig:outflows} to the large-scale outflow lobes from Fig.~1 in \citet{Beuther02outflow} suggests that MMS1a and MMS1b together form the large-scale outflow structure seen in their data with lower spatial resolution. In this picture, MMS1a launches a large scale and less collimated outflow, in contrast to the scenario in which MMS1b launches a collimated outflow. However, the jet-scenario for the cm emission does not support this outflow axis, as the VLA1--VLA3 axis is tilted by 70--80$\degr$ with respect to it (see Fig.~\ref{fig:kinematics}). This latter finding is in better agreement with cm~emission stemming from an ionized disk-like structure, but this will be discussed further in Sect.~\ref{sec:cm_emission}.

\subsubsection{Stability of Keplerian disks}
\label{sec:disk_stability}
Simulations of the formation of high-mass stars suggest that these objects accrete mass via massive disks, which tend to form spiral arms and fragment under self-gravity \citep[e.g.][]{Meyer17,Meyer18}. In this section, we now analyze the rotating structures in the dominant cores in terms of fragmentation by the gravitational collapse due to possible instabilities against axisymmetric perturbations. Implicitly, we assume that the rotating structures are disks in equilibrium and in Keplerian rotation, which is a reasonable assumption from the results in Sect.~\ref{sec:PVdiagrams}. For this analysis, we make use of the stability criterion for a self-gravitating disk, derived by \citet{Toomre64}:

\begin{equation}
	Q = \frac{c_\mathrm{s} \cdot \Omega_\mathrm{epi}}{ \pi G \cdot \Sigma}
	\label{eq:Toomre_Q}
\end{equation}
In this equation,  $ c_\mathrm{s} $ is the local speed of sound, $ \Omega_\mathrm{epi} $ is the epicyclic frequency, $ G $ is the gravitational constant and $ \Sigma $ is the disk surface density. \citet{Toomre64} found that rotating disks are unstable against axisymmetric perturbations for $ Q \lesssim Q_\mathrm{crit} = 1 $.
We note that the exact critical value is under current debate, as e.g. \citet{BinneyTremaine} report critical values up to $Q_\mathrm{crit} \sim 2$, when applying different assumptions on the disk temperature and density profiles.
In contrast to this, the studies by e.g. \citet{Behrendt15} showed that the critical value drops below 0.7 for sech$^2$ density profiles of the disk in $z$ direction. \citet{Takahashi16} revised the general picture towards spiral arm formation in protoplanetary disks for $Q \lesssim 1$, where these spiral arms fragment only if the arm-internal $Q$ drops below 0.6, see their Fig.~1. This value has been confirmed by the simulations of \citet{Klassen16} and \citet{Meyer18}, who found that only regions with $Q<0.6$ indeed start fragmentation.
With the discussion above in mind, we conduct the following Toomre~$Q$ stability analysis with $Q_\mathrm{crit}=1$ and a stable (unstable) regime for a larger (smaller) values of $Q$, where the unstable regime indicates potential spiral arm formation, eventually leading to fragmentation, if the local $Q$ drops below 0.6.

We calculate maps of the $Q$ parameter for the inner $ 2.2\arcsec \times 2.2\arcsec $ around the peaks of mm~continuum emission for the two major cores MMS1a~\&~b. Again, we exclude MMS2a from the analysis due to the absence of CH$_3$CN emission, resulting in insufficient information on the gas excitation temperature towards this core. We compute maps for each of the three variables, using the following equations:

\begin{align}
	&c_s = \sqrt{\gamma \frac{k_\mathrm{B}T}{\mu m_\mathrm{H}}} \\
	&\Sigma = \mu m_\mathrm{H} \cdot N_{\mathrm{H}_2} \\
	&\Omega_\mathrm{epi} \equiv \Omega_\mathrm{ang} = \frac{v_\mathrm{Kepler}(r)}{r} \approx \sqrt{\frac{GM_\star}{r^3}}
	\label{eq:epicyclic_frequency}
\end{align}
with the adiabatic coefficient $\gamma \approx 7/5$ for primarily diatomic gas, Boltzmann's constant $ k_\mathrm{B} $, the dust temperature $T$, the mean molecular mass $\mu m_\mathrm{H}$, the molecular hydrogen column density $N_{\mathrm{H}_2}$, the Keplerian angular velocity $\Omega_\mathrm{ang}$, and the Kepler orbital velocity $ v_\mathrm{Kepler}(r) $ from above. For the map of the speed of sound, we again use the temperature maps in Fig.~\ref{fig:temperature_maps} which we assume to be equal to the dust temperature, as discussed above. We obtain the surface density $ \Sigma $ by multiplying the H$_2$ column density map in Fig.~\ref{fig:column_density_maps} with the mean molecular weight $ \mu m_\mathrm{H} $, where we again emphasize that we computed the H$_2$ column density under the assumption of thermal coupling between the CH$_3$CN line emitting gas and the dust. The epicyclic frequency is identically equal to the angular velocity $\Omega_\mathrm{ang}$, in the case of Keplerian rotation \citep{Pringle07}, where this is computed from the mass estimates in Table~\ref{tab:best_fit} and a map of the orbital distance to the central HMPO. We note that the epicyclic frequency is treated as an lower limit, since the kinematic mass estimates are lower limits, as mentioned above, and that the surface density is treated as lower limits due to flux filtering effects.
Further considerations on the computation of the individual parameters are presented in Appendix~\ref{sec:appendix_ToomreQ}.

\begin{figure*}
	\centering
	\resizebox{.45\hsize}{!}{\includegraphics{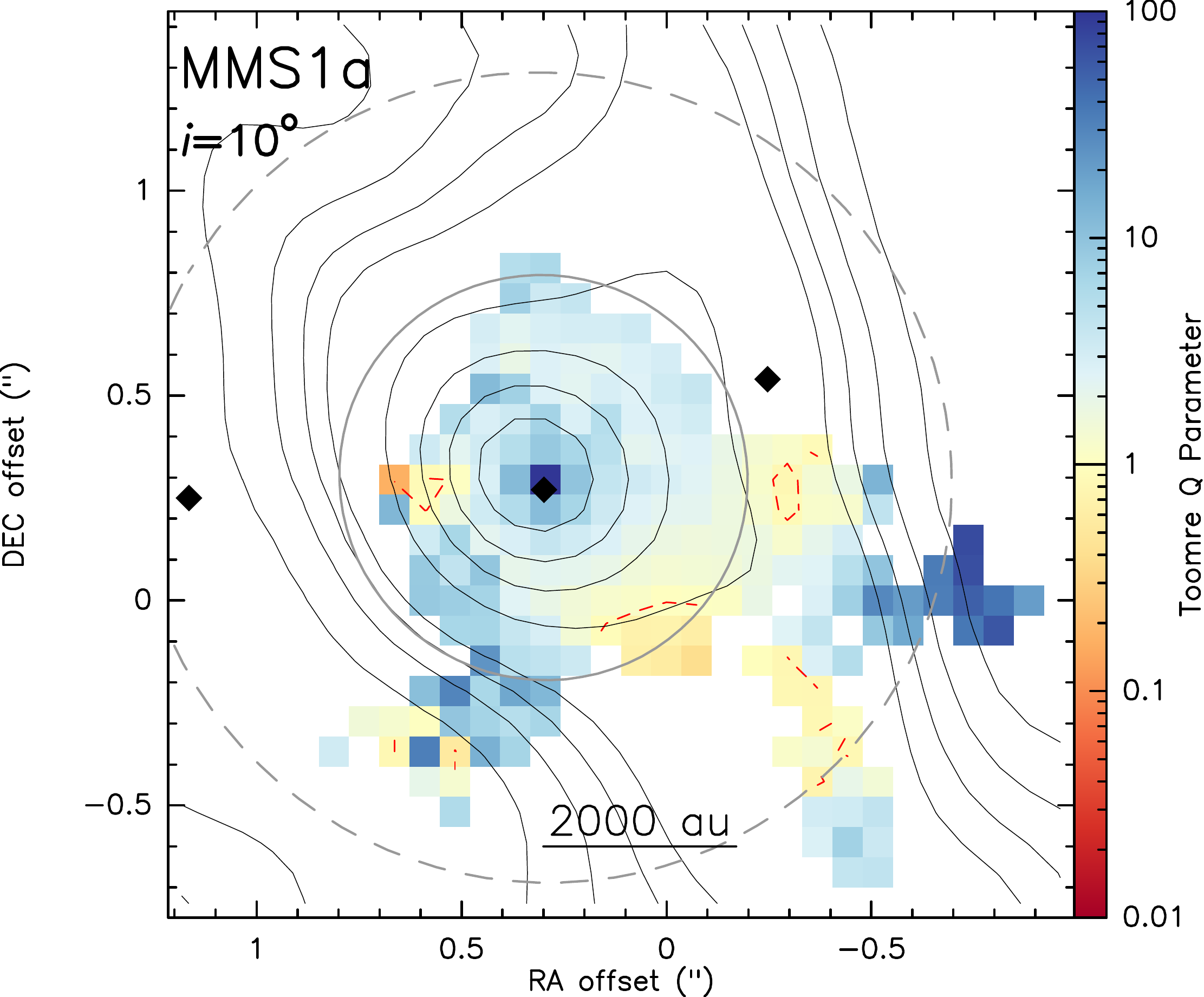}}
	\resizebox{.45\hsize}{!}{\includegraphics{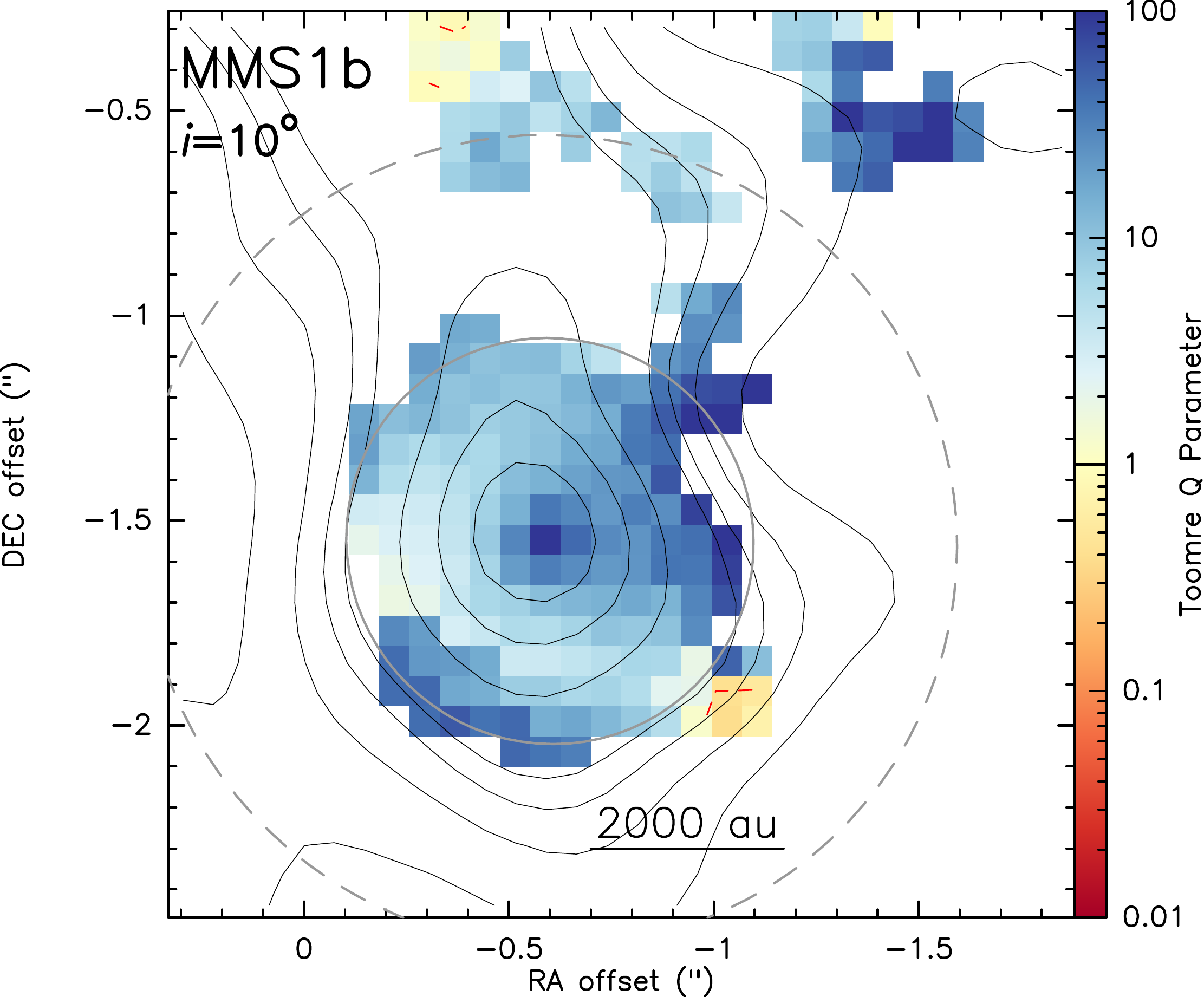}}
	\resizebox{.45\hsize}{!}{\includegraphics{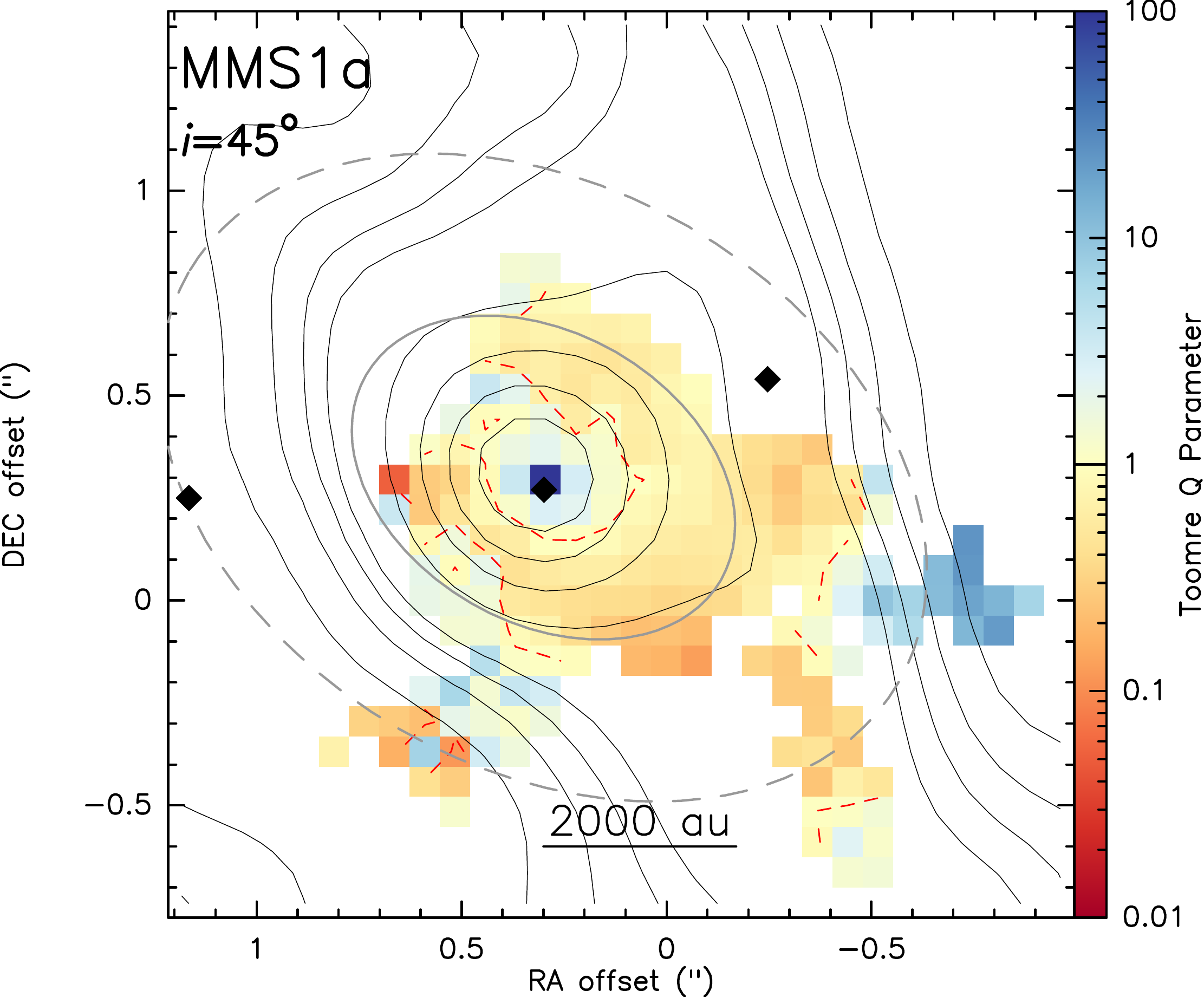}}
	\resizebox{.45\hsize}{!}{\includegraphics{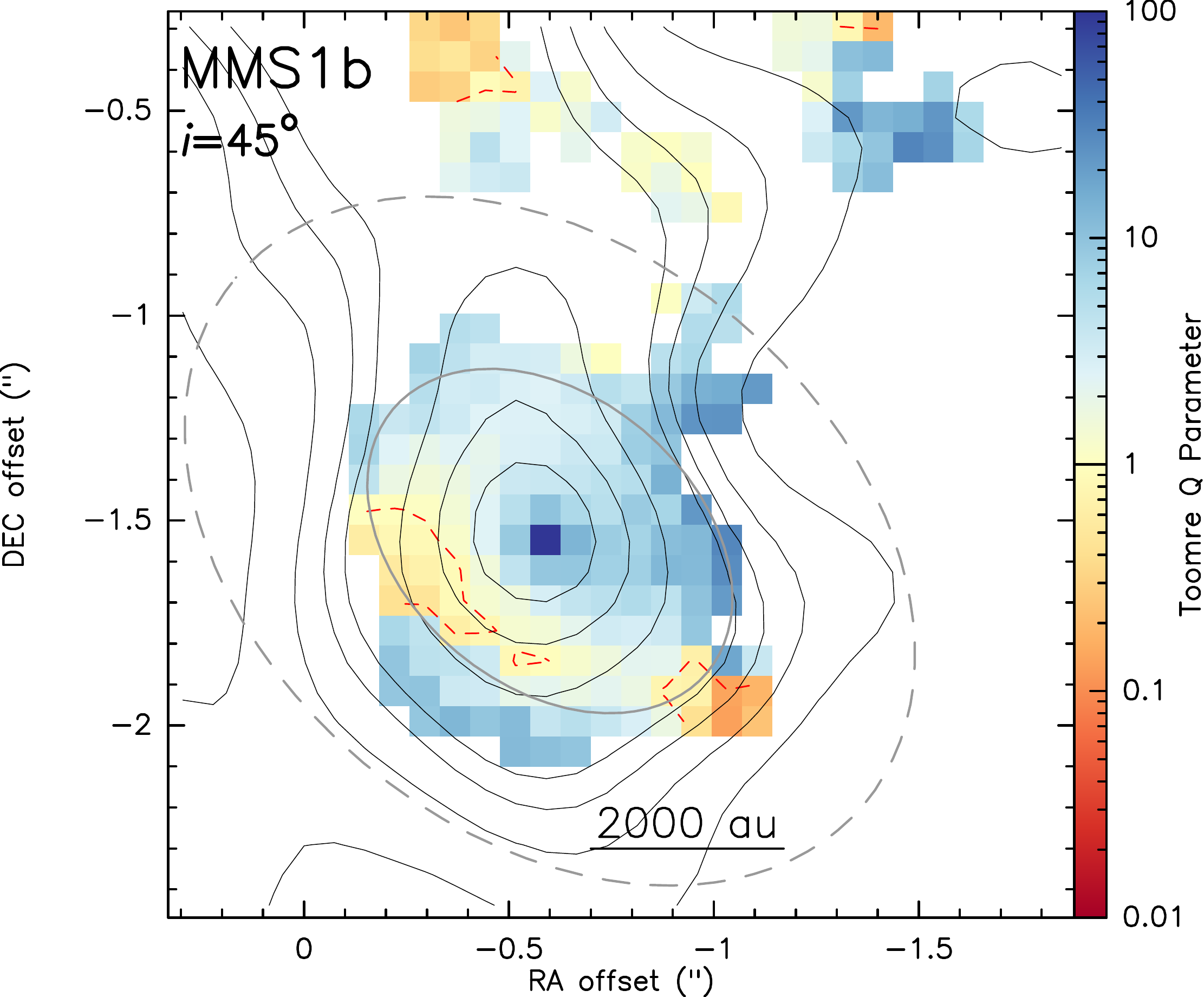}}
	\resizebox{.45\hsize}{!}{\includegraphics{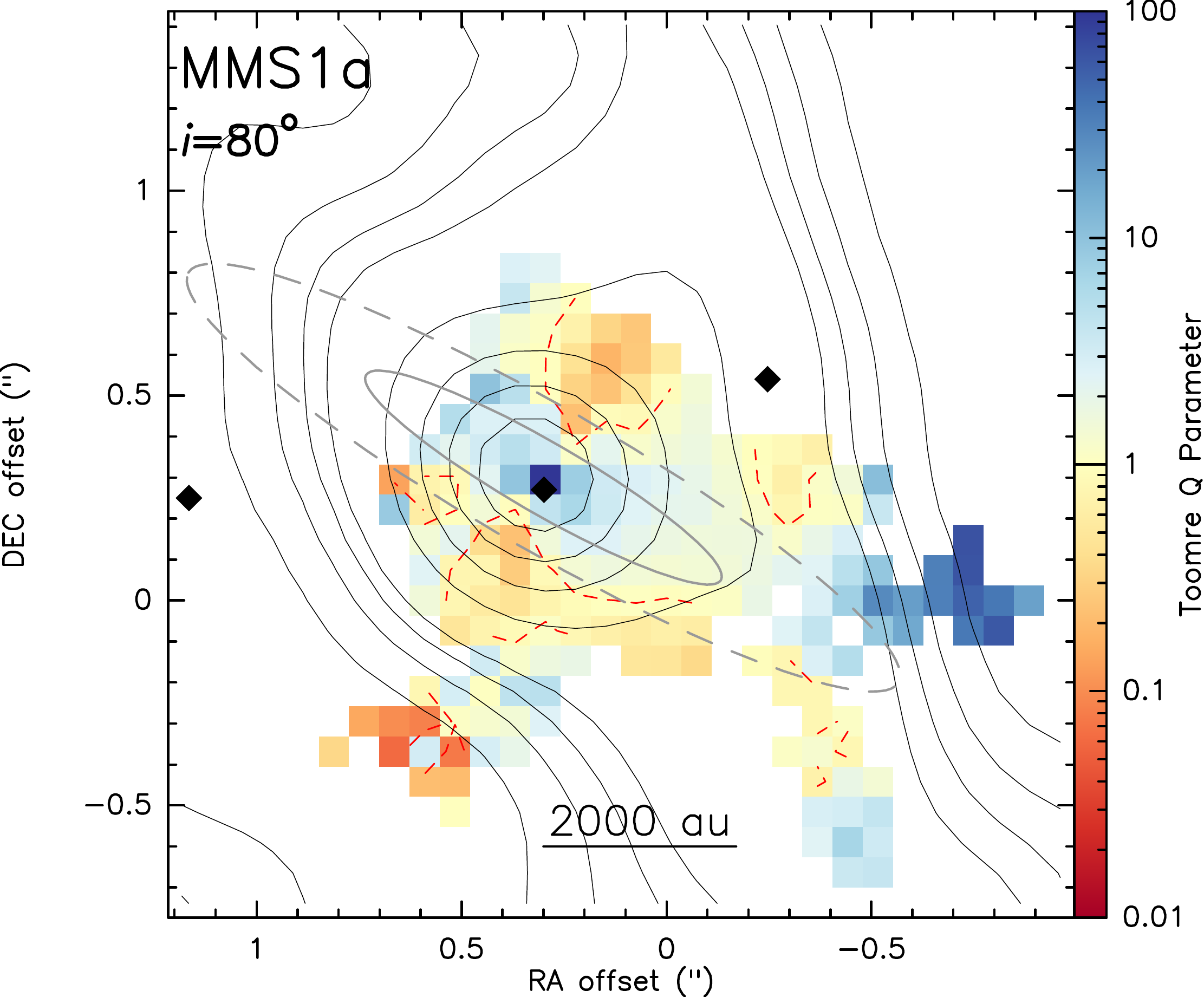}}
	\resizebox{.45\hsize}{!}{\includegraphics{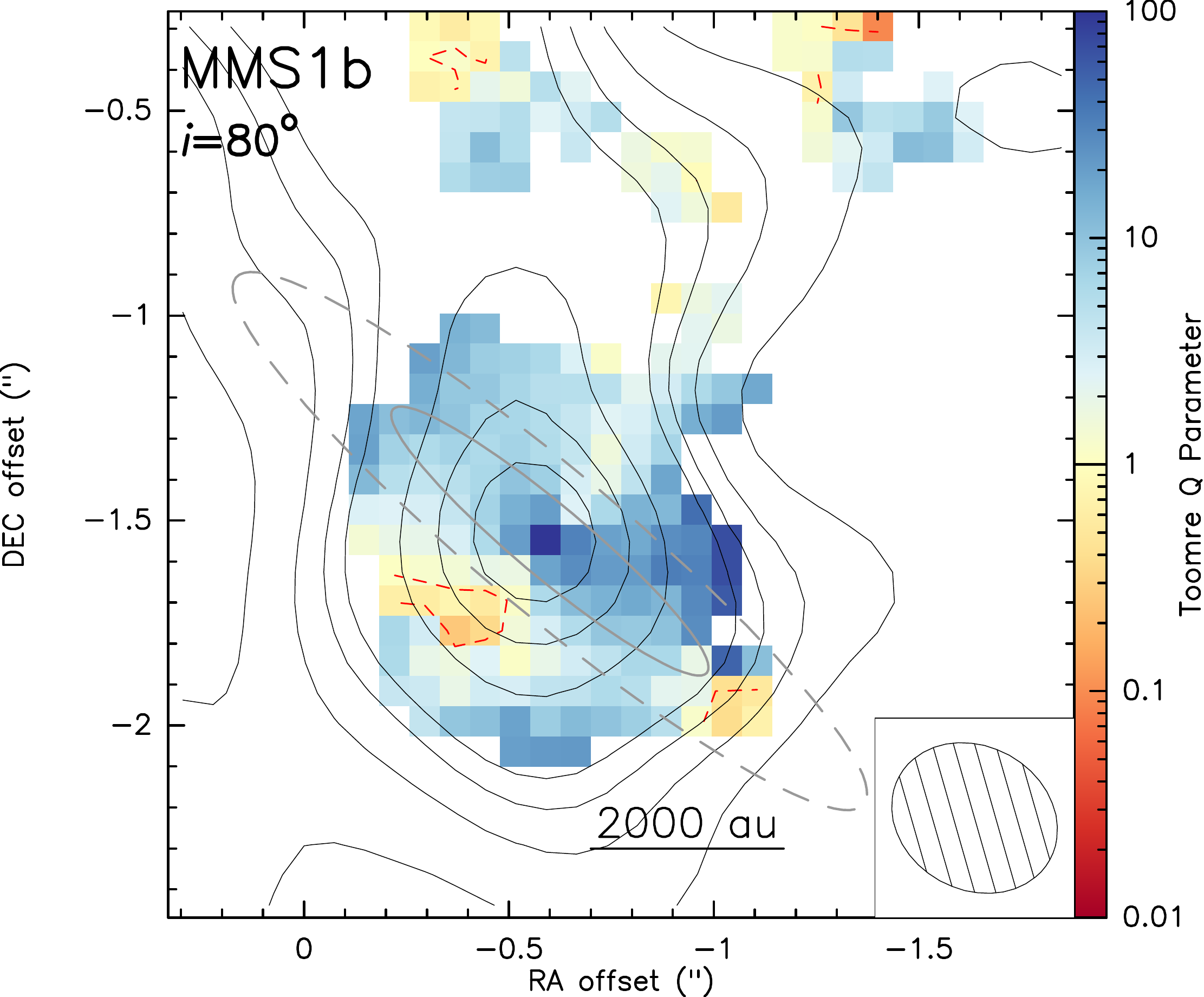}}
	\caption{Maps of Toomre's $ Q $ parameter towards MMS1a~\&~b. The color scale indicates the value of Toomre's $Q$ parameter, where blue is Toomre-stable, yellow is critical and red regions are gravitationally unstable against axisymmetric perturbations, where the level of $Q_\mathrm{crit}=1$ is indicated by the red dashed contours. The black contours indicate the 5, 10, 15, 20\,$\sigma$ continuum emission levels and increase further in steps of 20\,$\sigma$, where $\sigma$ = 0.28\,mJy\,beam$^{-1}$. The solid and dashed gray ellipses show the deprojected radii of the disks with diameter 4300\,au and 8600\,au, respectively. The black diamonds in the MMS1a panels mark the positions of the emission peaks at 3.6\,cm \citep{Rodriguez12}.
	The two \textit{top panels} show the $Q$ parameter maps for rotating disk structures seen almost face-on ($i=10\degr$) of the two cores. Motivated by the disk inclination estimate of $i_\mathrm{MMS1b} \sim 80\degr$ \citep{Rodriguez12}, we also present maps for such disk inclinations (\textit{bottom panels}) and intermediate values ($i=45\degr$, \textit{mid panels}). For the deprojection process, the angles of the disk major axes of $\phi_\mathrm{maj, MMS1a} \approx 240\degr$ and $\phi_\mathrm{maj, MMS1b} \approx 230\degr$ (relative to the north-south axis, see text) were used. The shaded ellipse in the lower right corner indicates the synthesized beam of the CH$_3$CN data of $0.42\arcsec \times 0.35\arcsec$ (PA~61\degr).}
	\label{fig:Toomre_Q}
\end{figure*}

In the next step, we plug all three maps into Eq.~\eqref{eq:Toomre_Q} and estimate a pixel-by-pixel map of Toomre's $ Q $ parameter for the two main mm~sources, as presented in Fig.~\ref{fig:Toomre_Q} for the three disk inclination angles $i = 10\degr, 45\degr$ and $80\degr$.
In the resulting Toomre $Q$ parameter maps (Fig.~\ref{fig:Toomre_Q}), the stable regions are colored blue, the critical values are shown in yellow and the unstable regions are shown in red. We note that the presented values are physically relevant only inside the dashed ellipses indicating the projected extent of the disk, as traced by rotation (cf. Fig.~\ref{fig:pv_diagrams}).

Considering the almost face-on disk scenario ($i=10\degr$, top panels in Fig.~\ref{fig:Toomre_Q}), the two cores already differ significantly. For the southern core MMS1b, the $Q$ values drop into the critical regime only in the outer parts of the candidate disk, see the solid and dashed ellipses indicating assumed disk radii of 2150 and 4300\,au, respectively. In contrast to this, the candidate disk around MMS1a shows $Q$ values in the yellowish critical regime of $ \sim 1$, especially the one towards the south-west, thereby indicating the potential of spiral arm formation.

Towards intermediate disk inclinations $i \sim 45\degr$, the $Q$ values tend to decrease globally, as the deprojection of the orbital distance and the HMPO mass cause the epicyclic frequency $\Omega_\mathrm{epi}$ to decrease towards larger orbital distances and especially faster along the disk minor axis due to the deprojection of $r$ (see mid panels in Fig.~\ref{fig:Toomre_Q}). In these maps, we identify the MMS1a disk to be essentially in a critical to unstable condition in the inner parts, and also for MMS1b we identify parts of the inner disk region to be partially in the critical regime.
However, the $Q$ parameters tend to increase again towards higher disk inclinations, $i \sim 80\degr$, as the deprojection of the column density reduces the disk surface density significantly, yielding higher values for $Q$. The maps are comparable to the low-$i$ maps as the region inside the inner ellipses only shows values in the critical to stable regime for MMS1b and, for MMS1a, we find a similar structure as for $i\sim10\degr$. However, due to the deprojection, most parts of both images are now located at radial distances beyond 4000\,au, making it unlikely that they are still part of a disk-like structure. We furthermore note that the outflow and the disk models for MMS1b of \citet{Rodriguez12} suggest the corresponding candidate disk to be highly inclined with $i\gtrsim80\degr$ and therefore make it likely that the candidate disk in this core is stable against such a fragmentation at the spatial resolution of our observations. Still, we note that the presented Toomre analysis is highly uncertain, especially towards nearly edge-on ($i > 80\degr$) disk scenarios due to the deprojection terms (see Eq.~\ref{eq:toomre_deprojection}).

We summarize that we find that only the rotating structure in MMS1a may be unstable to axisymmetric perturbations eventually forming spiral arm features and fragmenting due to local gravitational collapse under the assumptions that (1) it is a disk in equilibrium and in ordered, Keplerian-like rotation and (2) that the CH$_3$CN line emitting gas is well thermally coupled to the dust to trace well the dust temperature. However, though we identify potentially unstable regions, we do not find evidence for actually ongoing disk fragmentation in our data.

\subsection{On the origin of the cm~emission}
\label{sec:cm_emission}
\citet{Beuther02maser} and \citet{Rodriguez12} reported emission at 3.6\,cm towards MMS1a, with an angular resolution of $1.04\arcsec \times 0.62\arcsec$ and $0.31\arcsec \times 0.25\arcsec$, respectively. This radio emission structure is elongated in the east-west direction with a position angle of $\phi_\mathrm{cm} \approx 285\degr$ and \citeauthor{Rodriguez12} found it to be fragmented into three sources VLA1 -- 3. The intermediate source, VLA2, is coincident with the emission peak of MMS1a and \cite{Obonyo19} found that the spectral index of $\alpha_\mathrm{CQ}=0.33\pm0.14$ of this source matches well to thermal emission, where they derive the index by comparison of C (6\,cm) and Q-band (7\,mm) flux. The other two cm~sources VLA1 and 3, however, have slightly negative spectral indices of $\alpha_\mathrm{CQ}=-0.11\pm0.07$ and $-0.14\pm1.49$ \citep{Obonyo19}. The uncertainties are sufficiently high not to allow to solve the ambiguity whether these two sources trace an ionized jet \citep[as suggested by][]{Rodriguez12} or rather an ionized disk wind \citep[as suggested by the conversion of their flux densities to 8~GHz luminosities on the order of $\lesssim 1.1 \times 10^{12}$\,W\,Hz$^{-1}$ and a comparison of these values to other cm~sources from Fig.~6 of][]{Hoare07}. We note that these 8\,GHz luminosities of VLA1 are two orders of magnitude lower than typical \textsc{UCH\,ii} regions and thus make this option unlikely. Still, they indicate the presence of at least one high-mass YSO.

We furthermore note that the 3.6\,cm emission is not well aligned with the still uncertain outflow axis of MMS1a but well aligned with the velocity gradient seen in CH$_3$CN (see Fig.~\ref{fig:kinematics}). This suggests that either both trace circumstellar rotation with the cm~emission originating from the disk surface or both follow an acceleration, where the cm~emission originates from a jet and the CH$_3$CN gas is tracing potentially entrained gas \citep[as also found by][]{Leurini11,Busquet14,Palau17}. This scenario indicates that MMS1a may be launching two outflows, one seen in $^{13}$CO and the other one traced by the cm~emission, where this latter one is also capable of explaining the additional H$_2$ emission to the east, observed by \citet{Navarete15}. Even though the latter scenario appears to be the more likely one, we do not find clear evidence for excluding one of the two scenarios.

\subsection{Evolutionary stages}
\begin{figure*}
	\centering
	\resizebox{.9\hsize}{!}{\includegraphics{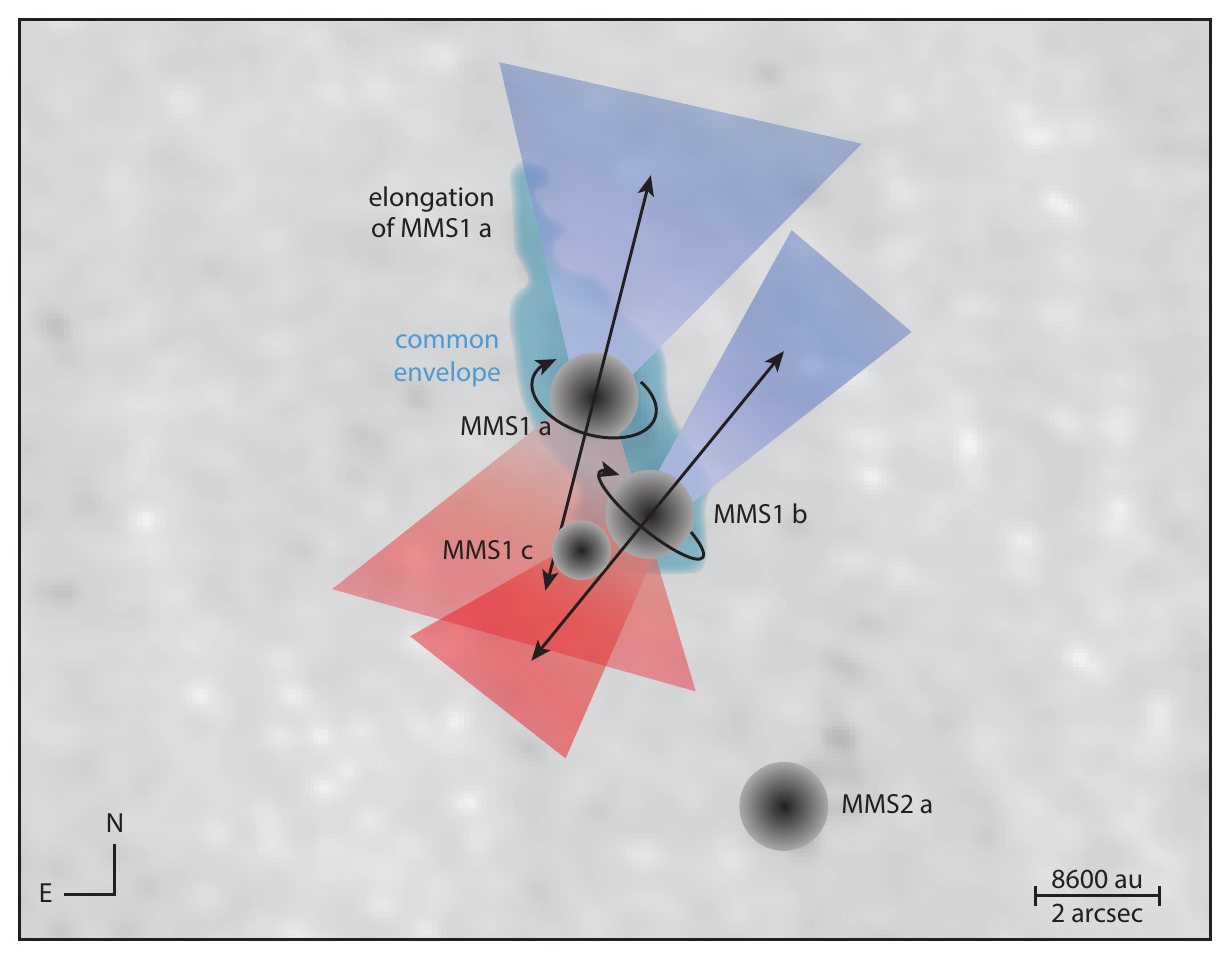}}
	\caption{Summary of the structure of IRAS~23033+5951. The fragmentation into four cores in two groups is depicted by the black spheres. Circumstellar rotation is indicated by arrows and the inferred outflow lobes are indicated by blue and red cones. The masses of the cores MMS1a--c and 2a are 23.5, 10.0, 1.2, and 11.4\,M$_\sun$. The three high-mass cores show evidence to be in more and more evolved stages, going from south to north.}
	\label{fig:summary}
\end{figure*}

Throughout the previous analysis, we found evidence for differences in the evolutionary stages of the three dominant cores. The chemical composition indicates an early, rather cool evolutionary stage of MMS2a, as we do not detect transition lines with upper state energies larger than 70\,K.
Along with missing signatures of outflowing material (cf. Sect.~\ref{sec:outflows_discussion} and the summary Fig.~\ref{fig:summary}), this suggests that MMS2a is still in a cooler, deeply embedded stage than the other two sources and presumably in the earliest evolutionary stage of the three major cores.

In contrast to this, we detect CH$_3$CN transitions towards MMS1b with upper state energy levels $\gtrsim 100$\,K. This core shows a richer chemical composition with N- and S-bearing molecules and a probable disk-outflow system, where the outflow is collimated, suggesting a stable circumstellar disk or rotating structure. \cite{Rodriguez12} traced this disk structure at the scales <~200\,au with H$_2$O (22\,GHz) maser emission.

The third core, MMS1a, again differs significantly from the other two as there is cm~emission and no maser emission. Even though we cannot clearly distinguish whether the cm~emission stems from a jet or disk winds, this detection suggests that the YSO within this core is in a later evolutionary stage than those in the other two cores. The detection of CH$_3$CN is weaker in this source than towards MMS1b presumably due to radiation destroying the more complex molecules, but may as well be explained by an overall lower core temperature.

The above mentioned sequence is reminiscent of the situation in the star-forming clump NGC 7538S \citep{Feng16}, one of the pathfinder sources for the CORE program. NGC 7538S was also resolved into three cores along one main direction with an apparent evolutionary age gradient.

Finally, we discuss the cores in IRAS~23033+5951 in the context of the evolution of outflow structure. In their Fig.~4, \citet{BeutherShepherd05} suggested that the appearance of outflows evolves along with the following two observables of the HMPO, the evolution of the surrounding \textsc{H\,ii} region, and the spectral type of the HMPO, where less compact \textsc{H\,ii} regions and earlier spectral types tend to have less collimated outflow cavities. This evolution, i.e. the outflow broadening with time is confirmed by simulations \citep[e.g.][]{Peters11,Vaidya11,Kuiper15,Kuiper16,Kuiper18,Kolligan18}, showing that strong, ionizing stellar flux makes it less likely to build up a stable disk and thereby a stable magnetic field which is crucial to the launching process of a jet. The low collimation of the molecular outflow may also be due to unresolved binaries or multiple-star systems, in which the massive bodies disturb the outflow-launching region with the same result \citep{Peters14}. These authors furthermore find that the collimated outflows from a group of YSOs, which share similar outflow axes due to the conservation of angular momentum from earlier phases, eventually add up to form what is observed as a poorly collimated outflow. Thereby, the finding of this outflow structure may point towards the presence of multiple YSOs in a given core.
In Sect.~\ref{sec:outflows_discussion}, we assign two outflows, one to each of the mm~sources MMS1a~\&~b. We discussed above that MMS1a with the less collimated large-scale outflow and with cm~emission is likely in a more evolved stage than the YSO(s) within MMS1b. This is consistent with the outflow evolution with time or evolutionary stage as suggested by \cite{BeutherShepherd05}.

\section{Conclusion}
\label{sec:conclusion}
We studied the high-mass star-forming region IRAS~23033+5951, addressing open questions to the formation of high-mass stars. As part of the CORE survey, the region was observed with {NOEMA} in the A, B and D configurations with short spacings from the {IRAM} 30-m single-dish telescope, in the 220\,GHz or 1.37\,mm spectral window.

The dust continuum emission was used for estimating core masses and H$_2$ column densities, and the molecular gas emission for analyzing the kinematic components of the cloud. For instance, the $^{13}$CO (2--1) transition was used for investigating the outflow structure and the H$_2$CO (3$_{0,3}$ -- 2$_{0,2}$) transition was used for studying rotating structures. The masses of embedded protostars were investigated by analyzing PV diagrams across these rotating objects. For the two most massive cores, we carried out radiative transfer modeling of a methyl cyanide ($12_K - 11_K$) cube with \textsc{xclass} and the resulting maps of rotational temperature were used for the calculation of pixel-by-pixel Toomre $ Q $ parameter maps.

The 1.37\,mm dust continuum observations reveal hierarchical fragmentation of the parental cloud into a group of at least four mm~sources, where the inferred structure is summarized in Fig.~\ref{fig:summary} along with rotation and outflow features of particular cores. The source MMS1c is low-mass ($1.2$\,M$_\sun$) and an order of magnitude less dense than the other cores ($3.2 \times 10^{23}$\,cm$^{-2}$). The other three pre- or protostellar cores have H$_2$ column densities from 1.62 to $4.28\times 10^{24}$\,cm$^{-2}$ and dominate the mm~emission. They indicate three different evolutionary stages within one maternal gas clump:

\begin{itemize}
	\item Towards the southern core MMS2a ($M=11.4$\,M$_\sun$), we detect only emission of spectral lines excited at low temperatures (< 70\,K). We identify no significant rotation of the core and no connection to a molecular outflow. This suggests that the core is in the earliest evolutionary stage of the three most massive cores.
	\item Compared to this, MMS1b ($M=10.0$\,M$_\sun$) is warmer, as spectral lines are detected with upper energy levels $\gtrsim$ 100\,K, and chemically rich, as also N- and S-bearing species are detected. The molecular emission indicates core rotation and the PV diagram shows a Keplerian-like rotation profile tracing a candidate disk, which is essentially Toomre-stable against gravitational fragmentation at the spatial scales traced by our observations. Furthermore, this object drives a rather collimated outflow, seen in $^{13}$CO and SO.
	\item The analysis of the presumably most evolved core, MMS1a ($M=23.5$\,M$_\sun$), suggests that the central object(s) is (are) driving an outflow with a wide opening angle, indicating the presence of one or more HMPOs. Associated with this core, the cm~emission likely indicates non-thermal emission in the close vicinity of the HMPO(s). The candidate disk indicates Toomre-instability in the inner 2000\,au.
\end{itemize}

The spatial resolution of the presented observations $\sim0.45\arcsec \approx 1900$\,au cannot reveal the fragmentation at smaller scales, relevant for the analysis of the typically high multiplicity of high-mass stars.
The results of this work support the picture that high-mass stars form by similar, but scaled-up processes, as their low-mass counterparts, where the mass accretion onto the protostars continues via disk-like rotating structures and the outflows decollimate with protostellar evolution.

\begin{acknowledgements}
We thank the referee for the suggestions, which significantly clarified the analysis and structure of the paper.
FB, HB, AA and JCM acknowledge support from the European Research Council under the Horizon 2020 Framework Program via the ERC Consolidator Grant CSF-648505.
Furthermore, the team thanks IRAM and its staff for conducting the large program CORE.
RK acknowledges financial support via the Emmy Noether Research Group on Accretion Flows and Feedback in Realistic Models of Massive Star Formation funded by the German Research Foundation (DFG) under grant no. KU 2849/3-1 and KU2849/3-2.
DS acknowledges support by the Deutsche Forschungsgemeinschaft through SPP 1833: ``Building a Habitable Earth'' (SE 1962/6-1).
ASM is partially supported by the German Research Foundation (DFG) through grant SFB956 (subproject A6).
AP acknowledges financial support from UNAM-PAPIIT IN113119 grant, M\'exico.
\end{acknowledgements}

\bibliographystyle{aa}
\bibliography{astronomy}

\begin{appendix}
\section{Position-velocity diagram fits}
\label{sec:discussion_pv_fit}

\subsection{Estimating extreme velocities}
The Keplerian velocity profile is described by Eq.~\eqref{eq:kepler_velocity} and it is an upper limit to the line-of-sight velocity at a given radial distance from the center of rotation, the supposed protostellar object. \citet{Seifried16} present an approach of utilizing this relation for an estimate of the protostellar mass. They compare different methods for estimating the maximum velocity at a given distance to the protostellar object from PV diagrams and find that the most robust one is the following (see their section 4.2):
\begin{enumerate}
	\item Estimate the noise level $\sigma$.
	\item Estimate the two (opposing) quadrants with the strongest emission.
	\item Iterate over positions.
	\begin{enumerate}
		\item Begin with the channel of highest (lowest) velocity while being in the quadrant of positions with higher (lower) velocities than the $v_\mathrm{LSR}$.
		\item Iterate towards lower (higher) velocity channels until the first pixel with flux above a chosen threshold, e.g. $4\,\sigma$, is found.
		\item Add the PV coordinates of this pixel to a list (for later analysis).
	\end{enumerate}
\end{enumerate}

\subsection{Fitting a Keplerian velocity profile}
We follow this method and collect a set of radial positions and corresponding maximum velocities, which we pass to a \textsc{python} least-squares optimization function.
To account for the uncertainties in the core position and the systemic velocity $v_\mathrm{LSR}$, we expand Eq.~\eqref{eq:kepler_velocity} by the positional shift $r_0$ and the velocity shift $v_0 = v_\mathrm{LSR}$. Furthermore, we introduce a sign-function, to account for the opposite velocity difference from the $v_\mathrm{LSR}$ of the blue and red-shifted emission on the respective sides of the emission peak, and a "$\pm$" representing the sign change due to the respective positive or negative velocity offset at the starting position. Thus we yield the following expression:

\begin{equation}
	v_\pm(r, M_\star, r_0, v_0) = \pm\mathrm{sign}(r - r_0) \cdot \sqrt{\frac{GM_\star}{\vert r - r_0 \vert}} + v_0
	\label{eq:Kepler_fit_function}
\end{equation}

We utilize the \textsc{python} package \textsc{astropy.modeling}\footnote{\textsc{python} package \textsc{astropy.modeling}, \url{http://docs.astropy.org/en/stable/modeling/}} \citep{Astropy13} and create a custom model for $v(r, M_\star, r_0, v_0)$ from Eq.~\eqref{eq:Kepler_fit_function}. This model is fitted to the PV data via Levenberg-Marquardt least-squares fitting, within the  \textsc{python} package \textsc{KeplerFit}\footnote{\textsc{KeplerFit}, \url{https://github.com/felixbosco/KeplerFit}}. We collect the best-fit parameters, the respective standard deviations and the integrated $\chi^2$ as a measure of the residuals.

\subsection{Analysis of uncertainties}
The outcome of this fitting procedure depends strongly on the input parameters. Therefore, we tested how the detection threshold, the different chemical species, the size of a flagged central region, and the imaging of the data themselves affect the fit estimates.

\subsubsection{Effect of detection threshold and chemical species}
\label{sec:test_chemical_species}
\begin{figure}
	\centering
	\resizebox{.9\hsize}{!}{\includegraphics{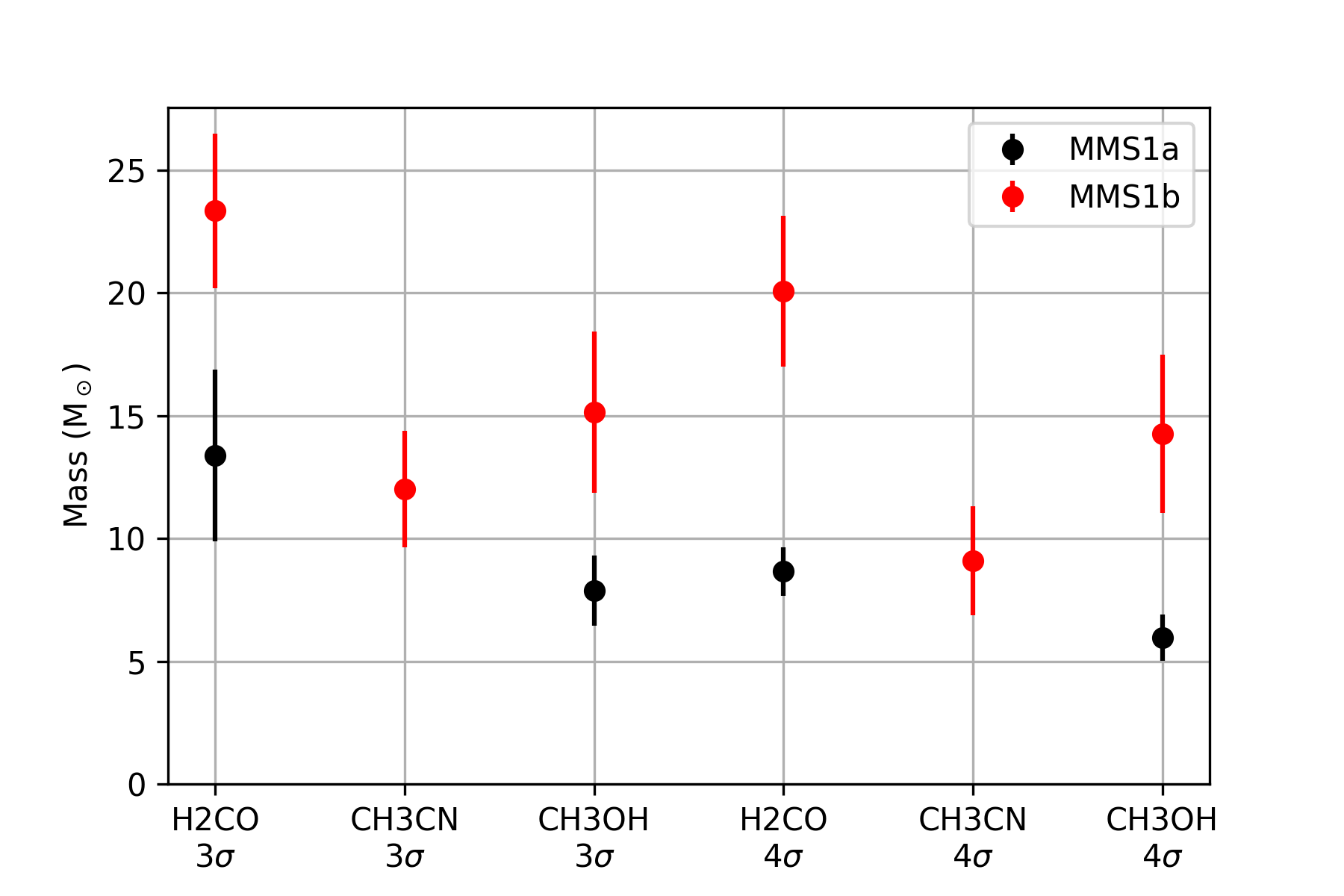}}
	\resizebox{.9\hsize}{!}{\includegraphics{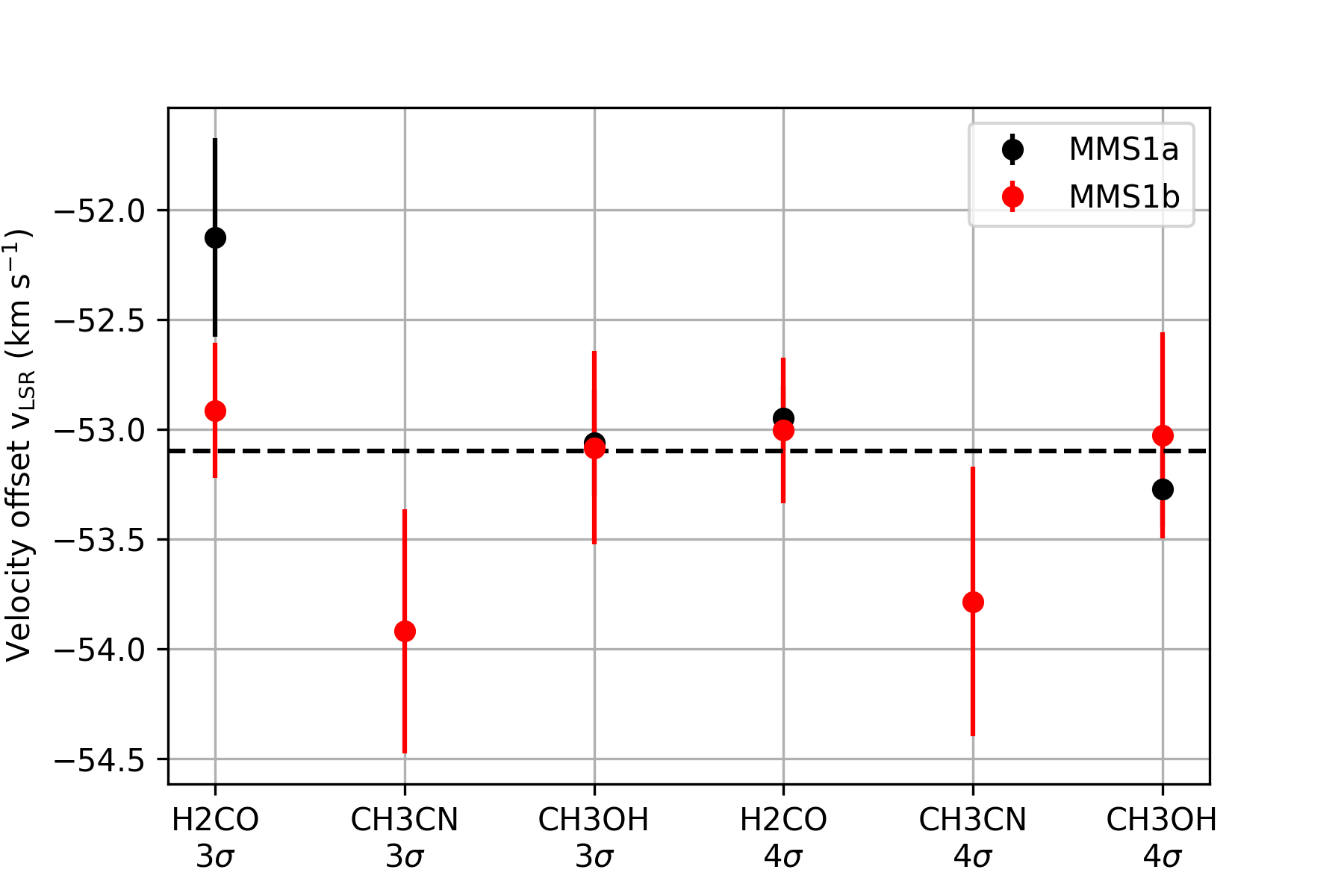}}
	\caption{Best-fit estimates of kinematic mass (\emph{top}) and systemic velocity (\emph{bottom}) for thresholds of 3 and $4\,\sigma$ for the YSO mass and velocity ($v_\mathrm{LSR}$), respectively. The error bars are the standard deviations from the lest-squares fits. The horizontal line in the velocity plot is the systemic velocity of the source IRAS~23033+5951.}
	\label{fig:test_threshold}
\end{figure}

The estimate of the extreme velocities is dependent on the detection threshold. Therefore, if we choose a lower detection threshold (e.g. 3$\sigma$) the algorithm estimates higher relative velocity offsets and the derived YSO mass is expected to be higher. In contrast to this, the absolute velocity offset $v_\mathrm{LSR}$ is supposed to be the same, since the relative velocity offsets should increase symmetrically. We see this effect in the parameters in Fig.~\ref{fig:test_threshold}, where the mass estimates for the $4\,\sigma$ detection threshold are smaller or similar to the corresponding $3\,\sigma$ estimates. The absolute velocity offsets agree well in general but, for MMS1a, we see a $\Delta v \sim 0.8$\,km\,s$^{-1}$ in the H$_2$CO transition. This difference may be due to the strong asymmetry between blue and red shifted emission for MMS1a.

Different molecular transition lines usually trace different regimes of density and temperature, see, for instance, Table~\ref{tab:obs_spec_lines}. Therefore, we compare the fit of the H$_2$CO (3$_{0,3}$ -- 2$_{0,2}$) low-temperature gas tracing line to the CH$_3$CN ($12_3$ -- $11_3$) high-temperature gas tracing line. We expect, that the H$_2$CO is detected towards larger distances from the YSO, whereas the CH$_3$CN line should trace the closer rotational structure. We see this detection constraint in the results, where CH$_3$CN gives mass estimates a factor of $\sim 2$ lower than the corresponding H$_2$CO fit. This is expected to be due to the overall lower signal-to-noise ratio in the CH$_3$CN data.

\subsubsection{Effect of flagging the innermost data points}

The PV diagrams in Fig.~\ref{fig:pv_diagrams} do not show the highest velocities close to the center of rotation which are expected from the assumption of Keplerian rotation. This may be due to filtering effects during the observation \citep{Krumholz07}. A second explanation for this non-detection may be the suppression of the highest velocity due to a presence of an unresolved binary or multiple YSO system \citep{Peters14} or due to higher optical thickness in line or continuum emission towards the central region. Therefore, we test how flagging these innermost data points affects the outcome of the fit, applying a detection threshold of $4\,\sigma$.

\begin{figure}
	\centering
	\resizebox{.9\hsize}{!}{\includegraphics{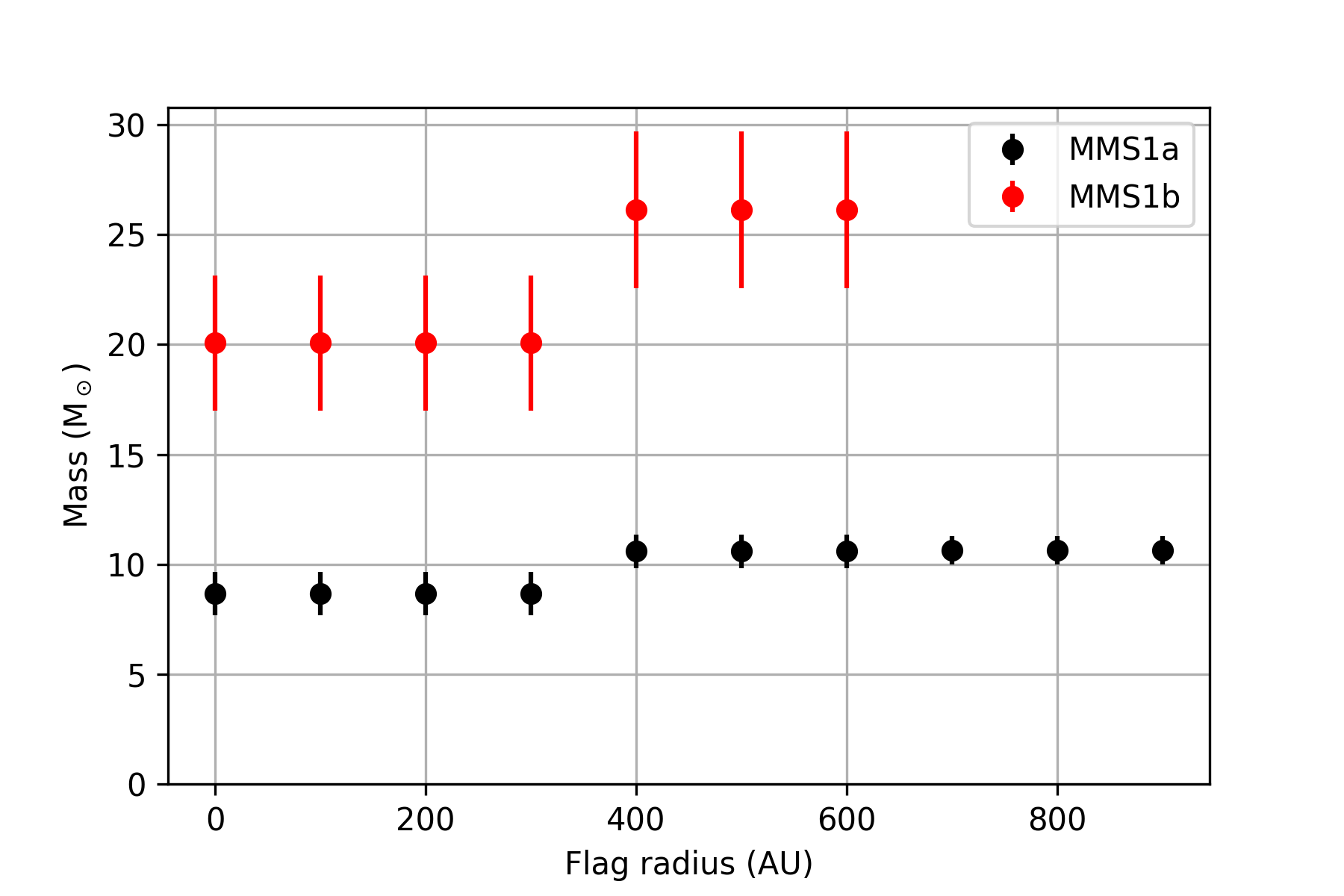}}
	\resizebox{.9\hsize}{!}{\includegraphics{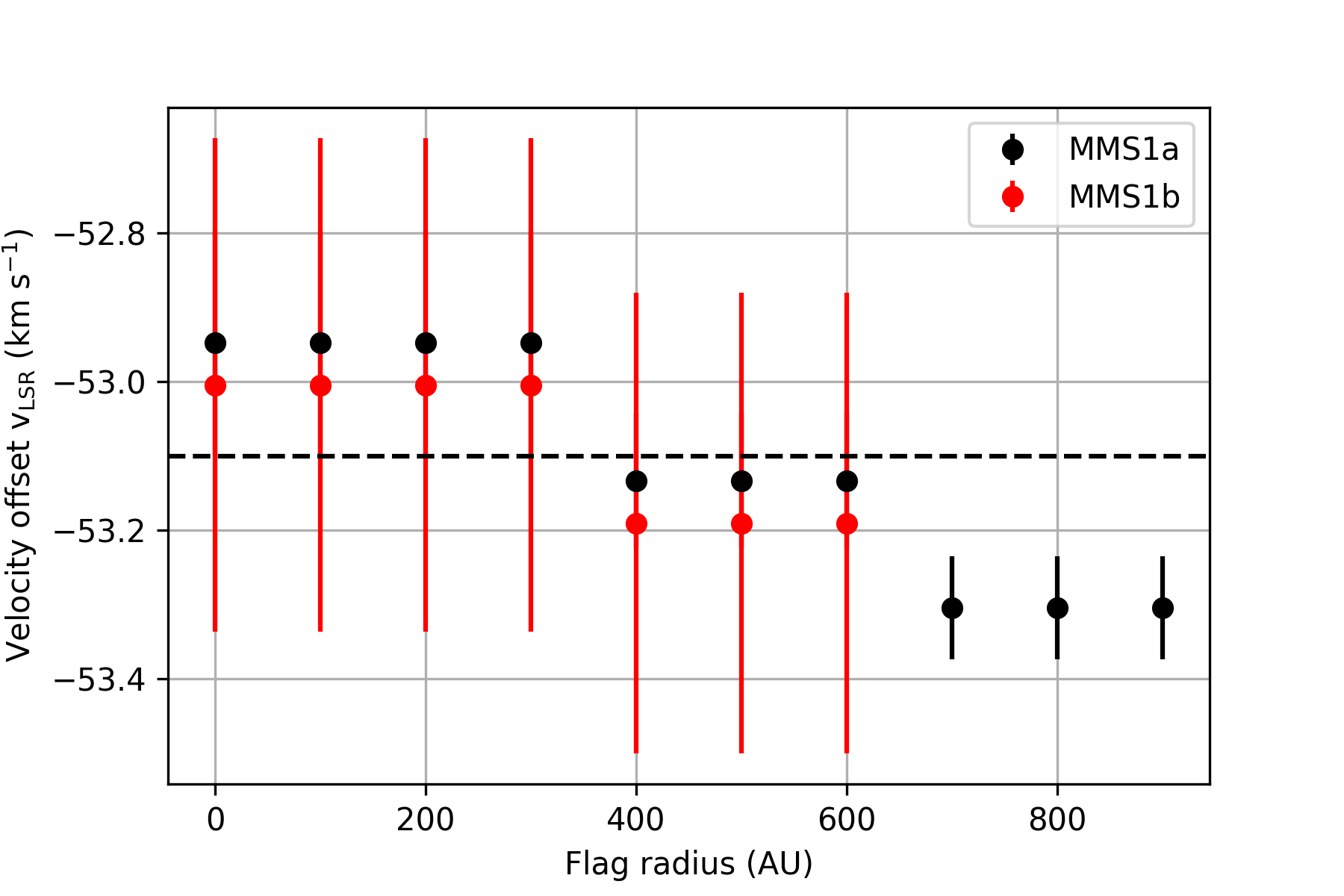}}
	\resizebox{.9\hsize}{!}{\includegraphics{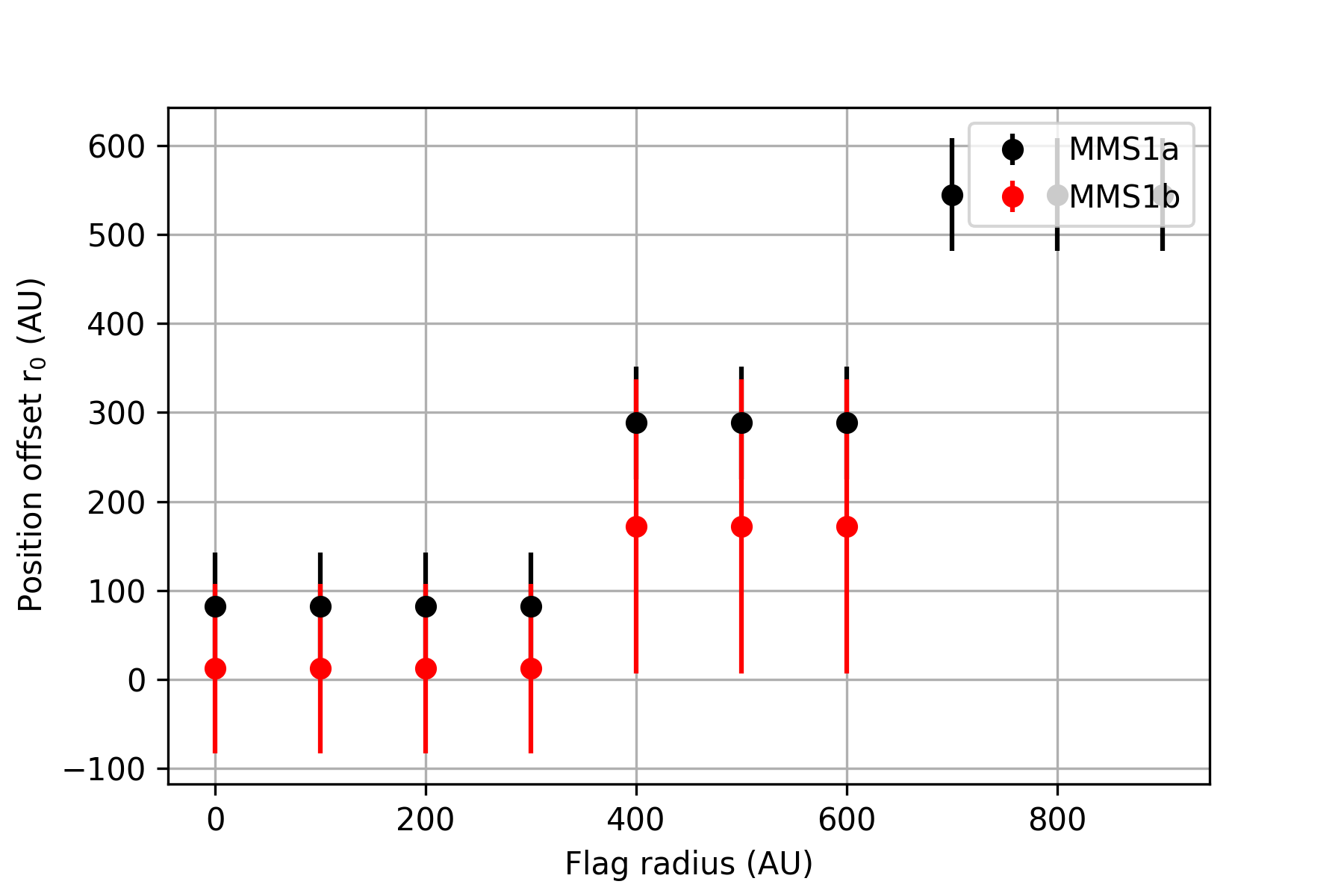}}
	\resizebox{.9\hsize}{!}{\includegraphics{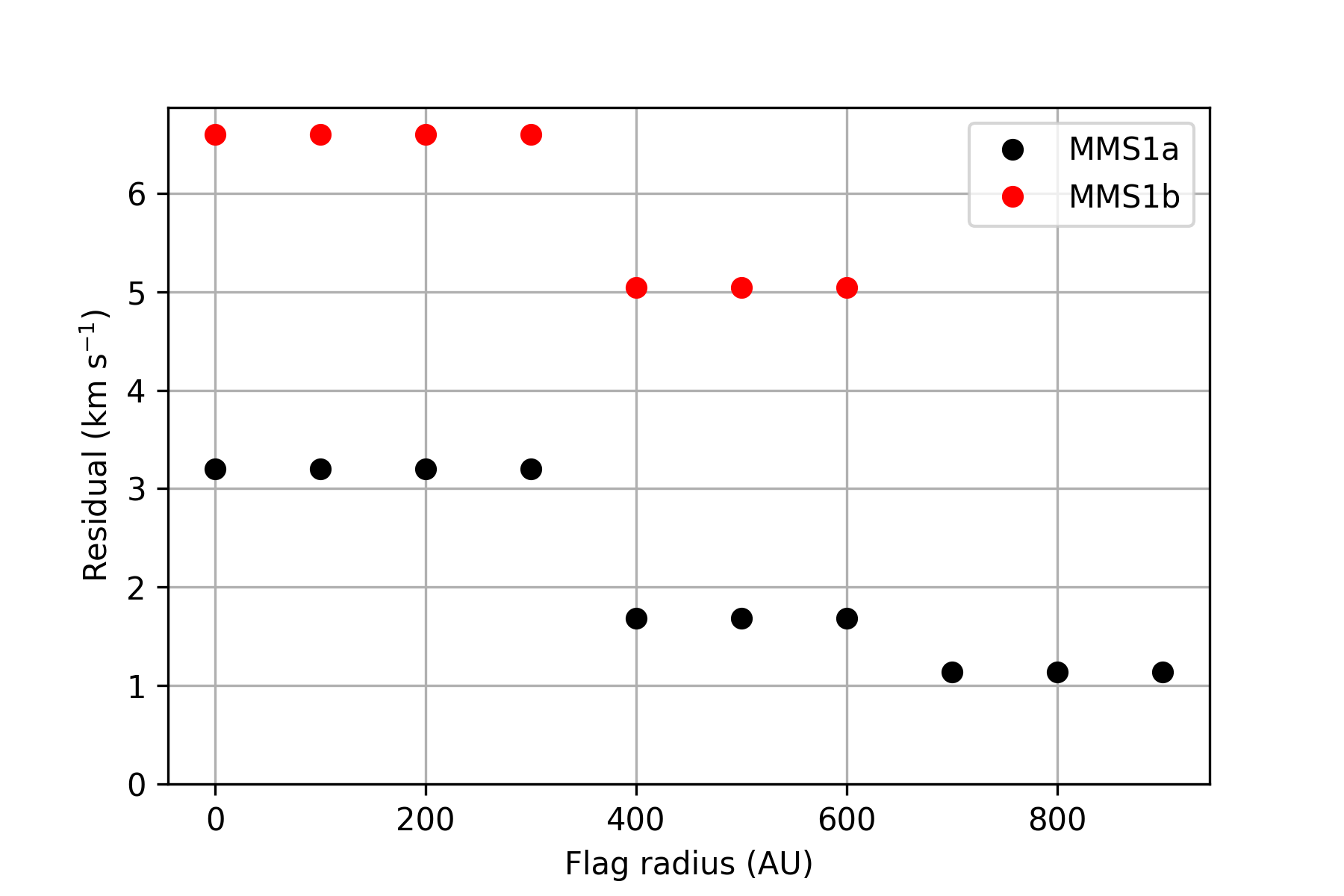}}
	\caption{Best fit results for the tests of flagging the inner region around the YSO. The routine was applied to the PV diagrams from the H$_2$CO (3$_{0,3}$ -- 2$_{0,2}$) image. The error bars are the standard deviations from the lest-squares fits. The horizontal line in the velocity plot is the systemic velocity of the source IRAS~23033+5951.}
	\label{fig:test_flag}
\end{figure}

While this yields a significant decrease in the fit residual (bottom panel in Fig.~\ref{fig:test_flag}), there is a trend of the mass estimate for both cores, as we obtain a higher mass estimate ($\gtrsim 20\%$). The velocity offset differs only on the order of 0.4\,km\,s$^{-1}$ and the position offset $r_0$ varies well below the resolution limit $\sim 1900$\,au. Therefore, we suggest to flag the innermost data points in the case of a constant extreme velocity in the center, while keeping in mind that the mass estimate from the non-flagged fit yields a lower limit.

\subsubsection{Effect of the merging and imaging process}

The H$_2$CO (3$_{0,3}$ -- 2$_{0,2}$) is also covered in the IRAM 30-m single dish data. Therefore, we are able to test how the merged data (in different weighting schemes) affects the outcome of the fit. In general, more natural weighting schemes result in larger beams sizes, which smooths out the resulting image, and in a higher signal-to-noise ratio. On the other hand, the more extended and lower-density structures are represented by the shorter baselines which are better represented in the more natural weighting scheme. We compare the results from three weighting schemes: natural and robust with the two weight thresholds of 1 and 0.1 (corresponding to uniform weighting) for both, the interferometric-only data set and the merged data set. The size of the synthesized beams and the rms noise are given in Table~\ref{tab:image_parameters} for all of these images.

\begin{table}
	\centering
	\caption{Parameters for synthesized beam and rms noise of the images from different weighting schemes and the merged data set. \emph{I} and \emph{S} represent the interferometric and the single-dish data, respectively.}
	\label{tab:image_parameters}
	\begin{tabular}{lcrr}
		\hline \hline
		Data & Weighting scheme & Beam  & rms noise \\
		& & ($\arcsec \times \arcsec$) &  (mJy\,beam$^{-1}$) \\
		\hline
		I+S & natural & 1.18 $\times$ 0.82 & 7.40 \\
		I+S & robust~1 & 0.52 $\times$ 0.44& 6.27 \\
		I+S & robust~0.1 & 0.43 $\times$ 0.36 & 6.92 \\
		I & natural & 0.72 $\times$ 0.65 & 6.79 \\
		I & robust~1 & 0.47 $\times$ 0.43 & 5.42 \\
		I & robust~0.1 & 0.43 $\times$ 0.35 & 5.57\\
		\hline
	\end{tabular}
\end{table}

\begin{figure}
	\centering
	\resizebox{.9\hsize}{!}{\includegraphics{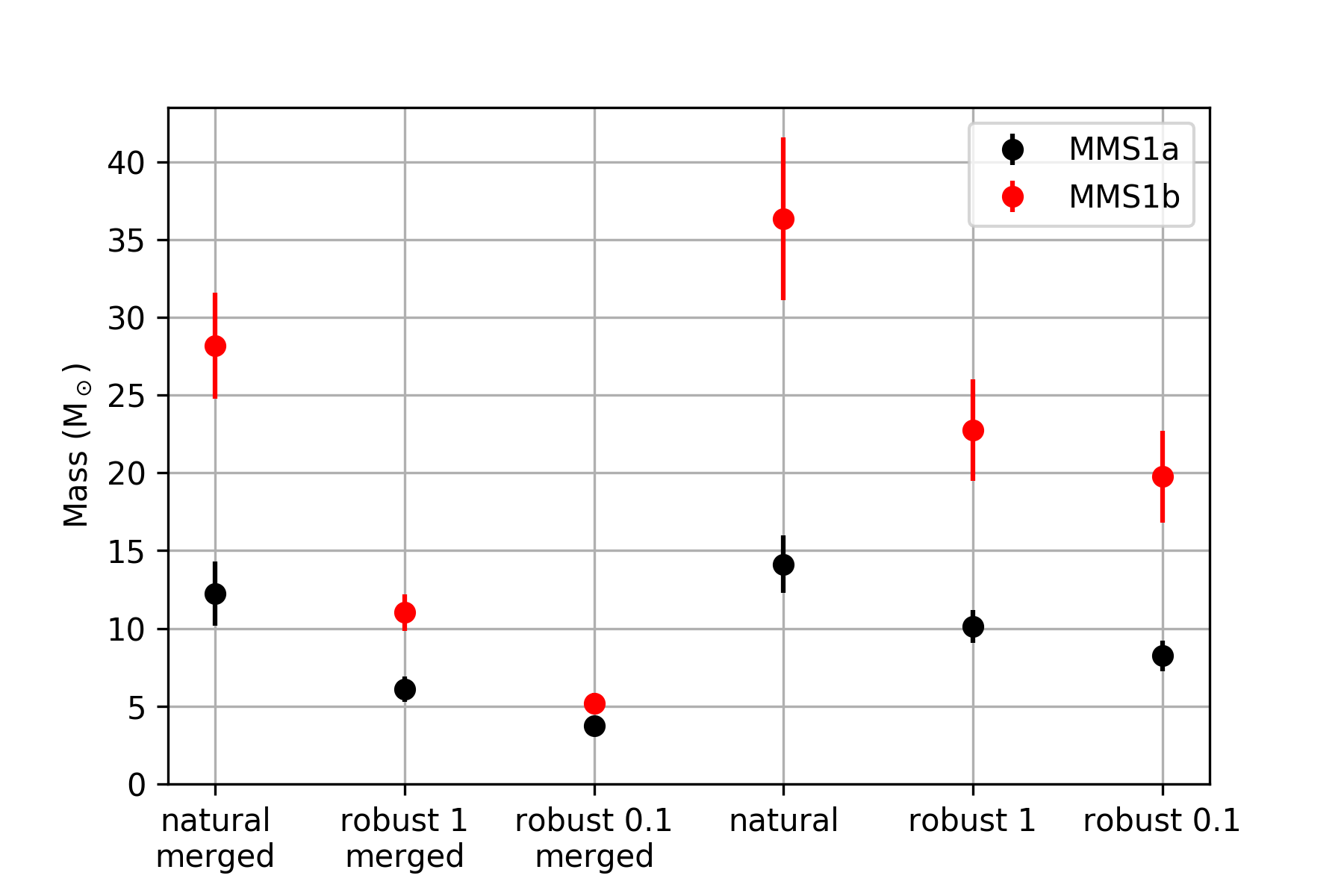}}
	\resizebox{.9\hsize}{!}{\includegraphics{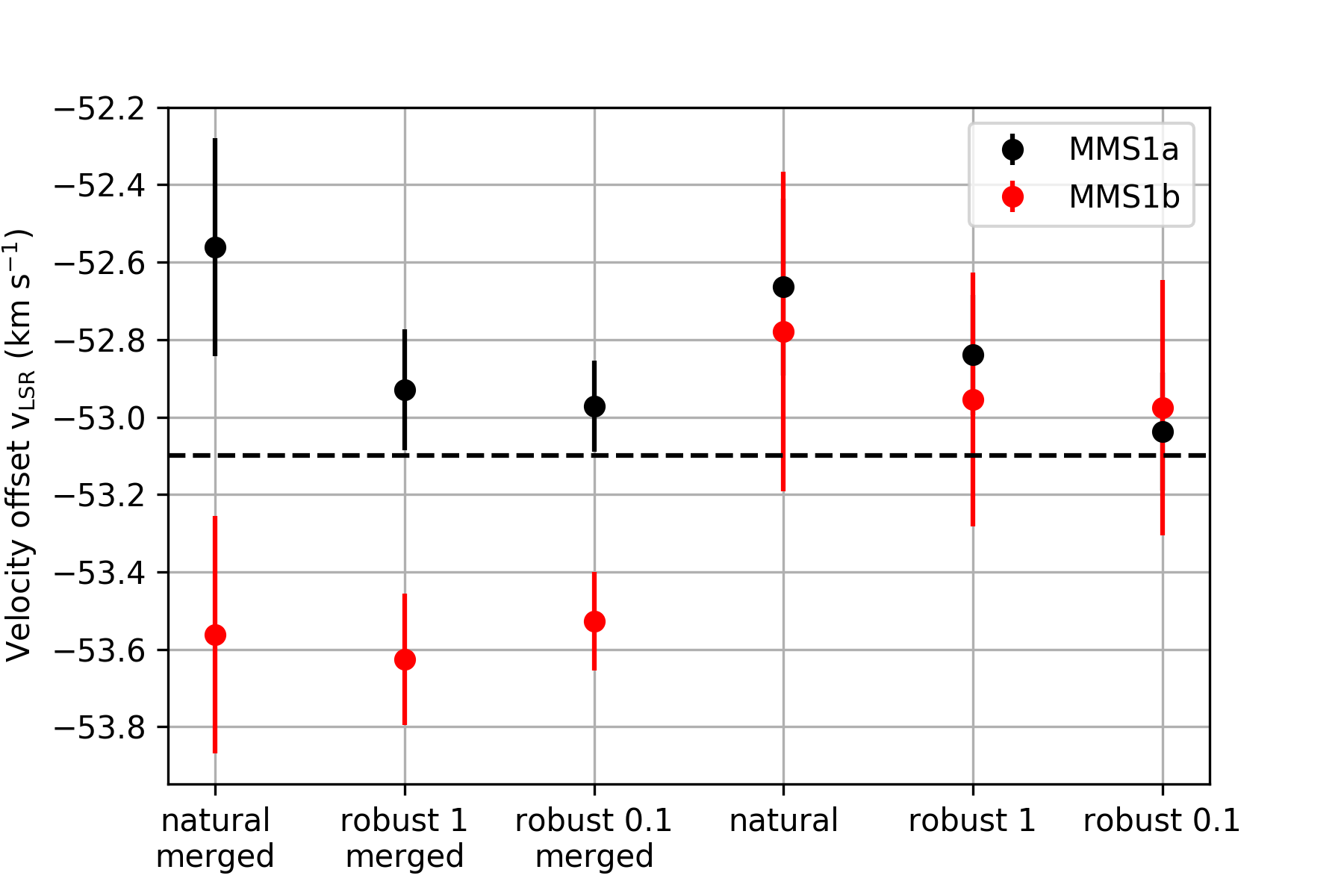}}
	\resizebox{.9\hsize}{!}{\includegraphics{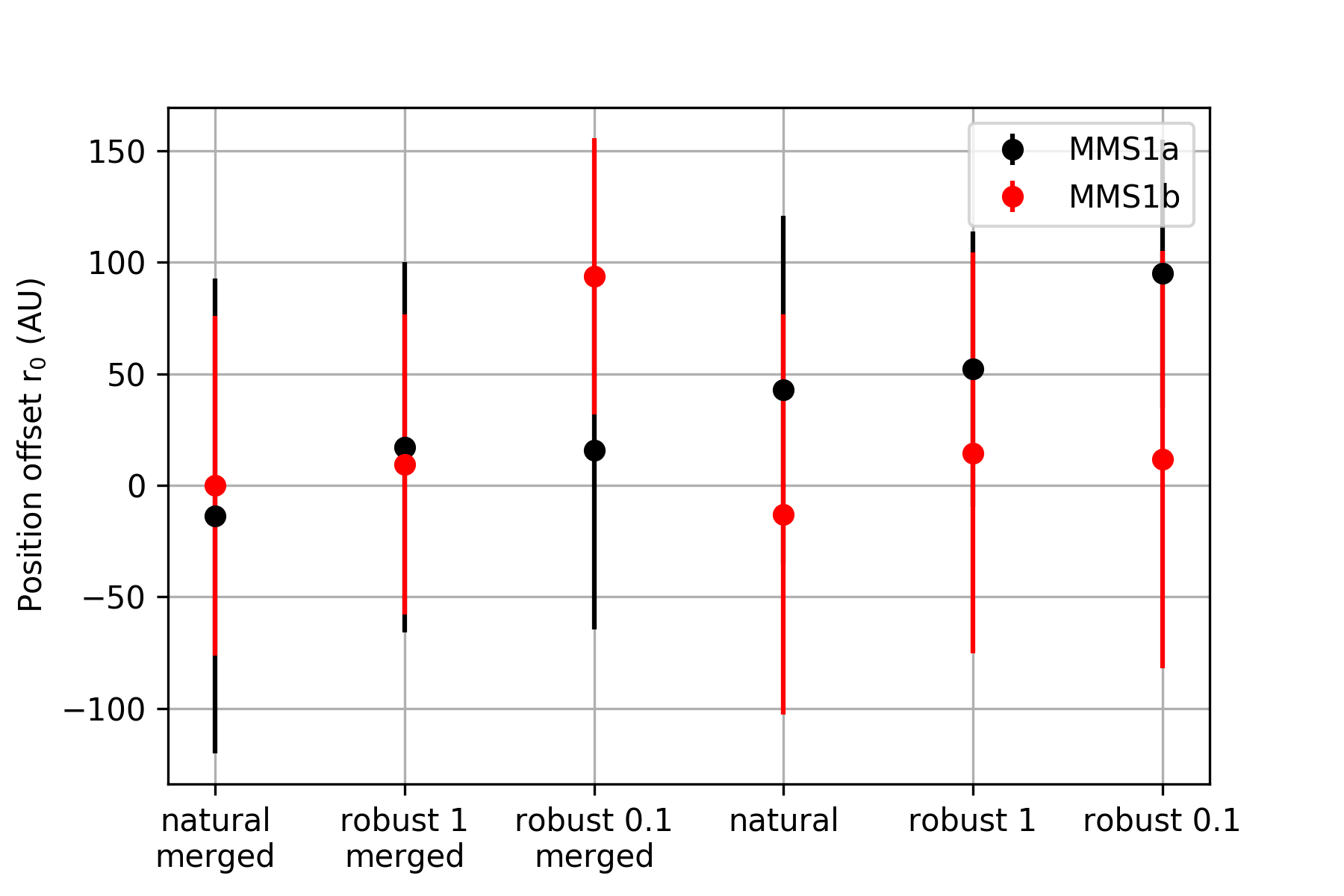}}
	\resizebox{.9\hsize}{!}{\includegraphics{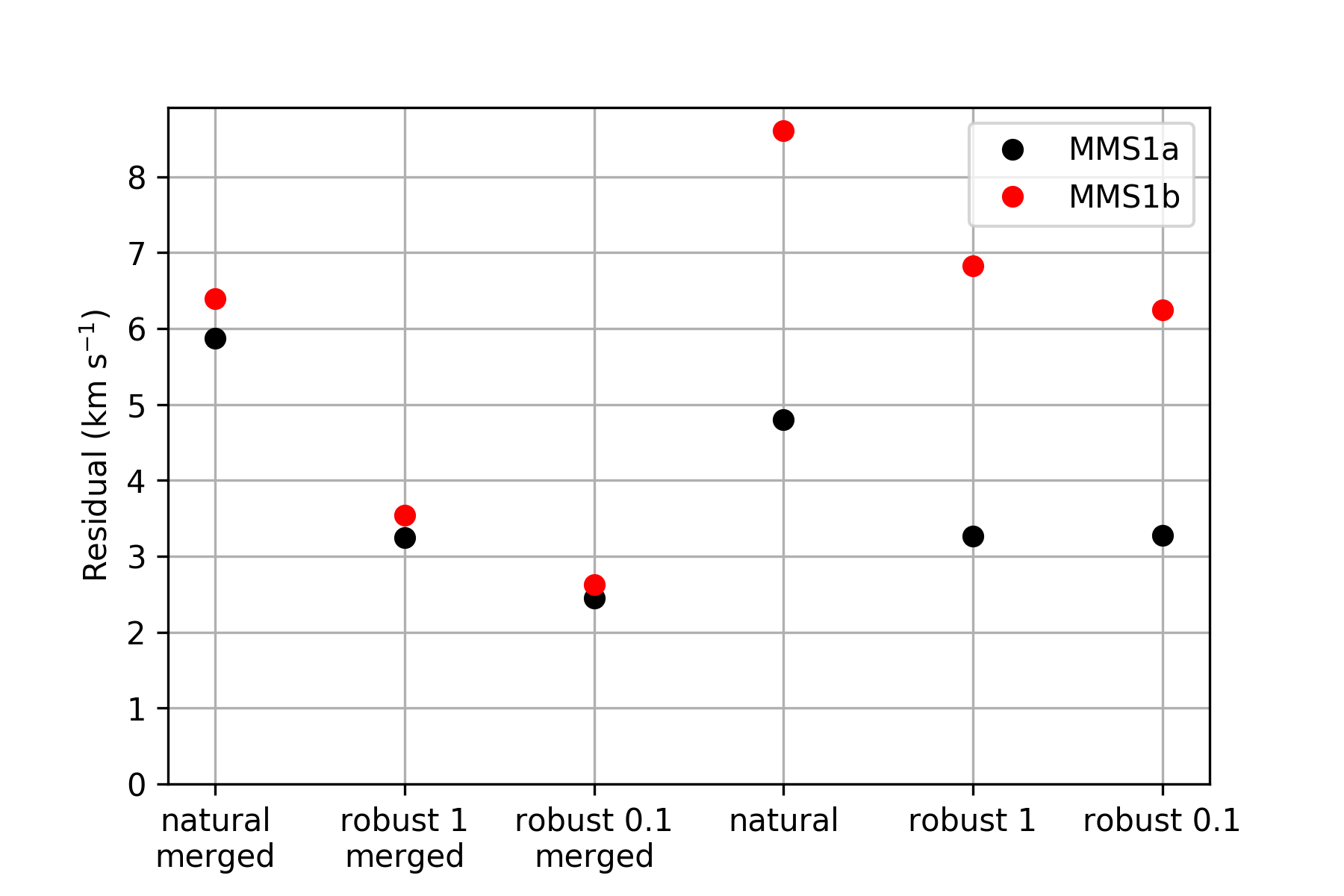}}
	\caption{Best fit results for PV diagrams from different weighting schemes. The results are presented for three weighting parameters of the merged data set and for the interferometric data. Within both series, the weighting schemes are natural, robust with a weight threshold of 1 and robust with a weight threshold of 0.1 (uniform), see Table~\ref{tab:image_parameters}. The error bars are the standard deviations from the least-squares fits. The horizontal line in the velocity plot is the systemic velocity of the source IRAS~23033+5951.}
	\label{fig:test_weighting_schemes}
\end{figure}

The results of this comparison are presented in Fig.~\ref{fig:test_weighting_schemes}. We see the expected trend from the variation of the weighting schemes, where the flux filtering effect and the lower signal-to-noise ratio should be the main reasons for the lower mass estimates towards the more robust weighting schemes. In the fits to the merged data sets, we usually do not identify the highest velocities close to the candidate YSO, since the flux is smoothed over the positions. This results in lower mass estimates when comparing the merged to the interferometric-only data.
In contrast to this, the deviations of the position and velocity offsets are weak and show no trend. Both stay well below or close to the resolution limits of 1900\,au and 0.5\,km\,s$^{-1}$. In the plot of the residuals, however, we see a clear trend towards lower $\chi^2$ values for the more robust weighting schemes, which is due to the detection of the higher velocities towards the center, whereas for the naturally weighted images, this non-detection yields a significant deviation between data and fit, close to $r_0$.

We suggest to prefer a robust weighting scheme to a natural one, since the Keplerian velocity distribution is only seen in the (dense-material tracing) robust weighting schemes. This yields a lower limit which we find to be a factor of $3\pm1.5$ below the corresponding larger estimates from the other weighting schemes.

\subsection{Dependency on disk inclination}
\label{sec:disk_inclination}
From the velocity field in the candidate disks, only the radial component is traced in red and blue shifted fractions of the emission lines. Therefore, the inclination acts as a limit to the mass estimate described above. In this section we derive, how this estimate depends on the disk inclination $i$ relative to the line of sight. For consistency, we will stick to the definition from Sect.~\ref{sec:disk_stability} and define the inclination of the disk to be $0\degr$ if the disk is seen face-on. In this case, the radial component $v_\mathrm{rad}$ of the velocity field decreases to almost zero, since the Keplerian velocity component acts only in the plane of rotation:

\begin{equation}
	v_\mathrm{rad}(r) = \cos(i) \cdot v_\perp(r) + \sin(i) \cdot v_\parallel(r) ,
\end{equation}
where the component $v_\parallel = v_\mathrm{Kepler}$ is the Keplerian velocity component in the disk plane and where the perpendicular component assumed to be negligible ($v_\perp \approx 0$). Thus, we indeed measure $M_\mathrm{fit} = M \cdot \sin^2 i$ with that method:

\begin{align}
	v_\mathrm{rad} &\approx \sin(i) \, \cdot v_\mathrm{Kepler} = \sin(i) \, \cdot \sqrt{\frac{GM_\star}{r}} \\
	v_\mathrm{rad} &\approx  \sqrt{\frac{G(M_\star \cdot \sin^2 i)}{r}}
	\label{eq:disk_inc_pa0}
\end{align}
We note that this mass scaling factor should not make a big difference for the estimate of protostellar mass for MMS1b, since the high-disk inclination causes only a factor of $\sin^2(80\degr) \approx 0.97$. For the northern source MMS1a, however, the mass may be underestimated by a factor of $1/\sin^2(45\degr)=2$ or even more.

\begin{align}
	M_\star &= M_{\star, \mathrm{fit}} / \sin^2 i \\
	M_{\star, \mathrm{MMS1b}} &= 18.8 \, \mathrm{M}_\sun / \sin^2(80\degr) = 19.4 \, \mathrm{M}_\sun \\
	M_{\star, \mathrm{MMS1a}} &\sim 5.8 \, \mathrm{M}_\sun / \sin^2(45\degr) = 11.6 \, \mathrm{M}_\sun
\end{align}

We summarize that this method yields only a lower limit to the YSO mass for cases in which the disk inclination is not known.

\section{Toomre $Q$ maps}
\label{sec:appendix_ToomreQ}

In this section, we describe in detail, how we compute the parameter maps for the Toomre $Q$ analysis taking into account the a priori unknown disk inclination.

The computation of disk surface density $\Sigma$ and epicyclic frequency $\Omega_\mathrm{epi}$ maps is less straight-forward than for the map for the speed of sound, as both quantities are affected by the yet poorly constrained disk inclination $i$, where we define $i$ to have value $0\degr$ for disks seen face-on and $90\degr$ for disks seen edge-on. The mass estimate to be plugged into Eq.~\eqref{eq:epicyclic_frequency} relates to the inclination $i$ as $ M_\star = M_\mathrm{fit} / \sin^2 i$. On the other hand, the projected orbital distance $r^2_\mathrm{proj} = \Delta x ^2 + \Delta y ^2$ to the YSO at the mm emission peak in the image can be computed from the distances $\Delta x$ and $\Delta y$ in pixel coordinates which are converted into a projected physical distance by the pixel size and the source distance of 4300\,pc.
\begin{figure}
	\centering
	\resizebox{.75\hsize}{!}{\includegraphics{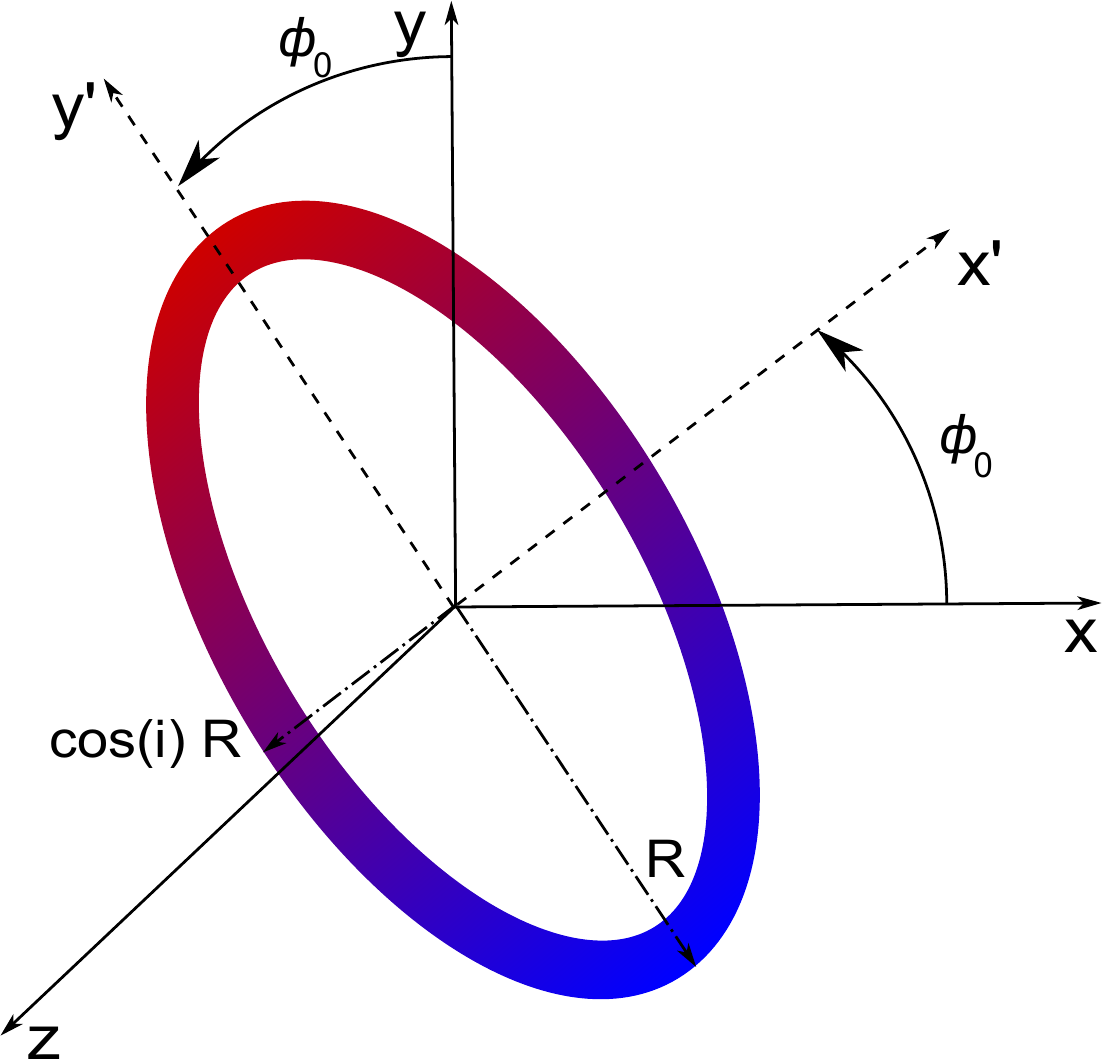}}
	\caption{Sketch of the inclined disk ellipse model for the initialization of the map of the deprojected radius, which was used to get a map of the epicyclic frequency $\Omega_\mathrm{epi}$.}
	\label{fig:inclined_disk_model}
\end{figure}
We use a disk model as sketched in Fig.~\ref{fig:inclined_disk_model}, where the position angle $\phi_0$ of the disk major axis is given with respect to the north-south axis ($y$-axis) and counterclockwise. We compute the deprojected orbital distance by scaling up the distance in the direction perpendicular to the disk major axis, $r_\mathrm{proj}^2(i, \phi') = r^2 \cdot (\cos^2 \phi' + \sin^2 \phi' \cdot \cos^2 i)$, where $\phi'$ is the position angle in the rotated system, with $\phi'_0 \equiv 0$. This expression yields $r=r_\mathrm{proj}$ for $i = 0\degr$ and diverges for $i\sim 90\degr$ towards directions perpendicular to the disk major axis, i.e. towards $\phi' \sim 90\degr$.

The computation of a map for the disk column density is uncertain as we cannot distinguish the contributions from the protostellar envelope and the disk material to the observed column density $\Sigma_\mathrm{obs}$. At this point we assume that the largest fraction originates from the rotating disk material. However, we note that this remains a source of large uncertainty in our disk stability analysis, as the $Q$ parameter increases with smaller disk column density, where this would result from removing the protostellar envelope contribution. We also note that the column density traces the disk column density best in a face-on disk scenario. We address this deprojection by applying the low-order approximation $\Sigma_\mathrm{disk} \approx \cos i \cdot \Sigma_\mathrm{obs}$, which was found to be a fair approximation for disks with typical density vertical and horizontal density structures \citep{Kee18}.
For large disk inclinations, however, the columns trace a range of disk radii at once and therefore only allow for rough estimates of the local disk column density, presumably rendering the analysis impossible for disks seen under an inclination $i > 80\degr$.

Summarizing the above considerations regarding the effect of the inclination $i$, we obtain the following relation:
\begin{align}
	Q_\mathrm{obs} &= \frac{c_\mathrm{s} \cdot \Omega_\mathrm{epi,obs}(i, \phi')}{\pi G \cdot \Sigma_\mathrm{obs}(i)} \\
    &= \frac{c_\mathrm{s} \cdot \Omega_\mathrm{epi}}{\pi G \cdot \Sigma}
    	\cdot \frac{\sin i \cdot \left( \cos^2 \phi' + \sin^2 \phi' \cos^2 i \right)^{-3/4}}{\cos^{-1} i} \\
    &= Q_\mathrm{real} \cdot \frac{\sin i \cdot \cos i}{\left( \cos^2 \phi' + \sin^2 \phi' \cos^2 i \right)^{3/4}}
     \label{eq:toomre_deprojection}
\end{align}

\end{appendix}

\end{document}